%% file: cafa2.tex
\documentclass[12pt]{article}
\usepackage[top=1in,bottom=1in,left=1in,right=1in]{geometry}

\usepackage{amsthm,amsmath}
\usepackage[utf8]{inputenc} 

\usepackage{graphicx}
\usepackage{lmodern}
\usepackage{multirow}
\usepackage{bbm}
\usepackage{units}
\usepackage{url}
\usepackage{authblk}
\usepackage{subfigure}
\usepackage{longtable}

\newcommand{\fmax}{F_{\max}}
\newcommand{\smin}{S_{\min}}

\begin{document}
\input{authors}

\title{An expanded evaluation of protein function prediction methods shows an improvement in accuracy}

\maketitle

\input{abstract.tex}

%



\input{introduction.tex}

\input{methods.tex}

\input{results.tex}

\input{conclusions.tex}


\section*{Competing interests}
The authors declare that they have no competing interests.

\section*{Authors' contributions}
PR and IF conceived of the CAFA experiment, supervised the project and significantly contributed to writing of the manuscript. YJ performed most analyses and significantly contributed to writing. IF, PR, CSG, WTC, ARB, DD and RL contributed to the analyses. SDM managed data acquisition. TRO developed the web interface, including the portal for submission and the storage of predictions. RPH, MJM and CO'D directed the biocuration efforts. EC-U, PD, REF, RH, DL, RCL, MM, ANM, PM-M, KP and AS performed biocuration. YM and PNR co-organized the human phenotype challenge. ML, AT, PCB, SEB, CO and BR steered the CAFA experiment and provided critical guidance. The remaining authors participated in the experiment, provided writing and data for their methods and contributed comments on the manuscript.

\section*{Acknowledgements}
We acknowledge the contributions by Maximilian Hecht, Alexander Grün, Julia Krumhoff, My Nguyen Ly, Jonathan Boidol, Rene Schoeffel, Yann Spöri, Jessika Binder, Christoph Hamm and Karolina Worf. This work was partially supported by the following grants: National Science Foundation (NSF) grants DBI-1458477 (PR), DBI-1458443 (SDM), DBI-1458390 (CSG), DBI-1458359 (IF), IIS-1319551 (DK), DBI-1262189 (DK) and DBI-1149224 (JC); National Institutes of Health (NIH) grants R01GM093123 (JC), R01GM097528 (DK), R01GM076990 (PP), R01GM071749 (SEB), R01LM009722 (SDM) and UL1TR000423 (SDM); the National Natural Science Foundation of China 3147124 (WT) and 91231116 (WT); the National Basic Research Program of China 2012CB316505 (WT); NSERC RGPIN 371348-11 (PP); FP7 ``infrastructures" project TransPLANT Award 283496 (ADJvD); Microsoft Research/FAPESP grant 2009/53161-6 and FAPESP fellowship 2010/50491-1 (DCAeS); Biotechnology and Biological Sciences Research Council (BBSRC) grants BB/L020505/1 (DTJ), BB/F020481/1 (MJES), BB/K004131/1 (AP), BB/F00964X/1 (AP) and BB/L018241/1 (CD); Spanish Ministry of Economics and Competiveness, grant number BIO2012-40205 (MT); KU Leuven CoE PFV/10/016 SymBioSys (YM); the Newton International Fellowship Scheme of the Royal Society grant NF080750 (TN). CSG was supported in part by the Gordon and Betty Moore Foundation's Data-Driven Discovery Initiative through Grant GBMF4552. Computational resources were provided by CSC — IT Center for Science Ltd, Espoo, Finland (TS). This work was supported by the Academy of Finland (TS). RCL, ANM were supported by British Heart Foundation grant RG/13/5/30112. PD, RCL, REF were supported by Parkinson's UK grant G-1307. Alexander von Humboldt Foundation through German Federal Ministry for Education and Research and Ernst Ludwig Ehrlich Studienwerk (ELES). Ministry of Education, Science and Technological Development of the Republic of Serbia (Grant no. 173001). This work was a Technology Development effort for ENIGMA - Ecosystems and Networks Integrated with Genes and Molecular Assemblies (http://enigma.lbl.gov), a Scientific Focus Area Program at Lawrence Berkeley National Laboratory is based upon work supported by the U.S. Department of Energy, Office of Science, Office of Biological \& Environmental Research (DE-AC02-05CH11231). ENIGMA only covers the application of this work to microbial proteins.

\section*{Additional Files}
    This submission contains three supplementary resources. (1) The Supplementary Information below provides additional CAFA2 analyses that are equivalent to those provided for the CAFA1 experiment in the CAFA1 supplement; (2) the large data repository provides all additional data, analyses as well as full prediction results for every method; (3) the entire library of code used in CAFA2 is available at \url{https://github.com/yuxjiang/CAFA2}.

  \subsection*{Supplementary Information}
    The section below contains supplementary figures, analyses, as well as information related to participating methods.

  \subsection*{Additional data}
   The following contains additional data analyses and raw data: \url{https://dx.doi.org/10.6084/m9.figshare.2059944}


\newpage
\input{suppl3.tex}

\bibliographystyle{bmc-mathphys} 

\bibliography{refdb}

\end{document}

%% file: authors.tex
\author[1]{Yuxiang Jiang}
\author[2]{Tal Ronnen Oron}
\author[3]{Wyatt T Clark}
\author[4]{Asma R Bankapur}
\author[5]{Daniel D'Andrea}
\author[5]{Rosalba Lepore}
\author[6]{Christopher S Funk}
\author[7]{Indika Kahanda}
\author[8,9]{Karin M Verspoor}
\author[7]{Asa Ben-Hur}
% ...
\affil[1]{Department of Computer Science and Informatics, Indiana University, Bloomington, IN, USA}
\affil[2]{Buck Institute for Research on Aging, Novato, CA, USA}
\affil[3]{Department of Molecular Biophysics and Biochemistry, Yale University, New Haven, CT, USA}
\affil[4]{Department of Microbiology, Miami University, Oxford, OH, USA}
\affil[5]{University of Rome, ``La Sapienza'', Rome, Italy}
\affil[6]{Computational Bioscience Program, University of Colorado School of Medicine, Aurora, CO, USA}
\affil[7]{Department of Computer Science, Colorado State University, Fort Collins, CO, USA}
\affil[8]{Department of Computing and Information Systems, University of Melbourne, Parkville, Victoria, Australia}
\affil[9]{Health and Biomedical Informatics Centre, University of Melbourne, Parkville, Victoria, Australia}
% ...
\setcounter{Maxaffil}{0}
\renewcommand\Authfont{\small}
\renewcommand\Affilfont{\itshape\footnotesize}

%% file: abstract.tex
\textbf{Background} 
The increasing volume and variety of genotypic and phenotypic data is a major defining characteristic of modern biomedical sciences. At the same time, the limitations in technology for generating data and the inherently stochastic nature of biomolecular events have led to the discrepancy between the volume of data and the amount of knowledge gleaned from it. A major bottleneck in our ability to understand the molecular underpinnings of life is the assignment of function to biological macromolecules, especially proteins. While molecular experiments provide the most reliable annotation of proteins, their relatively low throughput and restricted purview have led to an increasing role for computational function prediction. However, accurately assessing methods for protein function prediction and tracking progress in the field remain challenging.

\textbf{Methodology} 
We have conducted the second Critical Assessment of Functional Annotation (CAFA), a timed challenge to assess computational methods that automatically assign protein function. One hundred twenty-six methods from 56 research groups  were evaluated for their ability to predict biological functions using the Gene Ontology and gene-disease associations using the Human Phenotype Ontology on a set of 3,681 proteins from 18 species. CAFA2 featured significantly expanded analysis compared with CAFA1, with regards to data set size, variety, and assessment metrics. To review progress in the field, the analysis also compared the best methods participating in CAFA1 to those of CAFA2.

\textbf{Conclusions} 
The top performing methods in CAFA2 outperformed the best methods from CAFA1, demonstrating that computational function prediction is improving. This increased accuracy can be attributed to the combined effect of the growing number of experimental annotations and improved methods for function prediction. The assessment also revealed that the definition of top performing algorithms is ontology specific, that different performance metrics can be used to probe the nature of accurate predictions, and the relative diversity of predictions in the biological process and human phenotype ontologies. While we have observed methodological improvement between CAFA1 and CAFA2, the interpretation of results and usefulness of individual methods remain context-dependent.






%% file: introduction.tex
\section*{Introduction}
Computational challenges in the life sciences have a successful history of driving the development of new methods by independently assessing performance and providing discussion forums for the researchers \cite{Costello2013}. In 2010-2011, we organized the first Critical Assessment of Functional Annotation (CAFA) challenge to evaluate methods for the automated annotation of protein function and to assess the progress in method development in the first decade of the 2000s \cite{Radivojac2013}. The challenge used a time-delayed evaluation of predictions for a large set of target proteins without any experimental functional annotation. A subset of these target proteins accumulated experimental annotations after the predictions were submitted and was used to estimate the performance accuracy. The estimated performance was subsequently used to draw conclusions about the status of the field.


The CAFA1 experiment showed that advanced methods for the prediction of Gene Ontology (GO) terms \cite{Ashburner2000} significantly outperformed a straightforward application of function transfer by local sequence similarity. In addition to validating investment in the development of new methods, CAFA1 also showed that using machine learning to integrate multiple sequence hits and multiple data types tends to perform well. However, CAFA1 also identified nontrivial challenges for experimentalists, biocurators and computational biologists. These challenges include the choice of experimental techniques and proteins in functional studies and curation, the structure and status of biomedical ontologies, the lack of comprehensive systems data that is necessary for accurate prediction of complex biological concepts, as well as limitations of evaluation metrics \cite{Radivojac2013, Dessimoz2013, Gillis2013, Schnoes2013, Jiang2014}. Overall, by establishing the state-of-the-art in the field and identifying challenges, CAFA1 set the stage for quantifying progress in the field of protein function prediction over time.

In this study, we report on the major outcomes of the second CAFA experiment (CAFA2) that was organized and conducted in 2013-2014, exactly three years after the original experiment. We were motivated to evaluate the progress in method development for function prediction as well as to expand the experiment to new ontologies. 
The CAFA2 experiment also greatly expanded the performance analysis to new types of evaluation and included new performance metrics. 


%% file: methods.tex
\section*{Methods}

\subsection*{Experiment overview}

\begin{figure}[h!]
  \centering
  \includegraphics[width=\textwidth]{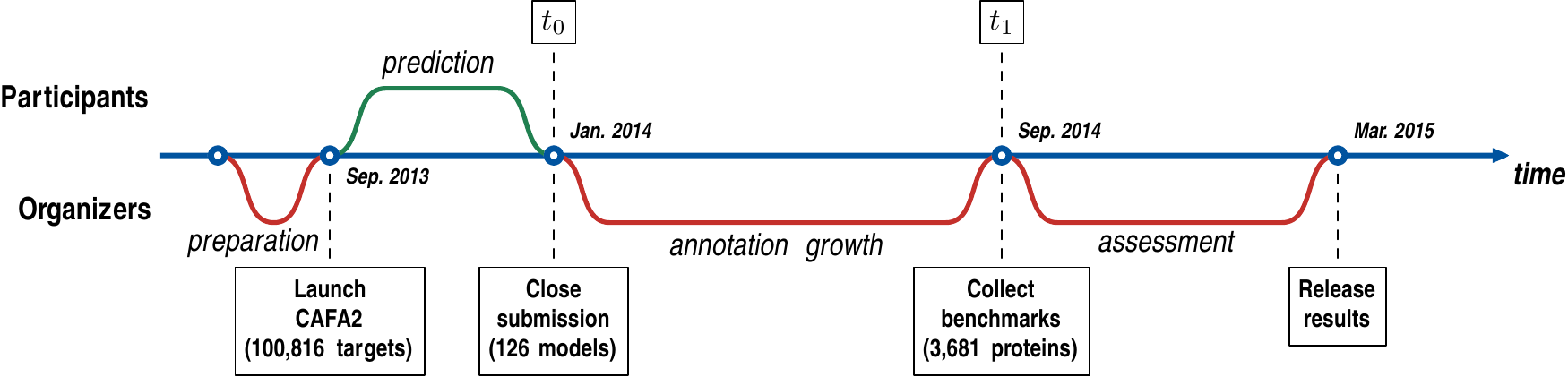}
  \caption{Timeline for the CAFA2 experiment.}
  \label{fig:timeline}
\end{figure}

The timeline for the second CAFA experiment followed that of the first experiment and is illustrated in Figure~\ref{fig:timeline}. Briefly, CAFA2 was announced in July 2013 and officially started in September 2013, when 100,816 \emph{target sequences} from 27 organisms were made available to the community. Teams were required to submit prediction scores within the $(0, 1]$ range for each protein-term pair they chose to predict on. The submission deadline for depositing these predictions was set for January 2014 (time point $t_0$). We then waited until September 2014 (time point $t_1$) for new experimental annotations to accumulate on the target proteins and assessed the performance of the prediction methods. We will refer to the set of all experimentally annotated proteins available at $t_0$ as the \emph{training set} and to a subset of target proteins that accumulated experimental annotations during $(t_0, t_1]$ and used for evaluation as the \emph{benchmark set}. It is important to note that the benchmark proteins and the resulting analysis vary based on the selection of time point $t_1$. For example, a preliminary analysis of the CAFA2 experiment was provided during the Automated Function Prediction Special Interest Group (AFP-SIG) meeting at the Intelligent Systems for Molecular Biology (ISMB) conference in July 2014. 

The participating methods were evaluated according to their ability to predict terms in Gene Ontology (GO) \cite{Ashburner2000} and Human Phenotype Ontology (HPO) \cite{Robinson2010}. In contrast with CAFA1, where the evaluation was carried out only for the Molecular Function Ontology (MFO) and Biological Process Ontology (BPO), in CAFA2 we also assessed the performance for the prediction of Cellular Component Ontology (CCO) terms in GO. The set of human proteins was further used to evaluate methods according to their ability to associate these proteins with disease terms from HPO, which included all sub-classes of the term HP:0000118, ``Phenotypic abnormality".

In total, 56 groups submitting 126 methods participated in CAFA2. From those, 125 methods made valid predictions on a sufficient number of sequences. One-hundred and twenty-one methods submitted predictions for at least one of the GO benchmarks, while 30 methods participated in the disease-gene prediction tasks using HPO.




\subsection*{Evaluation} 
The CAFA2 experiment expanded the assessment of computational function prediction compared with CAFA1. This includes the increased number of targets, benchmarks, ontologies, and method comparison metrics.


We distinguish between two major types of method evaluation. The first, \emph{protein-centric evaluation}, assesses performance accuracy of methods that predict all ontological terms associated with a given protein sequence. The second type, \emph{term-centric evaluation}, assesses performance accuracy of methods that predict if a single ontology term of interest is associated with a given protein sequence \cite{Radivojac2013}. The protein-centric evaluation can be viewed as a multi-label or structured-output learning problem of predicting a set of terms or a directed acyclic graph (a subgraph of the ontology) for a given protein. Because the ontologies contain many terms, the output space in this setting is extremely large and the evaluation metrics must incorporate similarity functions between groups of mutually interdependent terms (directed acyclic graphs). In contrast, the term-centric evaluation is an example of binary classification, where a given ontology term is assigned (or not) to an input protein sequence. These methods are particularly common in disease gene prioritization \cite{Moreau2012}. Put otherwise, a protein-centric evaluation considers a ranking of ontology terms for a given protein, whereas the term-centric evaluation considers a ranking of protein sequences for a given ontology term.

Both types of evaluation have merits in assessing performance. 
This is partly due to the statistical dependency between ontology terms, the statistical dependency among protein sequences and also the incomplete and biased nature of the experimental annotation of protein function \cite{Schnoes2013}. In CAFA2, we provide both types of evaluation, but we emphasize the protein-centric scenario for easier comparisons with CAFA1. We also draw important conclusions regarding method assessment in these two scenarios.

\subsubsection*{No-knowledge and limited-knowledge benchmark sets}
In CAFA1, a protein was eligible to be in the benchmark set if it had not had any experimentally-verified annotations in any of the GO ontologies at time $t_0$ but accumulated at least one functional term with an experimental evidence code between $t_0$ and $t_1$. In CAFA2, we refer to such benchmark proteins as \emph{no-knowledge} benchmarks. On the other hand, proteins with \emph{limited knowledge} are those that had been experimentally annotated in one or two GO ontologies, but not in all three, at time $t_0$. For example, for the performance evaluation in MFO, a protein without any annotation in MFO prior to the submission deadline was allowed to have experimental annotations in BPO and CCO. 

During the growth phase, the no-knowledge targets that have acquired experimental annotations in one or more ontologies became benchmarks in those ontologies. The limited-knowledge targets that have acquired additional annotations became benchmarks only for those ontologies for which there were no prior experimental annotations. The reason for using limited-knowledge targets was to identify whether the correlations between experimental annotations across ontologies can be exploited to improve function prediction. 

The selection of benchmark proteins for evaluating HPO-term predictors was separated from the GO analyses. There exists only a no-knowledge benchmark set in the HPO category. 

\subsubsection*{Partial and full evaluation modes}
Many function prediction methods apply only to certain types of proteins, such as proteins for which 3D structure data are available, proteins from certain taxa, or specific subcellular localizations. To accommodate these methods, CAFA2 provided predictors with an option of choosing a subset of the targets to predict on as long as they computationally annotated at least 5,000 targets, of which at least 10 accumulated experimental terms. We refer to the assessment mode in which the predictions were evaluated only on those benchmarks for which a model made at least one prediction at any threshold as \emph{partial evaluation mode}. In contrast, the \emph{full evaluation mode} corresponds to the same type of assessment performed in CAFA1 where all benchmark proteins were used for the evaluation and methods were penalized for not making predictions.

In most cases, for each benchmark category, we have two types of benchmarks, no-knowledge (NK) and limited-knowledge (LK), and two modes of evaluation, full-mode (FM) and partial-mode (PM). Exceptions are all HPO categories that only have no-knowledge benchmarks. The full mode is appropriate for comparisons of general-purpose methods designed to make predictions on any protein, while the partial mode gives an idea of how well each method performs on a self-selected subset of targets.

\subsubsection*{Evaluation metrics} 
Precision-recall ($pr$-$rc$) curves and remaining uncertainty-misinformation ($ru$-$mi$) curves were used as the two chief metrics in the protein-centric mode. We also provide a single measure evaluation in both types of curves as a real-valued scalar to compare methods; however, we note that any choice of a single point on those curves is somewhat arbitrary and may not match the intended application objectives for a given algorithm. Thus, a careful understanding of the evaluation metrics used in CAFA is necessary to properly interpret the results.


Precision~($pr$), recall~($rc$) and the resulting $\fmax$ are defined as 
\begin{eqnarray*}
  pr(\tau) &=& \frac{1}{m(\tau)}\sum_{i=1}^{m(\tau)} \frac{\sum_{f}
  \mathbbm{1}\left( f \in P_{i}(\tau) \wedge f \in T_{i}\right)}{\sum_{f}
  \mathbbm{1}\left( f \in P_{i}(\tau) \right)},\\
  rc(\tau) &=& \frac{1}{n_e}\sum_{i=1}^{n_e} \frac{\sum_{f} \mathbbm{1}\left( f \in
  P_{i}(\tau) \wedge f \in T_{i}\right)}{\sum_{f} \mathbbm{1}\left( f \in
  T_{i} \right)}, \\
  \fmax &=& \max_{\tau} \left\{ \frac{2\cdot pr(\tau)\cdot rc(\tau)}{pr(\tau) +
  rc(\tau)} \right\},
\end{eqnarray*}
\noindent where $P_{i}(\tau)$ denotes the set of terms that have predicted scores greater than or equal to $\tau$ for a protein sequence $i$, $T_{i}$ denotes the corresponding ground-truth set of terms for that sequence, $m(\tau)$ is the number of sequences with at least one predicted score greater than or equal to $\tau$, $\mathbbm{1}\left( \cdot \right)$ is an indicator function and $n_e$ is the number of targets used in a particular mode of evaluation. In the full evaluation mode $n_e = n$, the number of benchmark proteins, whereas in the partial evaluation mode $n_e = m(0)$, i.e. the number of proteins which were chosen to be predicted using the particular method. For each method, we refer to $\nicefrac{m(0)}{n}$ as the \emph{coverage} because it provides the fraction of benchmark proteins on which the method made any predictions.

The remaining uncertainty~($ru$), misinformation~($mi$) and the resulting minimum semantic distance~($\smin$) are defined as
\begin{eqnarray*}
  ru(\tau) &=& \frac{1}{n_e}\sum_{i=1}^{n_e} \sum_{f} ic(f) \cdot \mathbbm{1}\left( f
  \notin P_{i}(\tau) \wedge f \in T_{i} \right),\\
  mi(\tau) &=& \frac{1}{n_e}\sum_{i=1}^{n_e} \sum_{f} ic(f) \cdot \mathbbm{1}\left( f
  \in P_{i}(\tau) \wedge f \notin T_{i} \right), \\
  \smin &=& \min_{\tau}\left\{ \sqrt{ru(\tau)^{2} + mi(\tau)^{2}} \right\},
\end{eqnarray*}
\noindent where $ic(f)$ is the information content of the ontology term $f$ \cite{Clark2013}. It is estimated in a maximum likelihood manner as the negative binary logarithm of the conditional probability that the term $f$ is present in a protein's annotation given that all its parent terms are also present. Note that here, $n_e = n$ in the full evaluation mode and $n_e = m(0)$ in the partial evaluation mode applies to both $ru$ and $mi$.

In addition to the main metrics, we used two secondary metrics. Those were the weighted version of the precision-recall curves and the version of the $ru$-$mi$ curves normalized to the $[0,1]$ interval. These metrics and the corresponding evaluation results are shown in Supplementary Materials.

For the term-centric evaluation we used the area under the Receiver Operating Characteristic (ROC) curve (AUC). The AUCs were calculated for all terms that have acquired at least 10 positively annotated sequences, whereas the remaining benchmarks were used as negatives. The term-centric evaluation was used both for ranking models and to differentiate well and poorly predictable terms. The performance of each model on each term is provided in Supplementary Materials.


As we required all methods to keep two significant figures for prediction scores, the threshold $\tau$ in all metrics used in this study exhaustively runs from $0.01$ to $1.00$ with the step size of $0.01$.

\subsection*{Data sets}
Protein function annotations for the Gene Ontology assessment were extracted, as a union, from three major protein databases that are available in the public domain: Swiss-Prot \cite{Bairoch2005}, UniProt-GOA \cite{Huntley2015} and the data from the GO consortium web site \cite{Ashburner2000}. We used evidence codes EXP, IDA, IMP, IGI, IEP, TAS and IC to build benchmark and ground-truth sets. Annotations for the HPO assessment were downloaded from the Human Phenotype Ontology database~\cite{Robinson2010}.

\begin{figure}[h!]
  \centering
  \includegraphics[width=\textwidth]{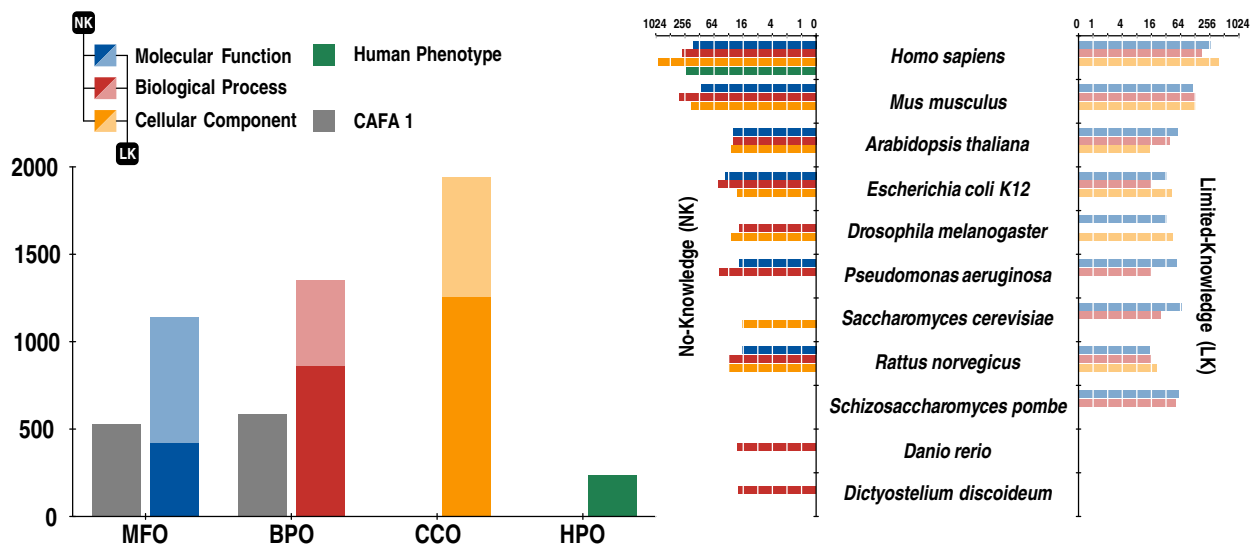}
\caption{CAFA2 benchmark breakdown. The left panel shows the benchmark size for each of the four ontologies. The right panel lists the breakdown of benchmarks for both types over 11 species (with no less than 15 proteins) sorted according to the total number of benchmark proteins. For both panels, dark colors (blue, red, yellow) correspond to no-knowledge~(NK) types, while their light color counterparts correspond to limited-knowledge~(LK) types. The size of CAFA~1 benchmarks are shown in gray.}
  \label{fig:benchmark}
\end{figure}

Figure~\ref{fig:benchmark} summarizes the benchmarks we used in this study. The left panel shows the benchmark sizes for each of the ontologies and compares these numbers to CAFA1. All species that have at least 15 proteins in any of the benchmark categories are listed in the right panel. 

\subsection*{Baseline models}
We built two baseline methods, Na\"ive and BLAST, and compared them with all participating methods. The Na\"ive method simply predicts the frequency of a term being annotated in a database \cite{Clark2011}. BLAST was based on search results using the Basic Local Alignment Search Tool (BLAST) software against the training database \cite{Altschul1997}. A term will be predicted as the highest local alignment sequence identity among all BLAST hits annotated with the term. Both of these two methods were ``trained'' on the experimentally annotated proteins available in Swiss-Prot at time $t_0$, except for HPO where the two baseline models were trained using the annotations from the $t_0$ release of the Human Phenotype Ontology.

%% file: results.tex
\section*{Results}

\subsection*{Overall performance}
\begin{figure}[htp!]
  \centering
  \includegraphics[width=0.45\textwidth,height=0.36\textwidth]{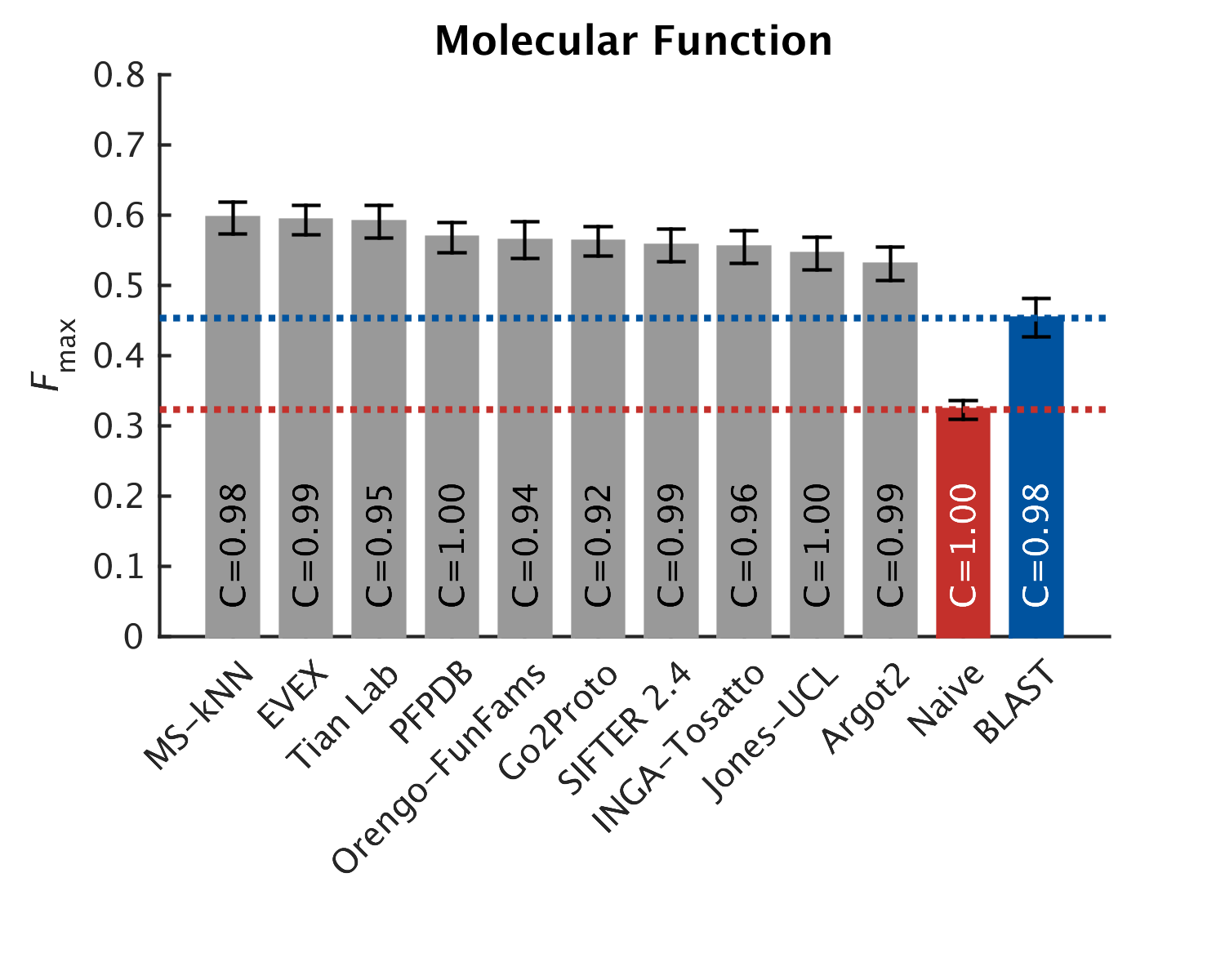}
  \includegraphics[width=0.45\textwidth,height=0.36\textwidth]{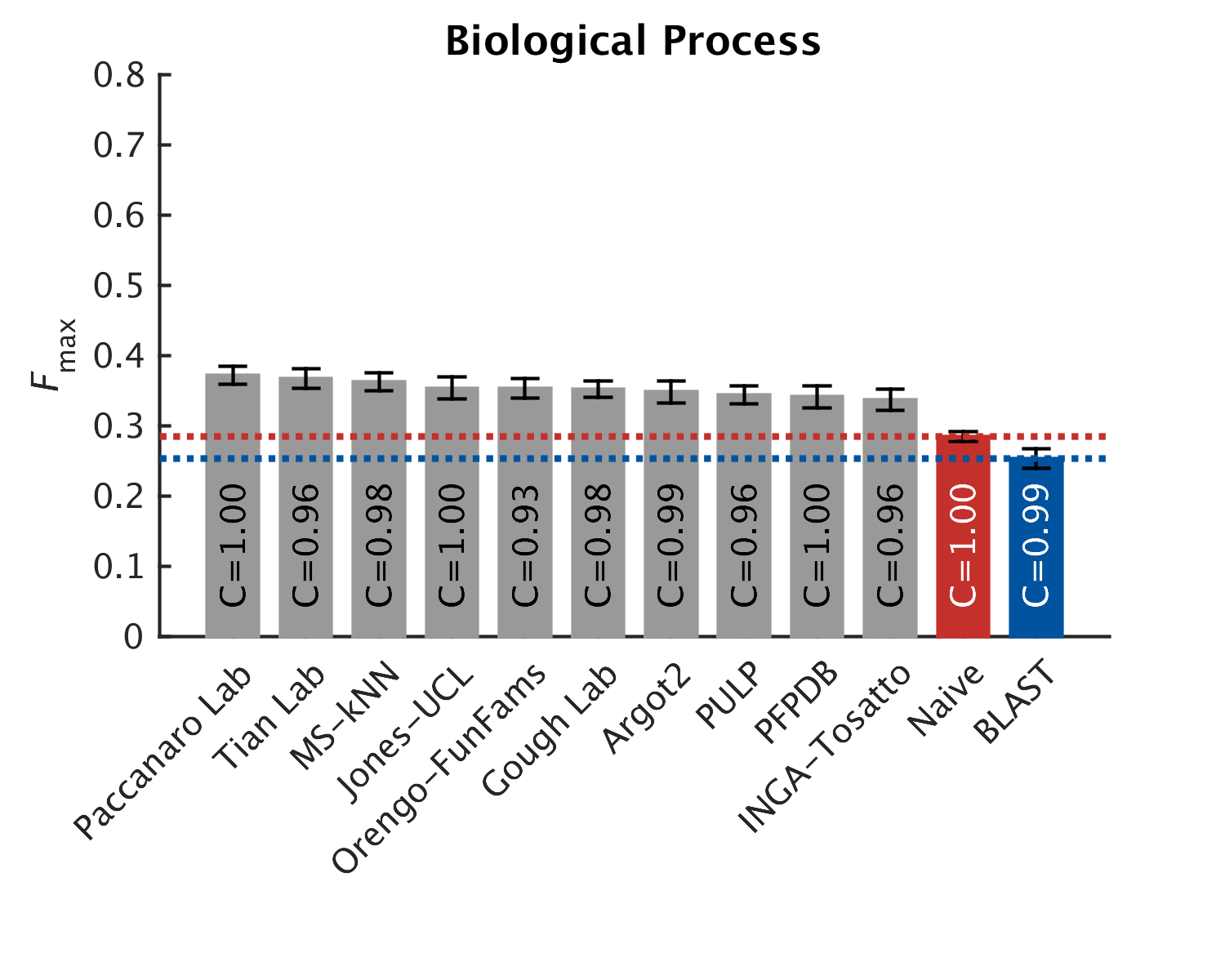}\\
  \includegraphics[width=0.45\textwidth,height=0.36\textwidth]{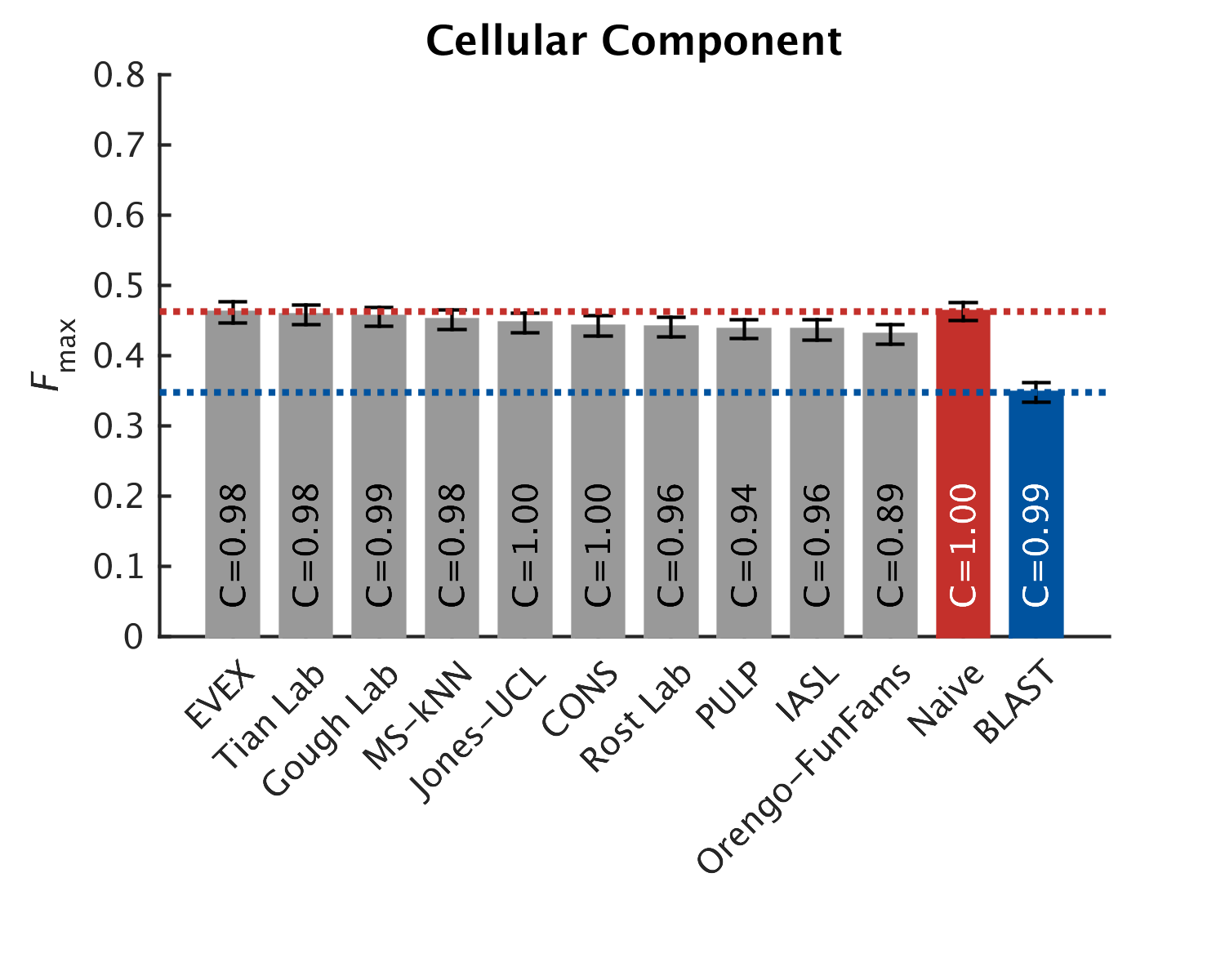}
  \includegraphics[width=0.45\textwidth,height=0.36\textwidth]{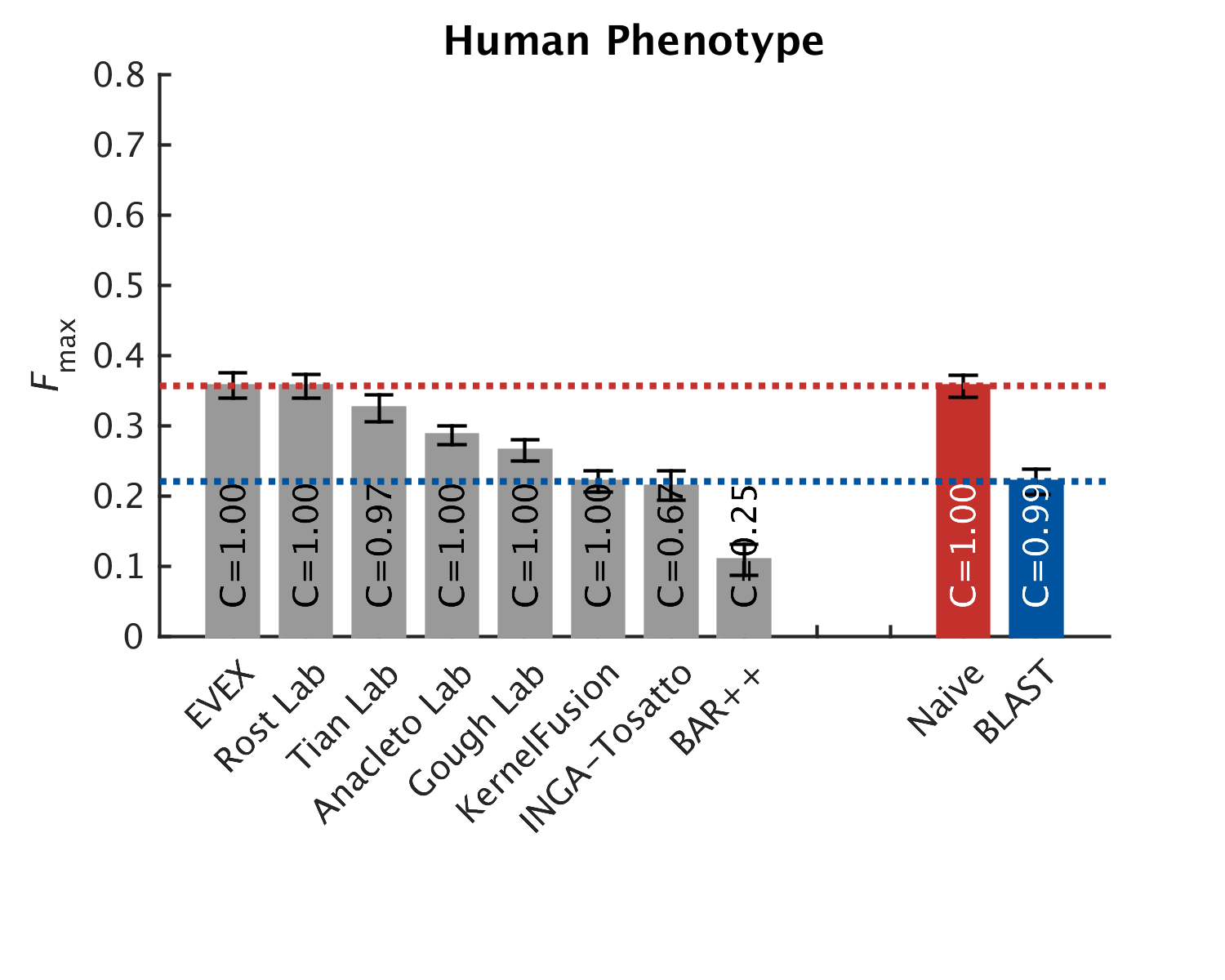}
\caption{Overall evaluation using the maximum F-measure, $\fmax$. Evaluation was carried out on no-knowledge benchmark sequences in the full mode. The coverage of each method is shown within its performance bar. A perfect predictor would be characterized with $\fmax = 1$. Confidence intervals (95\%) were determined using bootstrapping with 10,000 iterations on the set of benchmark sequences. For cases in which a principal investigator participated in multiple teams, only the results of the best-scoring method are presented. Details for all methods are provided in Supplementary Materials.}
  \label{fig:main_result_fmax}
\end{figure}

\begin{figure}[htp!]
  \centering
  \includegraphics[width=0.45\textwidth,height=0.36\textwidth]{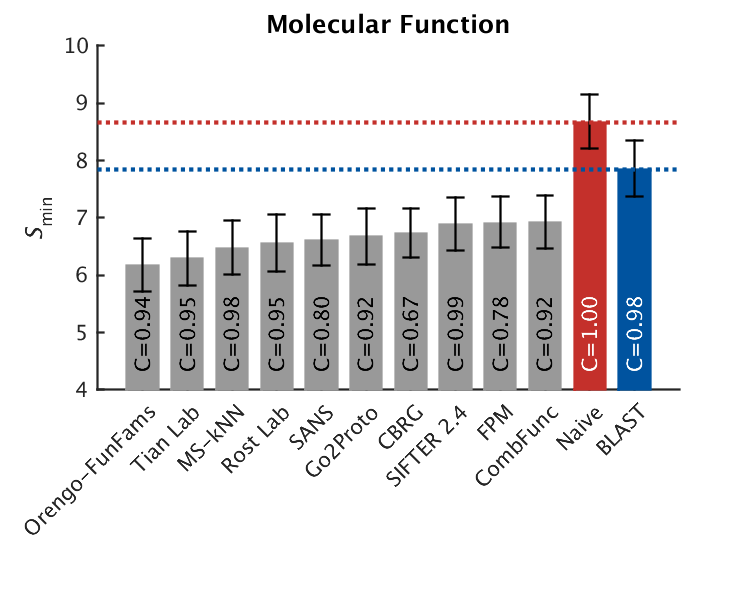}
  \includegraphics[width=0.45\textwidth,height=0.36\textwidth]{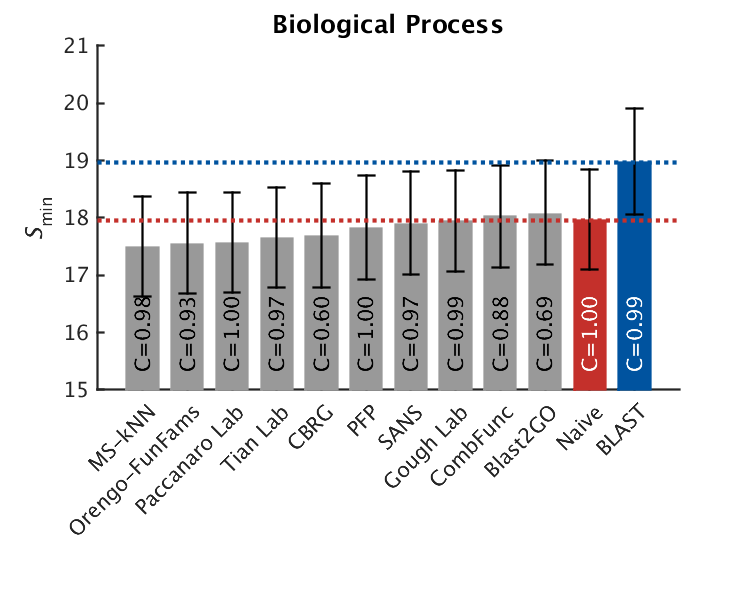}\\
  \includegraphics[width=0.45\textwidth,height=0.36\textwidth]{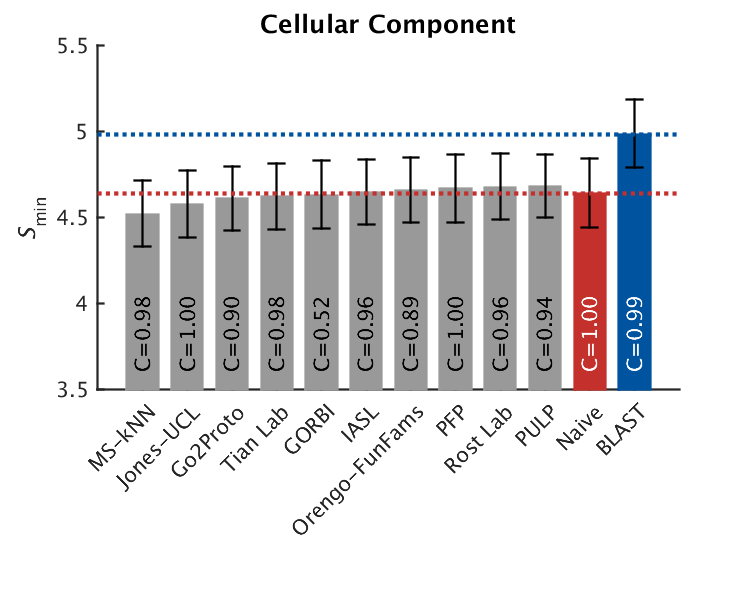}
  \includegraphics[width=0.45\textwidth,height=0.36\textwidth]{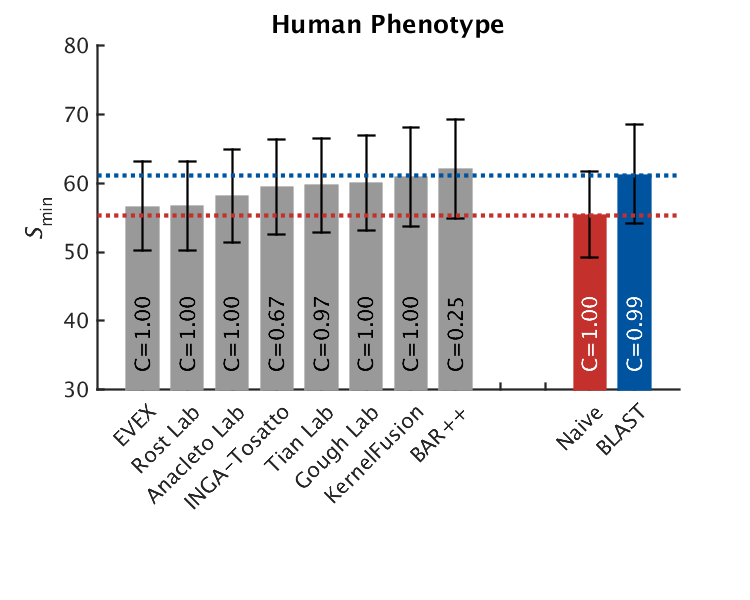}
  \caption{Overall evaluation using the minimum semantic distance, $\smin$. Evaluation was carried out on no-knowledge benchmark sequences in the full mode. The coverage of each method is shown within its performance bar. A perfect predictor would be characterized with $\smin = 0$. Confidence intervals (95\%) were determined using bootstrapping with 10,000 iterations on the set of benchmark sequences. For cases in which a principal investigator participated in multiple teams, only the results of the best-scoring method are presented. Details for all methods are provided in Supplementary Materials.}
  \label{fig:main_result_smin}
\end{figure}

The performance accuracies of the top~10 
methods are shown in Figures~\ref{fig:main_result_fmax} and \ref{fig:main_result_smin}. The $95\%$ confidence intervals were estimated using bootstrapping on the benchmark set with $B = 10,000$ iterations \cite{Efron1993}. The results provide a broad insight into the state of the art. 

Predictors performed very differently across the four ontologies. Various reasons contribute to this effect including: (1) the topological properties of the ontology such as the size, depth, and branching factor; (2) term predictability; for example, the BPO terms are considered to be more abstract in nature than the MFO and CCO terms; (3) the annotation status, such as the size of the training set at $t_0$ as well as various annotation biases \cite{Schnoes2013}.

In general, CAFA2 methods perform better in predicting MFO terms than any other ontology. Top methods achieved the $\fmax$ scores around $0.6$ and considerably surpassed the two baseline models. Maintaining the pattern from CAFA1, the performance accuracies in the BPO category were not as good as in the MFO category. The best-performing method scored slightly below $0.4$.

For the two newly-added ontologies in CAFA2, we observed that the top predictors performed no better than the Na\"ive method under $\fmax$, whereas they slightly outperformed the Na\"ive method under $\smin$ in CCO. One possible reason for the competitive performance of the Na\"ive method in the CCO category is the fact that a small number of relatively general terms are frequently used, and those relative frequencies do not diffuse quickly enough with the depth of the graph. For instance, the annotation frequency of ``organelle'' (GO:0043226, level~2), ``intracellular part'' (GO:0044424, level~3) and ``cytoplasm'' (GO:0005737, level~4) are all above the best threshold for the Na\"ive method ($\tau_{\mathrm{optimal}} = 0.32$). Correctly predicting these terms increases the number of ``true positives'' and thus boosts the performance of the Na\"ive method under the $F_{\max}$ evaluation. However, once the less informative terms are down-weighted (using the $S_{\min}$ measure), the Na\"ive method becomes significantly penalized and degraded. The weighted $\fmax$ and normalized $\smin$ evaluations can be found in Supplementary Materials.

However, high frequency of general terms does not seem to be the major reason for the observed performance in the HPO category. One possible explanation for this effect  would be that the average number of HPO terms associated with a human protein is much larger than in GO. The mean number of annotations per protein in HPO is  84, while the MFO, BPO and CCO the mean number of annotations per protein are 10, 39, and 14 respectively. The high number of annotations per protein makes  prediction using HPO terms significantly more difficult. In addition, unlike for GO terms, the HPO annotations cannot be transferred from other species based on homology and other available data. Successfully predicting the HPO terms in the protein-centric mode is a difficult problem. 

\subsection*{Term-centric evaluation}
Protein-centric view, despite its power in showing the strengths of a predictor, does not gauge a predictor's performance for a specific function. We therefore also assessed predictors in the term-centric manner by calculating AUCs for individual terms. Averaging those AUCs over terms provides a metric for ranking predictors, whereas averaging performances over terms provides insights into how well this term can be predicted computationally by the community. 

\begin{figure}[h!]
  \centering
  \includegraphics[width=0.45\textwidth,height=0.36\textwidth]{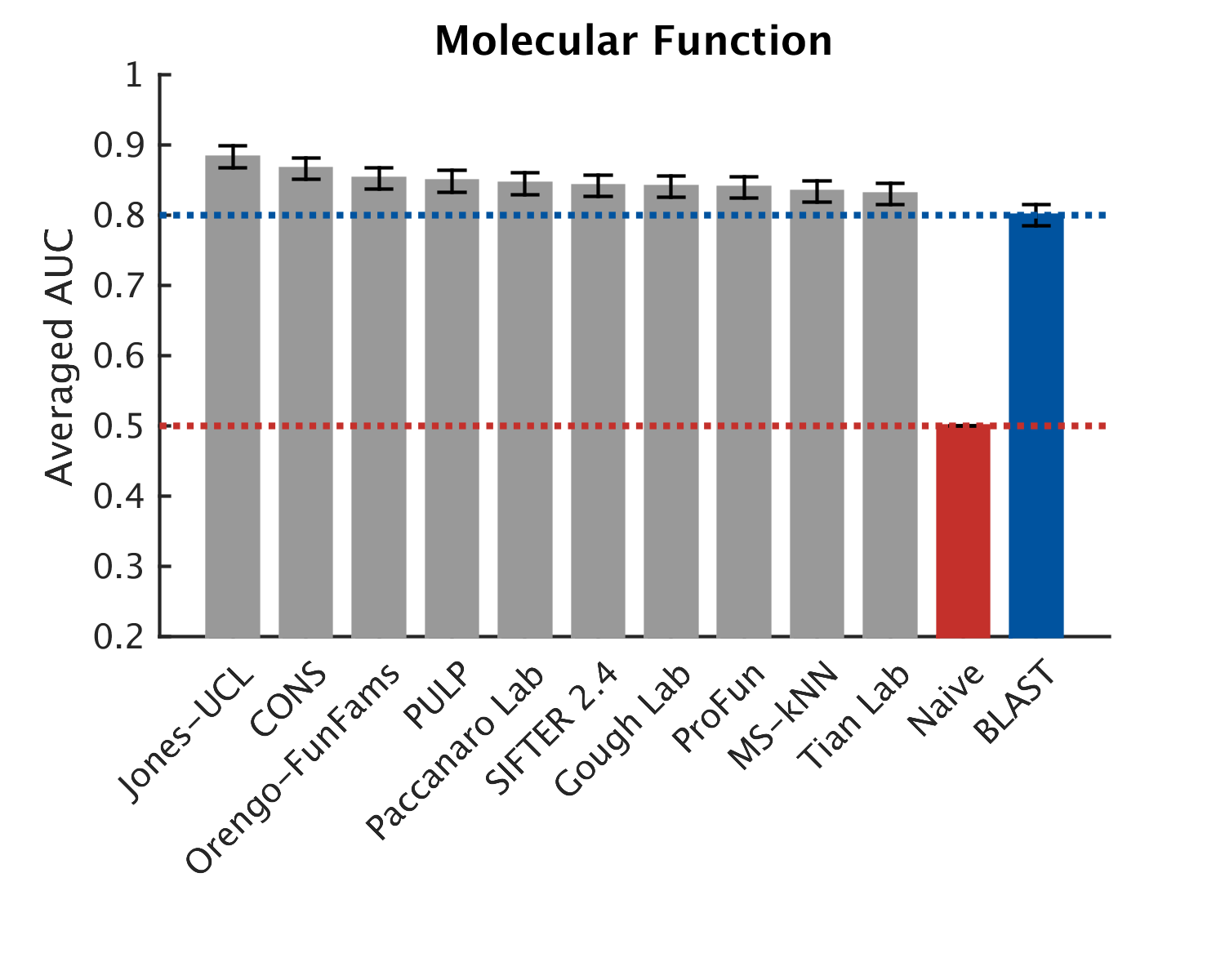}
  \includegraphics[width=0.45\textwidth,height=0.36\textwidth]{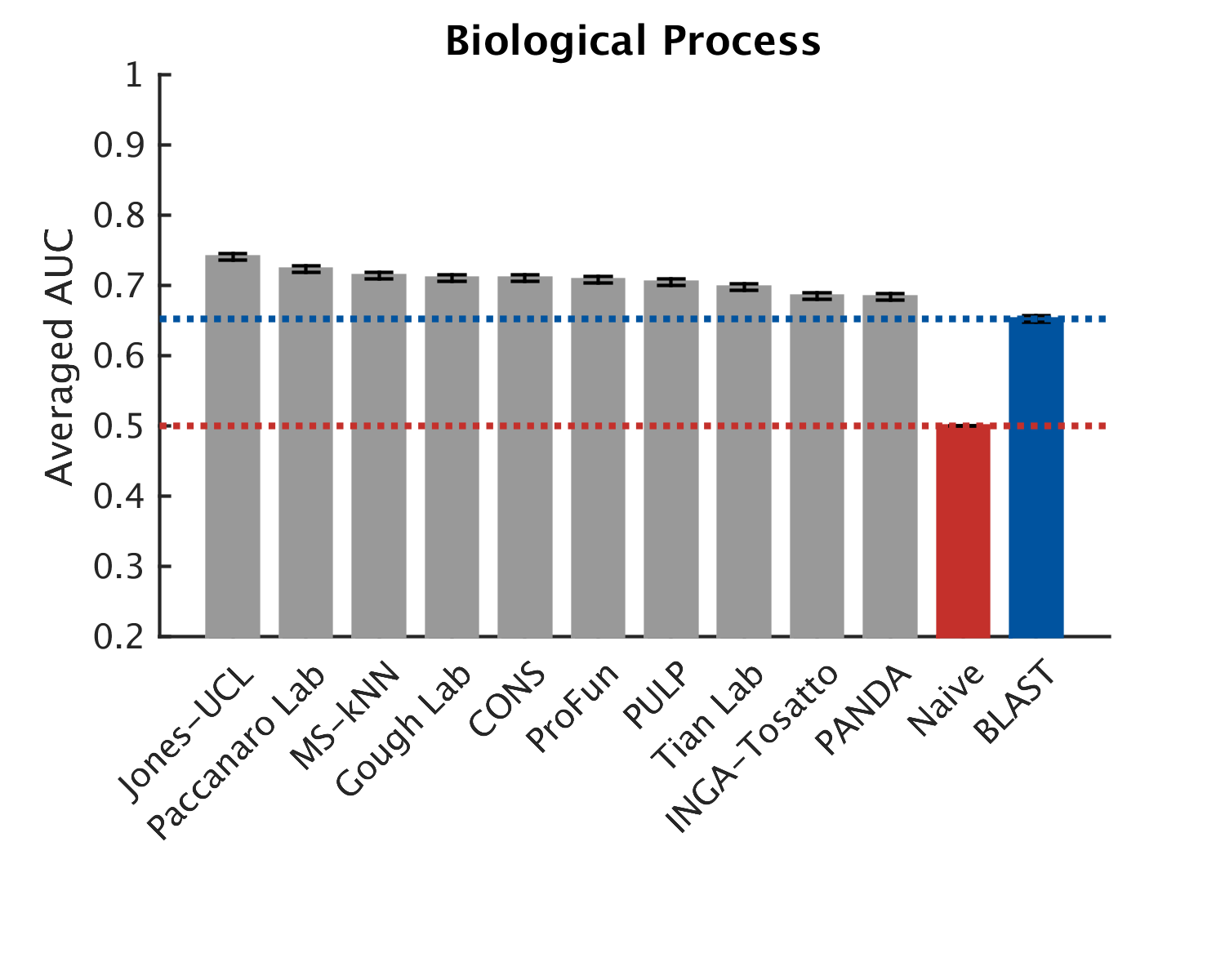}\\
  \includegraphics[width=0.45\textwidth,height=0.36\textwidth]{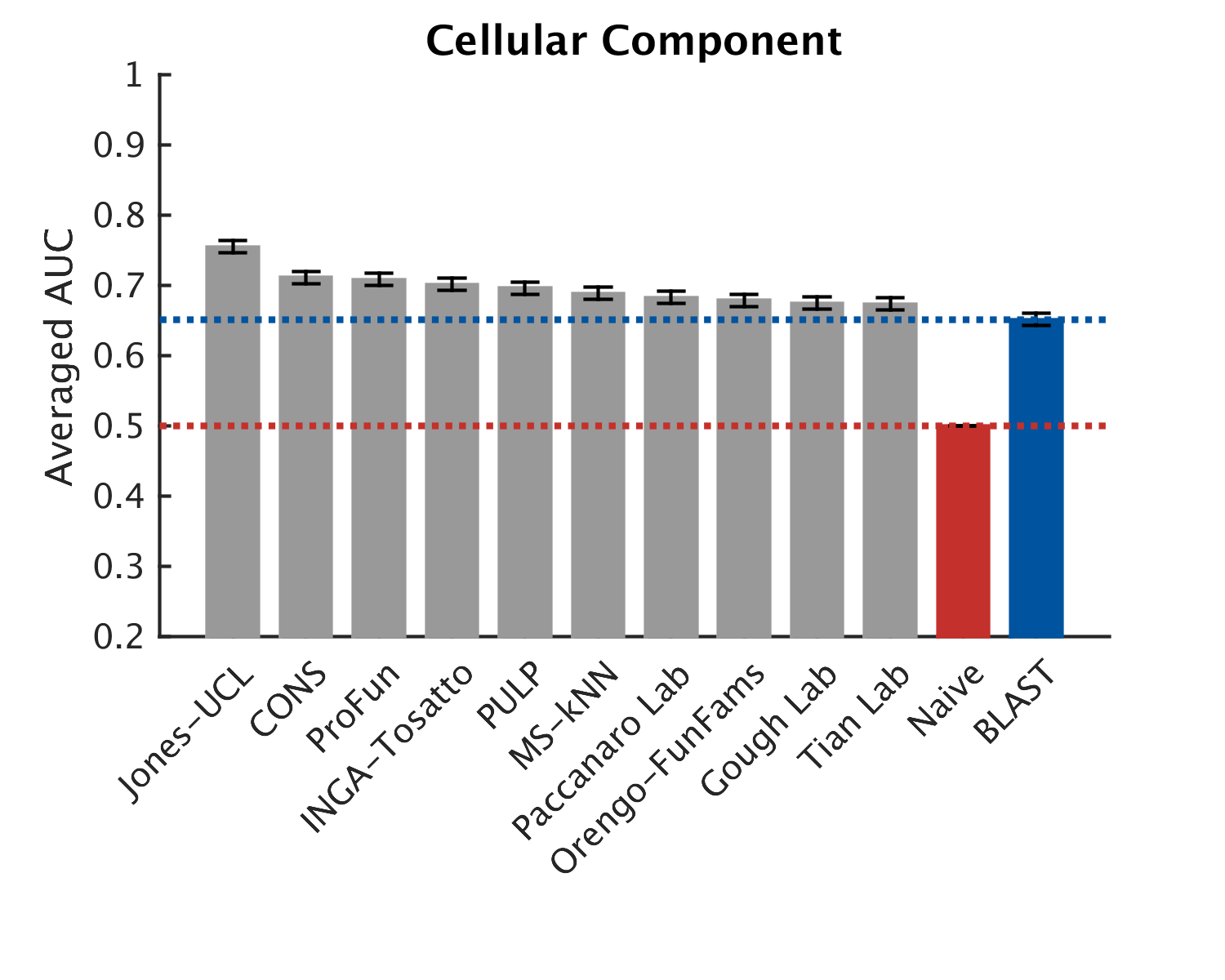}
  \includegraphics[width=0.45\textwidth,height=0.36\textwidth]{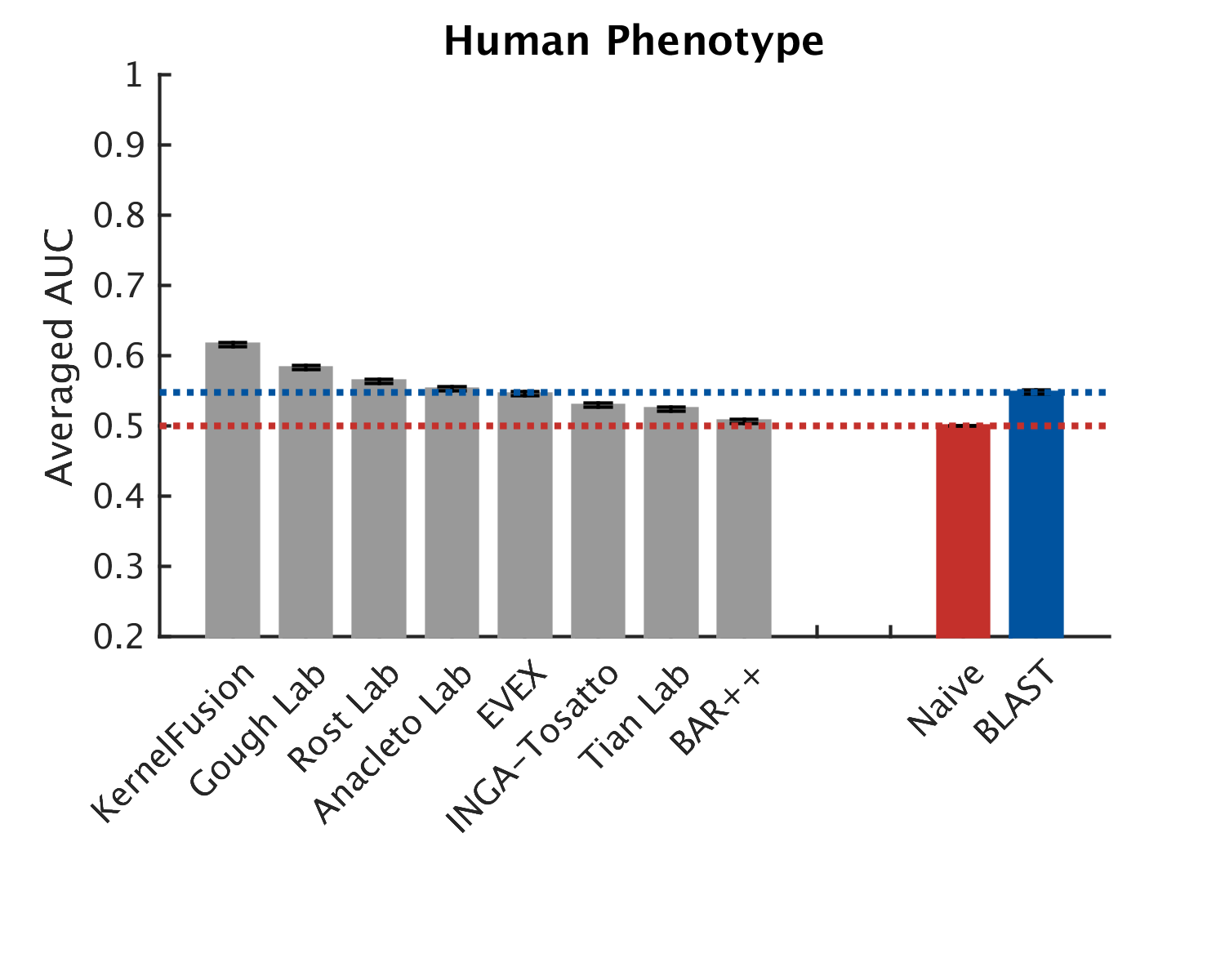}
  \caption{Overall evaluation using the averaged AUC over terms with no less than 10 positive annotations. Evaluation was carried out on no-knowledge benchmark sequences in the full mode. Error bars indicate the standard error in averaging AUC over terms for each method. For cases in which a principal investigator participated in multiple teams, only the results of the best-scoring method are presented. Details for all methods are provided in Supplementary Materials.}
  \label{fig:averaged_auc}
\end{figure}

\begin{figure}[h!]
  \centering
  \includegraphics[width=0.45\textwidth]{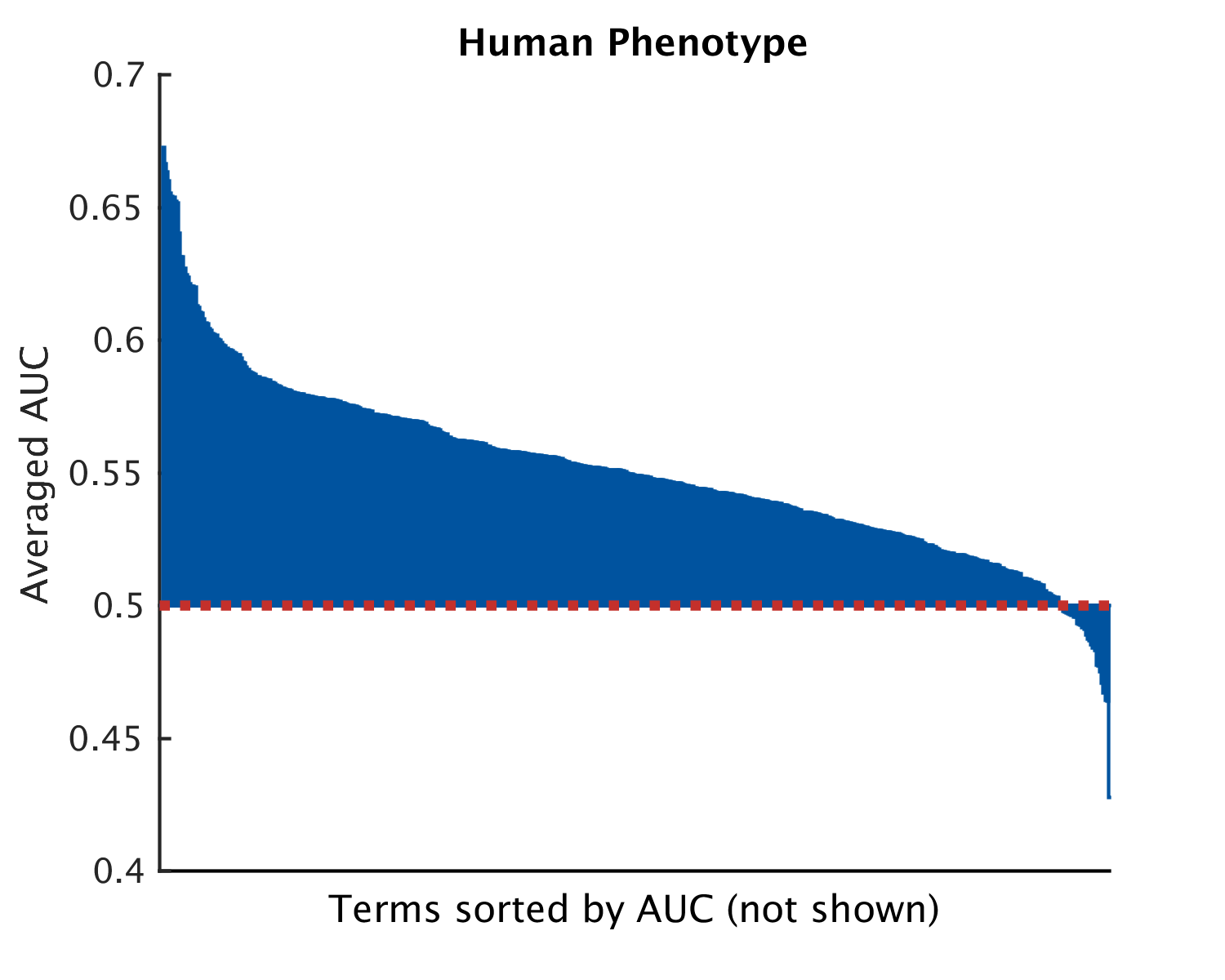}
  \includegraphics[width=0.45\textwidth]{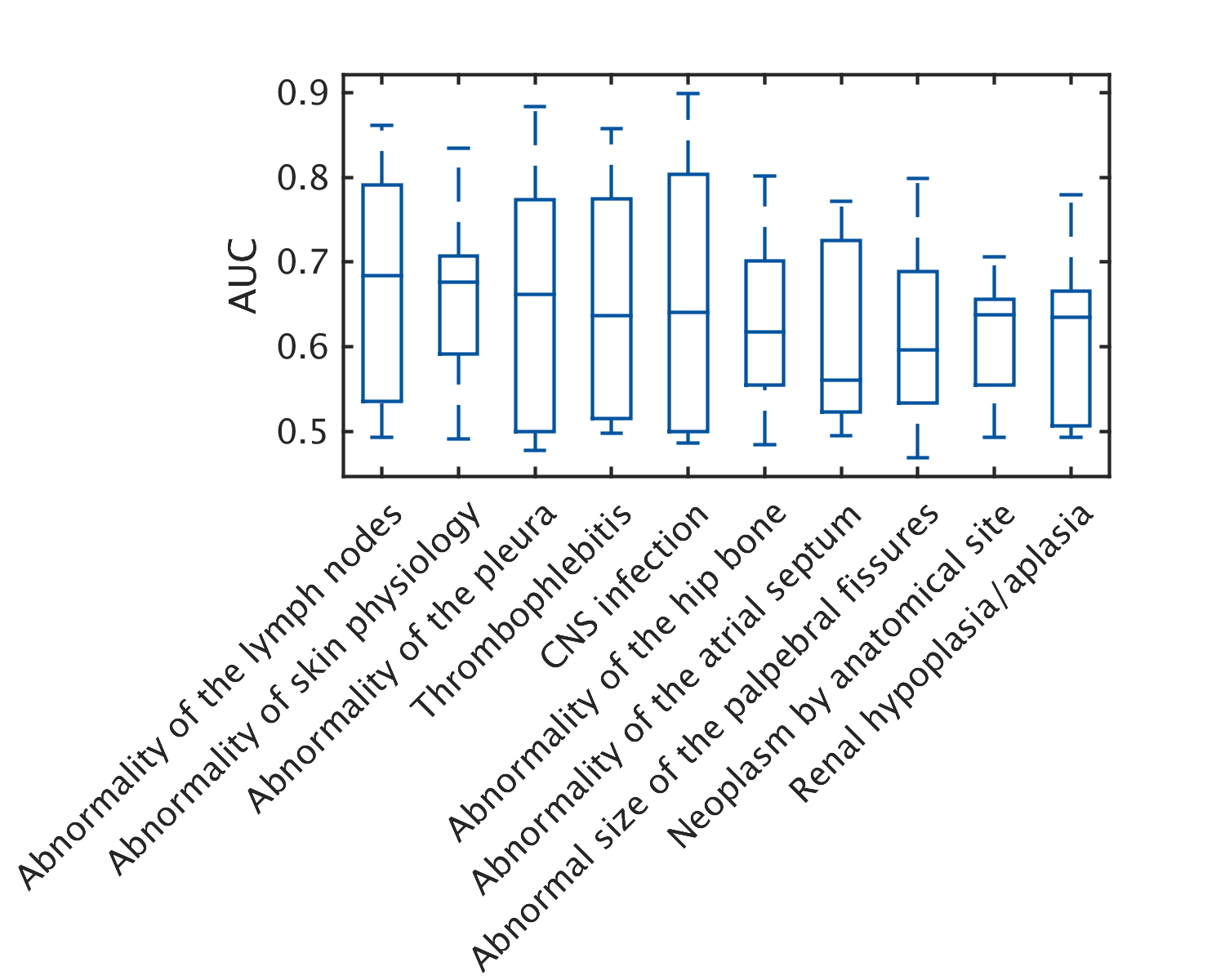}
  \caption{Averaged AUC per term for Human Phenotype ontology. Left panel: Terms are sorted based on AUC, dashed red line indicates the performance of the Na\"ive method. Right panel: The top~10 accurately predicted terms without overlapping ancestors (except for the root).}
  \label{fig:auc_hpo}
\end{figure}

Figure \ref{fig:averaged_auc} shows the performance evaluation where the AUCs for each method were averaged over all terms for which at least ten positive sequences were available. Proteins without predictions were counted as predictions with a score of 0. As shown in Figures \ref{fig:main_result_fmax}-\ref{fig:main_result_smin}, correctly predicting CCO and HPO terms for a protein might not be an easy task according to the protein-centric results. However, the overall poor performances could also result from the dominance of poorly predictable terms. Therefore, a term-centric view can help differentiate prediction quality across terms. As shown in Figure~\ref{fig:auc_hpo}, most of the terms in HPO obtain AUC greater than the Na\"ive model, with some terms on average achieving reasonably well AUCs around $0.7$. Depending on the training data available for participating methods, well predicted phenotype terms range from mildly specific such as ``Lymphadenopathy'' and ``Thrombophlebitis'' to general ones such as ``Abnormality of the Skin Physiology''. 

%
%

\subsection*{Performance on various categories of benchmarks}

\subsubsection*{Easy vs. difficult benchmarks}
As in CAFA1, the no-knowledge GO benchmarks were divided into ``easy'' versus ``difficult'' categories based on their maximal global sequence identity with proteins in the training set. Since the distribution of sequence identities roughly forms a bimodal shape (Supplementary Materials), a cutoff of $60\%$ was manually chosen to define the two categories. The same cutoff was used in CAFA1. Unsurprisingly, across all three ontologies, the performance of the BLAST model was substantially impacted for the difficult category because of the lack of high sequence identity homologs and as a result, transferring annotations was relatively unreliable. However, we also observed that most top methods were insensitive to the types of benchmarks, which provides us with encouraging evidence that state-of-the-art protein function predictors can successfully combine multiple potentially unreliable hits, as well as multiple types of data, into a reliable prediction.

\subsubsection*{Species-specific categories}
The benchmark proteins were split into even smaller categories for each species as long as the resulting category contained at least 15 sequences. However, because of space limitations, we only show the breakdown results on eukarya and prokarya benchmarks in Figure~\ref{fig:evp} (the species-specific results are provided in Supplementary Materials). It is worth noting that the performance accuracies on the entire benchmark sets were dominated by the targets from eukarya due to their larger proportion in the benchmark set and annotation preferences. The eukarya benchmark rankings therefore coincide with the overall rankings, but the smaller categories typically showed different rankings and may be informative to more specialized research groups.

\begin{figure}[h!]
  \centering
  \includegraphics[width=0.45\textwidth]{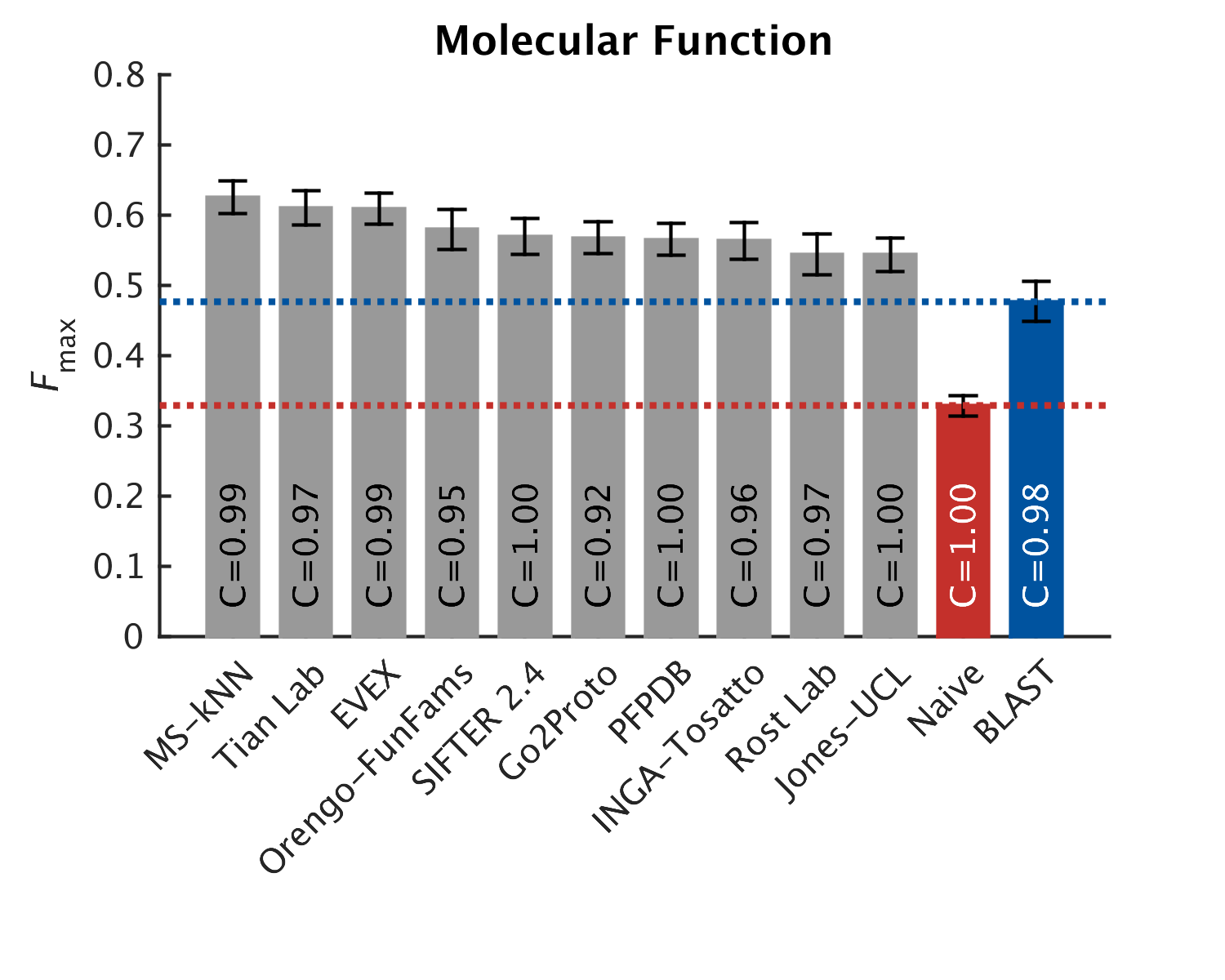}
  \includegraphics[width=0.45\textwidth]{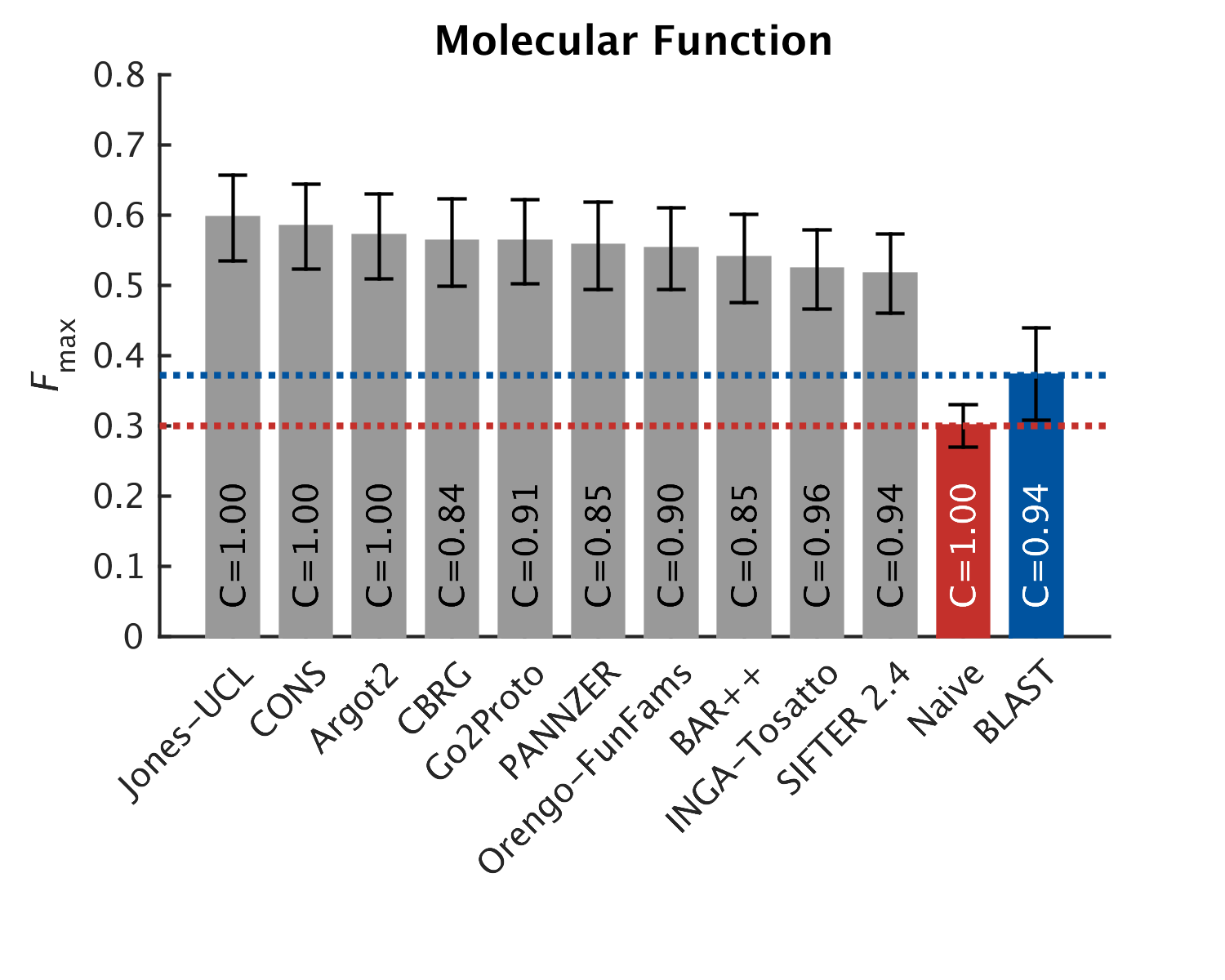}\\
  \includegraphics[width=0.45\textwidth]{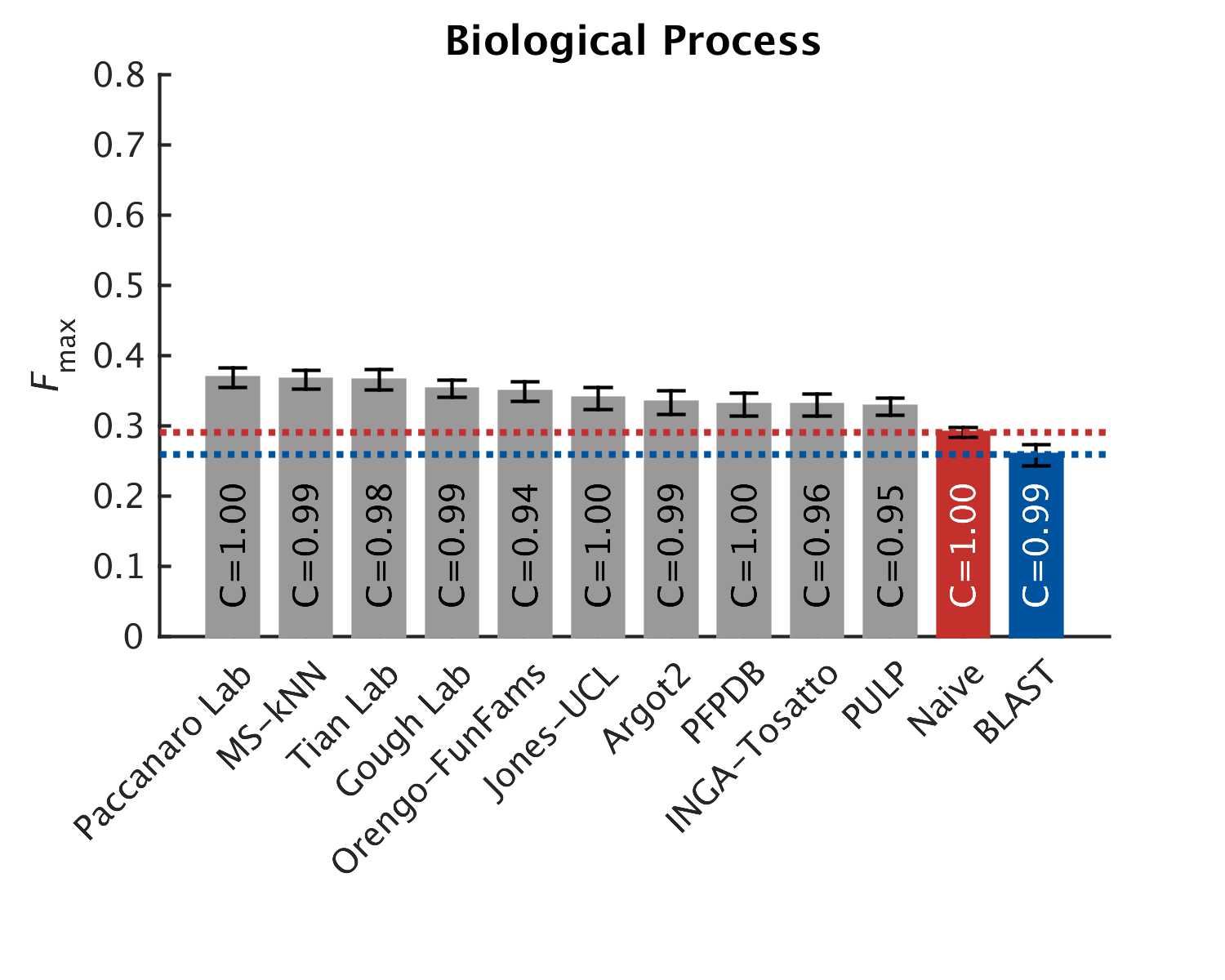}
  \includegraphics[width=0.45\textwidth]{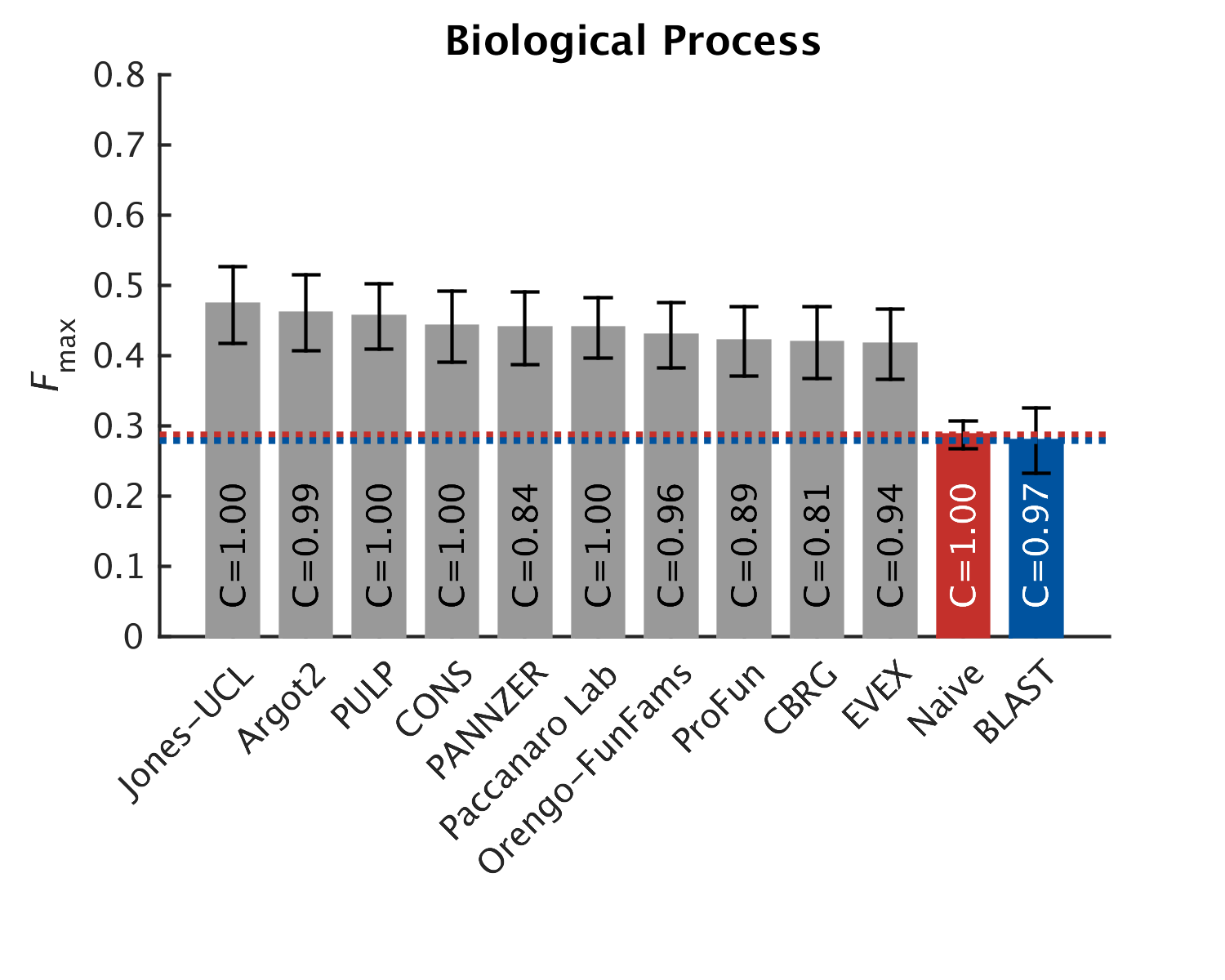}\\
  \includegraphics[width=0.45\textwidth]{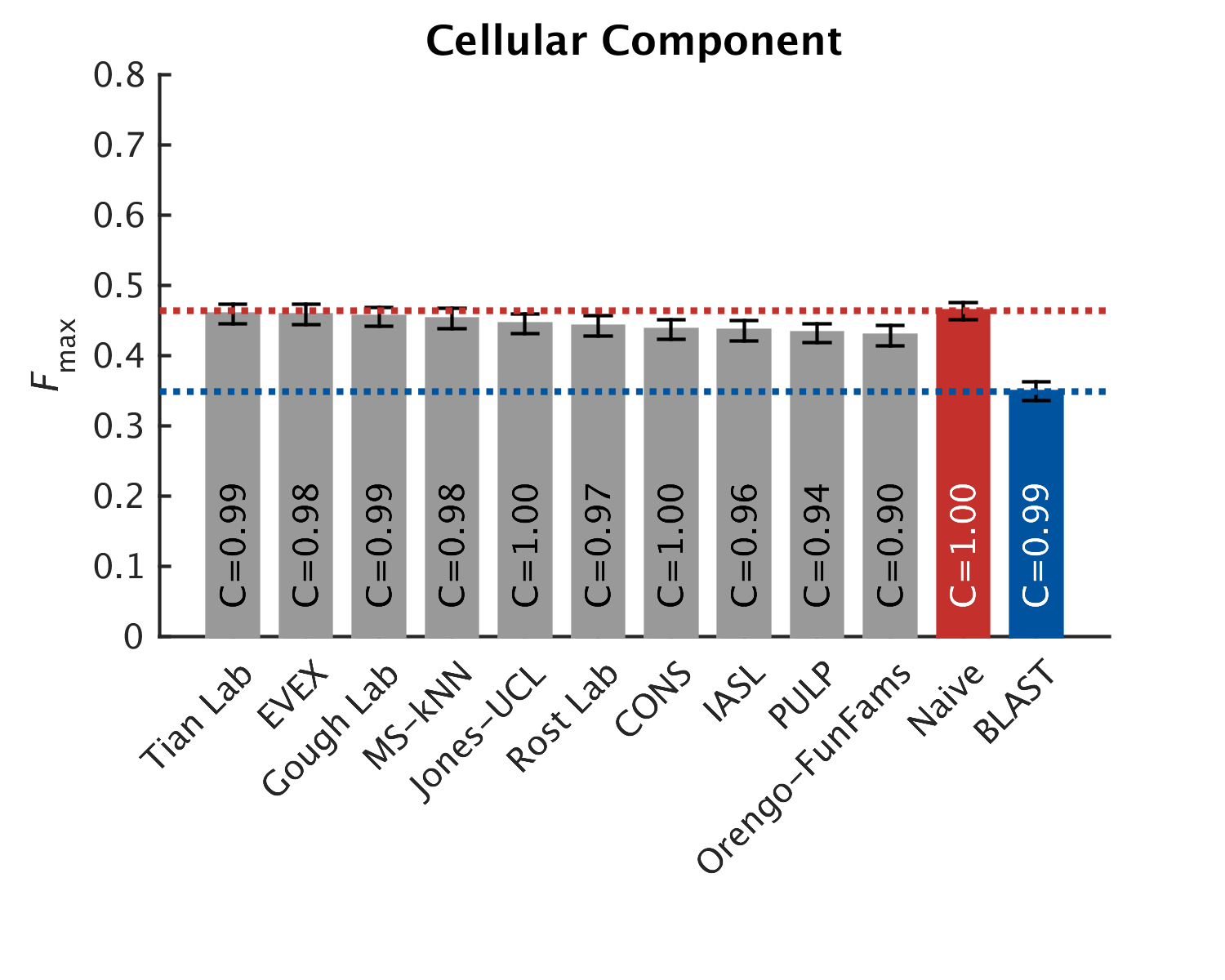}
  \includegraphics[width=0.45\textwidth]{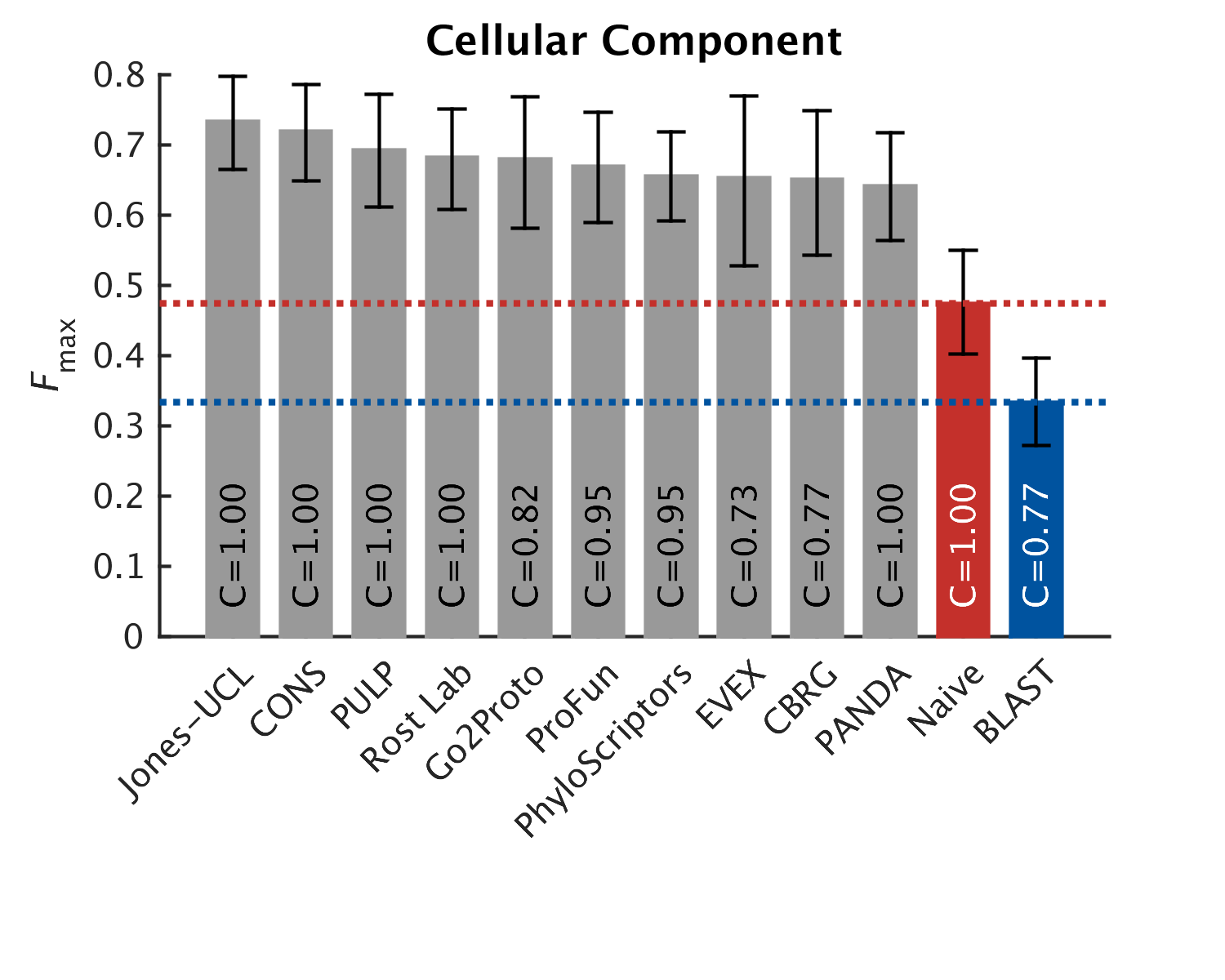}
  \caption{Performance evaluation using the maximum F-measure, $\fmax$, on eukaryotic~(left) versus prokaryotic~(right) benchmark sequences. Evaluation was carried out on no-knowledge benchmark sequences in the full mode. The coverage of each method is shown within its performance bar. Confidence intervals (95\%) were determined using bootstrapping with 10,000 iterations on the set of benchmark sequences. For cases in which a principal investigator participated in multiple teams, only the results of the best-scoring method are presented. Details for all methods are provided in Supplementary Materials.}
\label{fig:evp}
\end{figure}

For all three GO ontologies, no-knowledge prokarya benchmark sequences collected over the annotation growth phase mostly (over $80\%$) came from two species: \emph{E.~coli} and \emph{P.~aeruginosa}~(for CCO, 21 out of 22 proteins were from \emph{E.~coli}). Thus, one should keep in mind that the prokarya benchmarks essentially reflect the performance on proteins from these two species. Methods predicting the MFO terms for prokaryotes are slightly worse than those for eukaryotes. In addition, direct function transfer by homology for prokaryotes did not work well using this ontology. However, the performance was better using the other two ontologies, especially CCO. It is not very surprising that top methods achieved good performance for \emph{E.~coli} as it is a well-studied model organism.

\subsection*{Top methods have improved since CAFA1}
The second CAFA experiment was conducted three years after the first one. As our knowledge of protein function has increased since then, it was worthwhile to assess whether computational methods have also been improved and if so, to what extent. Therefore, to monitor the progress of the community over time, we revisit some of the top methods in CAFA1 and compare them with their successors.

The comparison was done on an overlapping benchmark set created from CAFA1 targets and CAFA2 targets. More precisely, we used the stored predictions on the target proteins from CAFA1 and compared them with the new predictions from CAFA2 on the overlapping set of CAFA2 benchmarks and CAFA1 targets (a sequence had to be a no-knowledge target in both experiments to be eligible in this evaluation). For this purpose, we used a hypothetical ontology by taking the intersection of the two Gene Ontology snapshots (versions from January 2011 and June 2013) so as to mitigate the influence of ontology changes. We thus collected 356 benchmark proteins for MFO comparisons and 698 for BPO comparisons. The two baseline methods were trained on respective Swiss-Prot annotations for both ontologies so that they serve as controls for database change. In particular, SwissProt2011 (for CAFA1) contained 29,330 and 31,282 proteins for MFO and BPO, while SwissProt2014 (for CAFA2) contained 26,907 and 41,959 proteins for the two ontologies.

\begin{figure}[ht!]
  \centering
  \textbf{Molecular Function} \\
  \includegraphics[width=0.8\textwidth]{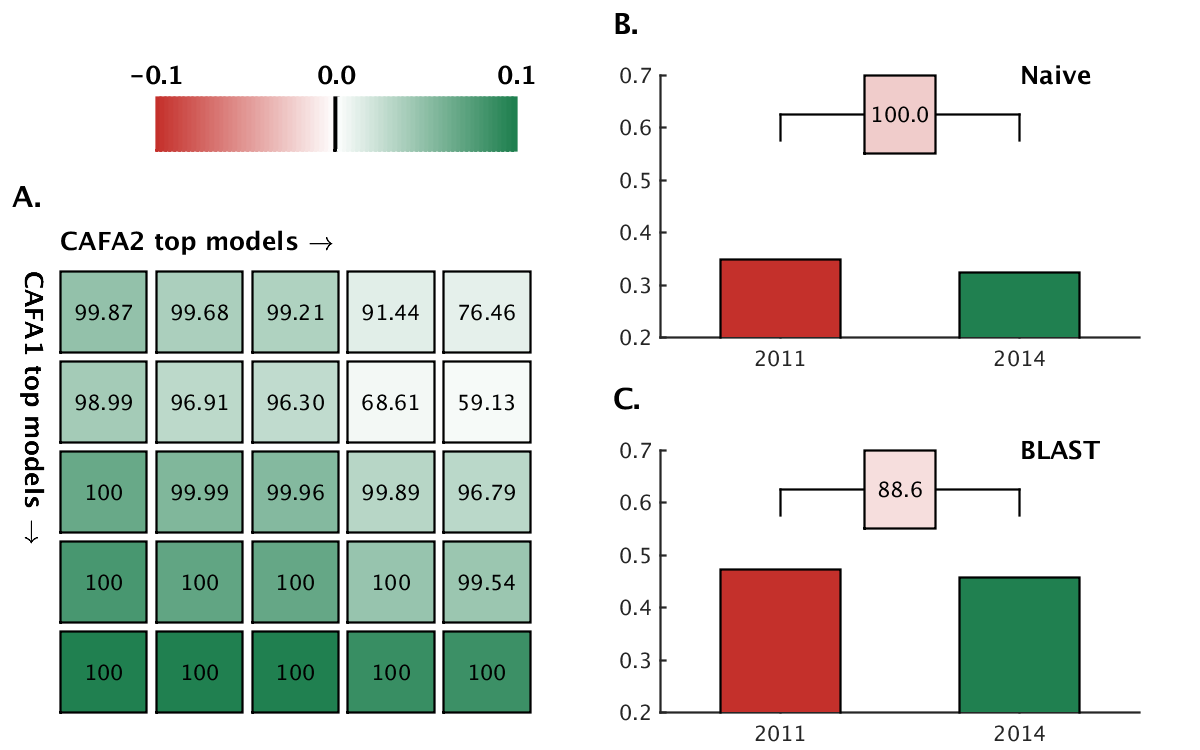} \\
  \vspace{0.5cm}
  \textbf{Biological Process} \\
  \includegraphics[width=0.8\textwidth]{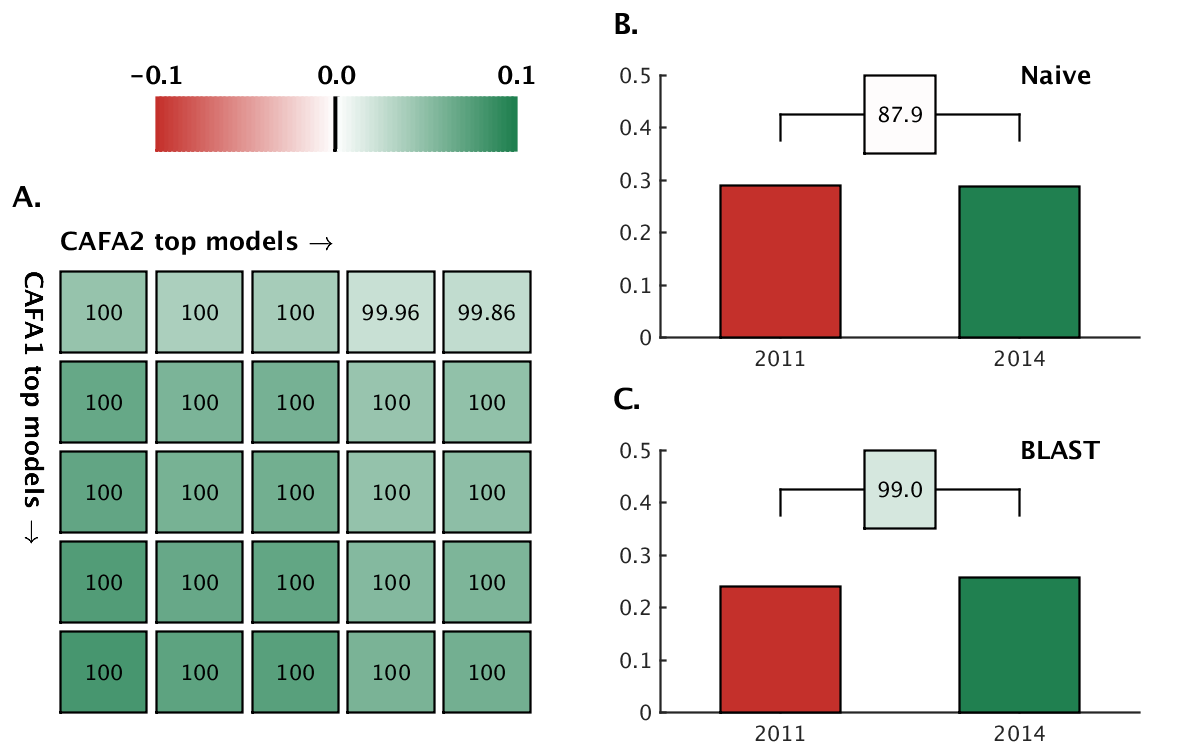}
\caption{CAFA1 versus CAFA2 (top methods).  A comparison in $\fmax$ between top~5 CAFA1 models against top~5 CAFA2 models. Colored boxes encode the results such that (1) colors indicate margins of a CAFA2 method over a CAFA1 method in $\fmax$ and (2) numbers in the box indicate the percentage of wins. For both MFO and BPO results, \textbf{A}. CAFA1 top~5 models (rows, from top to bottom) against CAFA2 top~5 models (columns, from left to right) \textbf{B}. Comparison of Na\"ive baselines trained respectively on SwissProt2011 and SwissProt2014. \textbf{C}. Comparison of BLAST baselines trained on SwissProt2011 and SwissProt2014.}
  \label{fig:head2head}
\end{figure}

To conduct a ``head-to-head'' analysis between any two methods, we generated $B = 10,000$ bootstrap samples and let methods compete on each such benchmark set. The average performance metric as well as the number of wins were recorded. Figure~\ref{fig:head2head} summarizes the results of this analysis. We use a color code from green to red to indicate the performance improvement $\delta$ from CAFA1 to CAFA2,
\begin{eqnarray*}
  \delta(m_{2}, m_{1}) = 
  \frac{1}{n}\sum_{i=1}^{n}\fmax^{(i)}(m_{2}) -
  \frac{1}{n}\sum_{i=1}^{n}\fmax^{(i)}(m_{1})
\end{eqnarray*}
\noindent where $m_{1}$ and $m_{2}$ stand for methods from CAFA1 and CAFA2, respectively, and $\fmax^{(i)}(\cdot)$ represents the $\fmax$ of a method evaluated on the $i$-th bootstrapped benchmark set. The selection of top methods for this study was based on their performance in each ontology on the entire benchmark sets. Panels B and C in Figure~\ref{fig:head2head} show the comparison between baseline methods trained on different data sets. We see no improvements of these baselines except for BLAST on BPO where it is slightly better to use the newer version of Swiss-Prot as the reference database for the search. On the other hand, all top methods in CAFA2 outperformed their counterparts in CAFA1. For predicting molecular functions, even though transferring functions from BLAST hits does not give better results, the top models still managed to perform better. It is possible that the newly acquired annotations since CAFA1 enhanced BLAST, which involves direct function transfer, and perhaps lead to better performances of those ``downstream'' methods that rely on sequence alignments. However, this effect does not completely explain the extent of performance improvement achieved by those methods. This is promising evidence that top methods from the community have improved since CAFA1 and that the improvement was not simply due to updates of curated databases.

\subsection*{Diversity of methodology}
We analyzed the extent to which methods generated similar predictions within each ontology. We calculated the pairwise Pearson correlation between methods on a common set of gene-concept pairs and then visualized these similarities as networks (Supplementary Materials).


In the molecular function ontology, where we observed the highest overall performance of prediction methods, eight of ten top methods were in the largest connected component. In addition, we observed a high connectivity between methods, suggesting that the participating methods are leveraging similar sources of data in similar ways. Predictions for the biological process ontology showed a contrasting pattern. In this ontology, the largest connected component contained only two of the top ten methods. The other top methods were contained in components made up of other methods produced by the same lab. This suggests that the approaches that participating groups have taken generate more diverse predictions for this ontology and that there are many different paths to a top performing biological process prediction method. Results for the human phenotype ontology were more similar to the biological process ontology, while results for cellular component were more similar in structure to molecular function. 

Taken together, these results suggest that ensemble approaches that aim to include independent sources of high quality predictions may benefit from leveraging the data and techniques used by different research groups and that such approaches that effectively weigh and integrate disparate methods may demonstrate more substantial improvements over existing methods in the process and phenotype ontologies where current prediction approaches share less similarity.

%% file: conclusions.tex
\section*{Conclusions}


Accurately annotating the function of biological macromolecules is difficult, and requires the concerted effort of experimental scientists, biocurators, and computational biologists. We conducted the second CAFA challenge to assess the status of computational function prediction of proteins and to quantify the progress in the field. Following the success of CAFA1 three years ago, we decided to significantly expand the number of protein targets, the number of biomedical ontologies used for annotation, the number of analysis scenarios, as well as the metrics used for evaluation. We believe the results of the CAFA2 experiment provide useful information on the status of the state-of-the-art in protein function prediction, can guide the development of new concept annotation methods, and help experimental studies through prioritization. Understanding the function of biological macromolecules brings us closer to understanding life at the molecular level and improving human health. 

\subsection*{The field has moved forward}
Three years ago, in CAFA1, we concluded that the top methods for function prediction outperform straightforward function transfer by homology. In CAFA2, we observe that the methods for function prediction have improved compared to those from CAFA1. As part of the CAFA1 experiment, we stored all predictions from all methods on 48,298 target proteins from 18 species. We used those stored predictions and compared them to the newly deposited predictions from CAFA2 on the overlapping set of benchmark proteins and CAFA1 targets. The head-to-head comparisons among top five CAFA1 methods against top five CAFA2 methods reveal that the top CAFA2 methods outperformed all top CAFA1 methods. 

Although it is difficult to disentangle the contributions of larger training sets from those of methodological novelties, the fact that the BLAST algorithm using the data from 2011 and data from 2014 showed little difference, led us to conclude that a larger share of the contribution likely belongs to the new methods. The experiences from CAFA1 and continuous AFP-SIG meetings every year during the ISMB conference where many new developments are readily shared may have contributed to this outcome \cite{Wass2014}.

\subsection*{Evaluation metrics}
A fair performance assessment in protein function prediction is far from straightforward. 
Although various evaluation metrics have been proposed under the framework of multi-label and structured-output learning, the evaluation in this subfield also needs to be interpretable to a broad community of researchers as well as the public. To address this, we used several metrics in this study as each provides useful insights and complements the others. Understanding the strengths and weaknesses of current metrics and developing better metrics remains important.

One important observation with respect to metrics is that the protein-centric and term-centric views may give different perspectives to the same problem. For example, while in the MFO and BPO we generally observe positive correlation between the two, in CCO and HPO these different metrics might lead to entirely different interpretations of the experiment. Regardless of the underlying cause, as discussed in Results, it is clear that some ontological terms are predictable with high accuracy and can be reliably used in practice even in these ontologies. In the meantime, more effort will be needed to understand the problems associated with statistical and computational aspects of method development.

In CAFA2 we introduced minimum semantic distance as another protein-centric metric \cite{Clark2013}. The investigation of the BLAST baseline reveals that the best local sequence identity cutoff for transferring experimental annotations from sequence hits occurs around $0.5$ for all three GO ontologies and just under ($0.35$) for HPO, if $\fmax$ is used as the evaluation metric. However, for $\smin$, these cutoffs are substantially higher to over $0.6$ for MFO, $0.7$ for HPO and surprisingly over $0.9$ for both BPO and CCO. We believe these higher thresholds provide biologically interesting results and have thus decided to use both $pr$-$rc$ curves and $ru$-$mi$ curves in protein-centric performance assessments. 

\subsection*{Well-performing methods}
We observe that participating methods usually specialize in one or a few categories of protein function prediction and have been developed with their own application objectives in mind. Therefore, performance rankings of methods often change from one benchmark set to another. There are complex factors that influence the final ranking including the selection of the ontology, types of benchmark sets and evaluation, as well as evaluation metrics, as discussed earlier. Most of our assessment results show that the performances of top-performing methods are generally comparable to each other. Thus, although a small group of methods could be considered as generally good, there is no single method that dominates over all benchmarks.

We also observed that when provided a chance to select a reliable set of predictions, the methods generally perform better (partial evaluation mode vs. full evaluation mode). Although most methods seem not to have been actively developed for the partial evaluation mode, this outcome is very encouraging. On the other hand, the limited-knowledge category of assessment seems to have not provided any boost in terms of performance accuracy. However, this was a new prediction category in CAFA and so few methods may have been optimized for prediction in the limited-knowledge scenario. Many important comparisons can be found in Supplementary Materials.

\subsection*{Final notes}
The automated functional annotation remains an exciting yet challenging task with implications relevant to the entirety of biomedical sciences. Three years after CAFA1, the top methods from the community have shown encouraging progress in both MFO and BPO categories. However, in terms of raw scores, there is still significant room for improvement in all ontologies, and particularly in BPO, CCO and HPO. There is also a need to develop an experiment-driven, as opposed to curation driven, component of the evaluation to address limitations for term-centric evaluation. In the future CAFA experiments, we will continue to monitor the performance over time and invite a broad range of computational biologists, computer scientists, statisticians and others to address these engaging problems of concept annotation for biological macromolecules through CAFA.


%% file: suppl3.tex
\renewcommand{\thesubfigure}{(\Alph{subfigure})}
\newcommand{\headstyle}{\bfseries\sffamily\LARGE}
\newcommand{\titlestyle}{\Large}
\newcommand{\authorstyle}{}
\newcommand{\journalstyle}{\itshape}
\newcommand{\labelstyle}{\bfseries\sffamily}
\newcommand{\speciesstyle}{\itshape}
\newcommand{\tablestyle}{\footnotesize}
\renewcommand{\figurename}{{\labelstyle Supplementary Figure}}
\renewcommand{\tablename}{{\labelstyle Supplementary Table}}
\bibliographystyle{plain}
%
%
\noindent{\headstyle Supplementary Information} \\[4ex]
\noindent{\titlestyle ``An expanded evaluation of protein function prediction methods shows an improvement in accuracy''} 


\vspace{.5in}

\noindent Content:
\begin{itemize}
  \item Supplementary Figures.
    \begin{enumerate}
      \item [1.] Benchmark annotation depth distribution.
      \item [2.] Top predictors, precision-recall curves.
      \item [3.] Benchmark sequence identity histogram.
      \item [4.] Top predictors, easy vs. difficult, precision-recall curves.
      \item [5.] Top predictors, eukarya vs. prokarya, precision-recall curves.
      \item [6.] Top predictors, species breakdown, $\fmax$ bars.
      \item [7.] Top predictors, weighted precision-recall curves.
      \item [8.] Top predictors, normalized remaining uncertainty-misinformation.
      \item [9.] Similarity networks between methods.
     
    \end{enumerate}
  \item Supplementary Table 1. Participating teams.
  \item Supplementary Table 2. Keywords.
\end{itemize}

\vspace{.5in}

\noindent Additional supplementary data (297MB) provides all additional data, analyses and full prediction results for every method. It is available at:

\vspace{.125in}

\url{https://dx.doi.org/10.6084/m9.figshare.2059944}

\vspace{.5in}

\noindent Code used in CAFA2 is available at: 

\vspace{.125in}

\url{https://github.com/yuxjiang/CAFA2}


%
%
%
%

\newpage

\paragraph{\labelstyle Supplementary Figure 1}
Distribution of depths of the leaf annotations, over all benchmarks in (A) Molecular Function ontology, (B) Biological Process ontology, (C) Cellular Component ontology and (D) Human Phenotype ontology. A leaf term for a benchmark protein is defined as any term whose descendant nodes (more specific nodes) are not among the experimentally determined terms for that protein.\\

\vspace{.5in}
\begin{figure}[hp!]
  \centering
  \subfigure[]{\includegraphics[width=.48\textwidth]{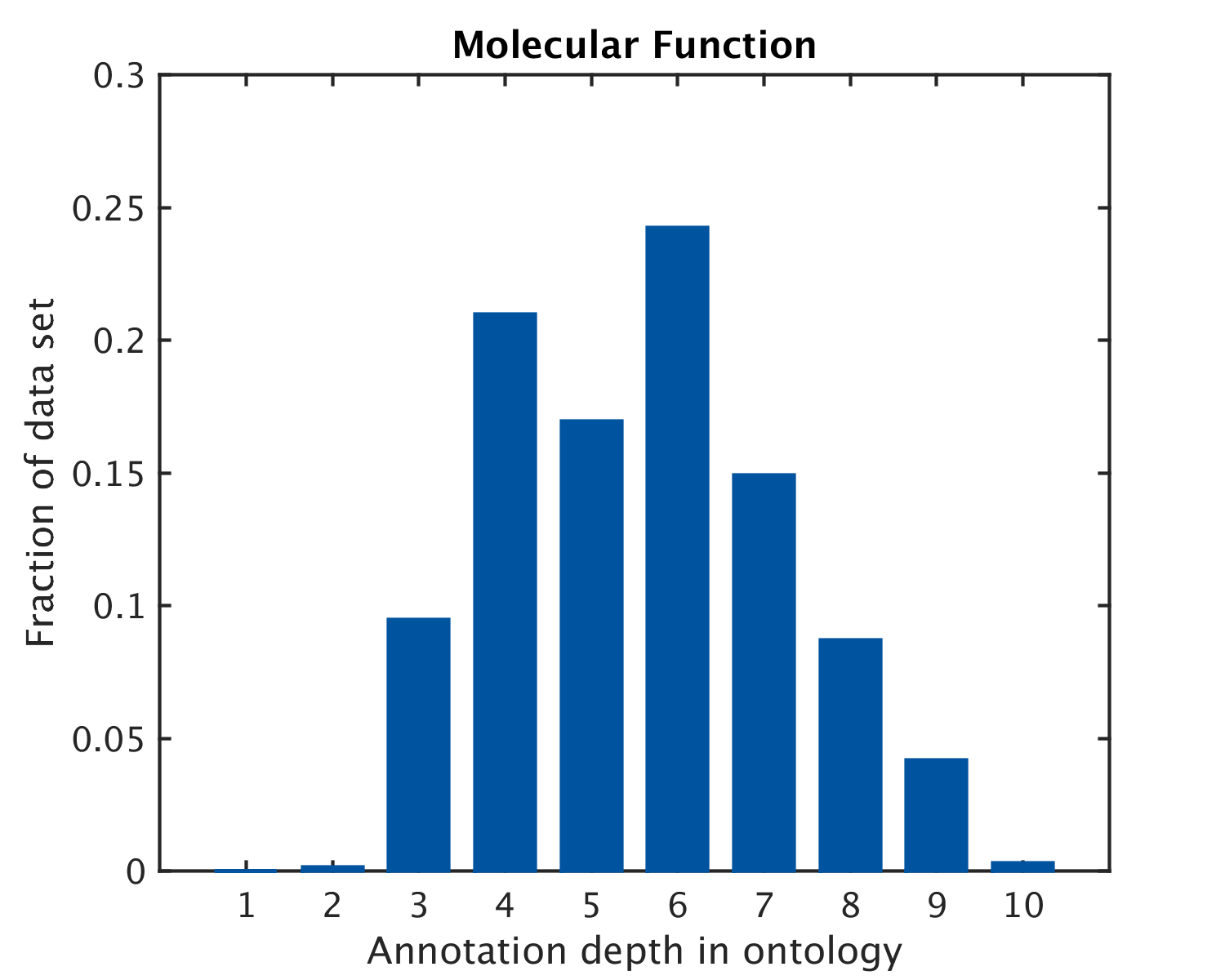}}
  \subfigure[]{\includegraphics[width=.48\textwidth]{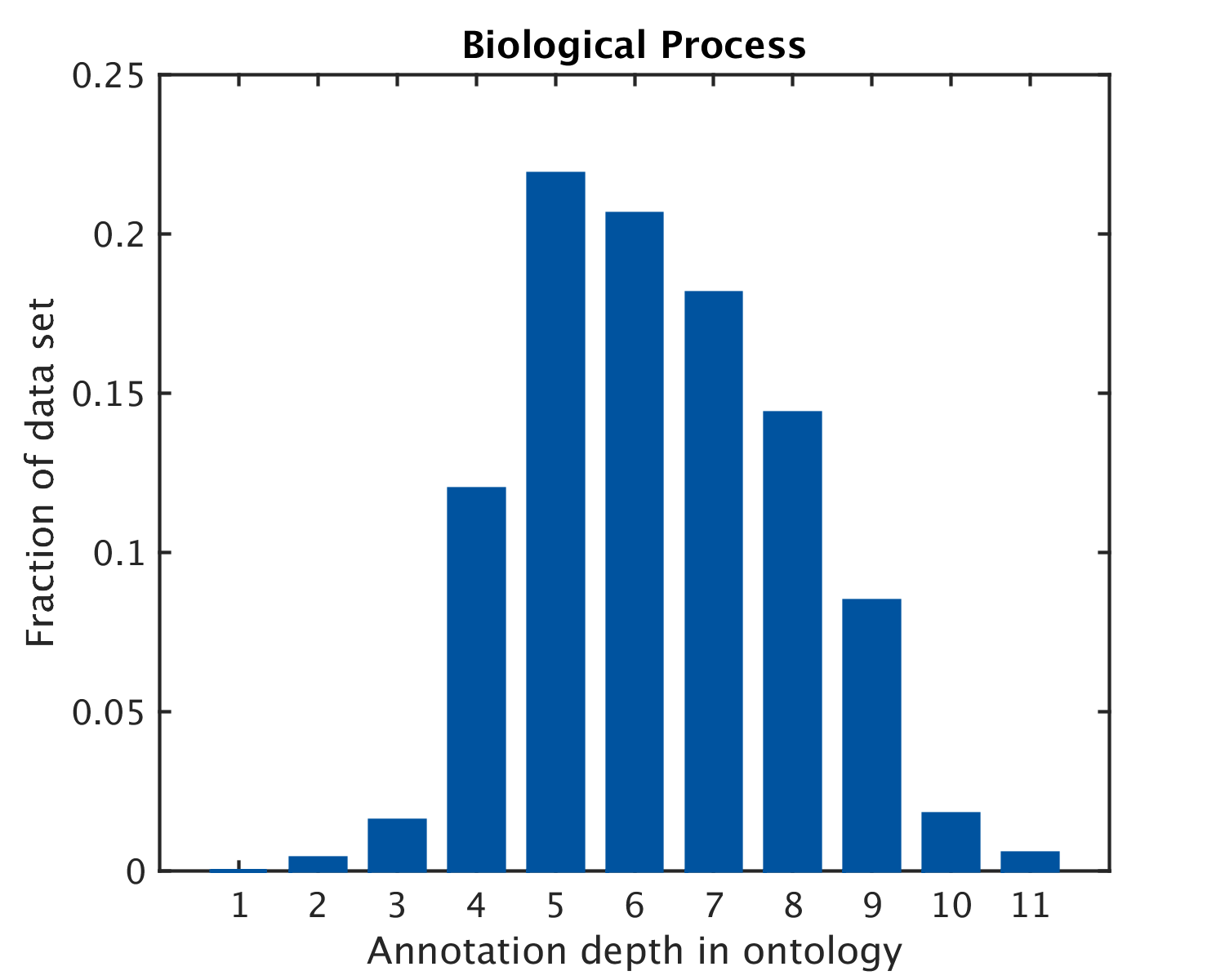}}\\
  \subfigure[]{\includegraphics[width=.48\textwidth]{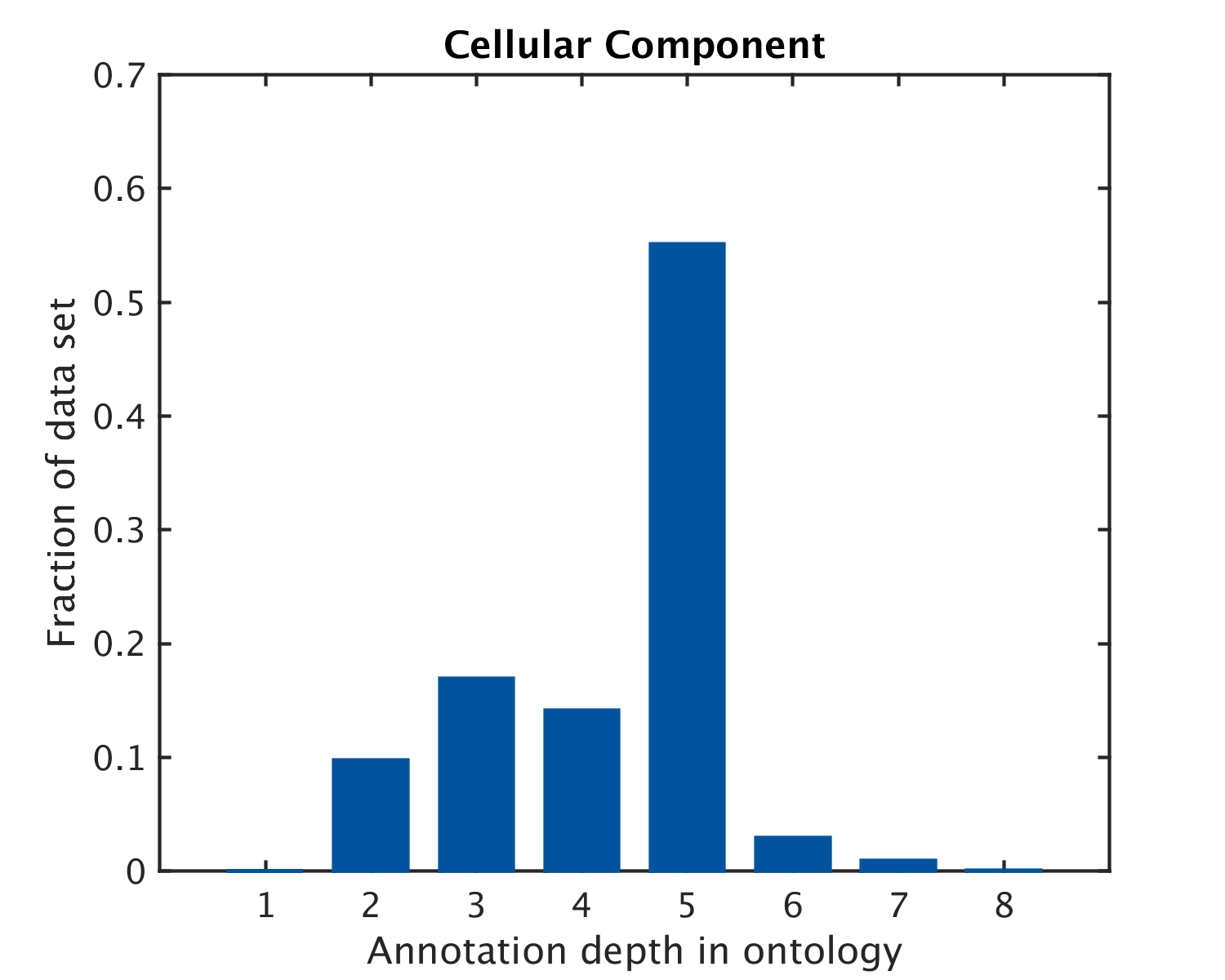}}
  \subfigure[]{\includegraphics[width=.48\textwidth]{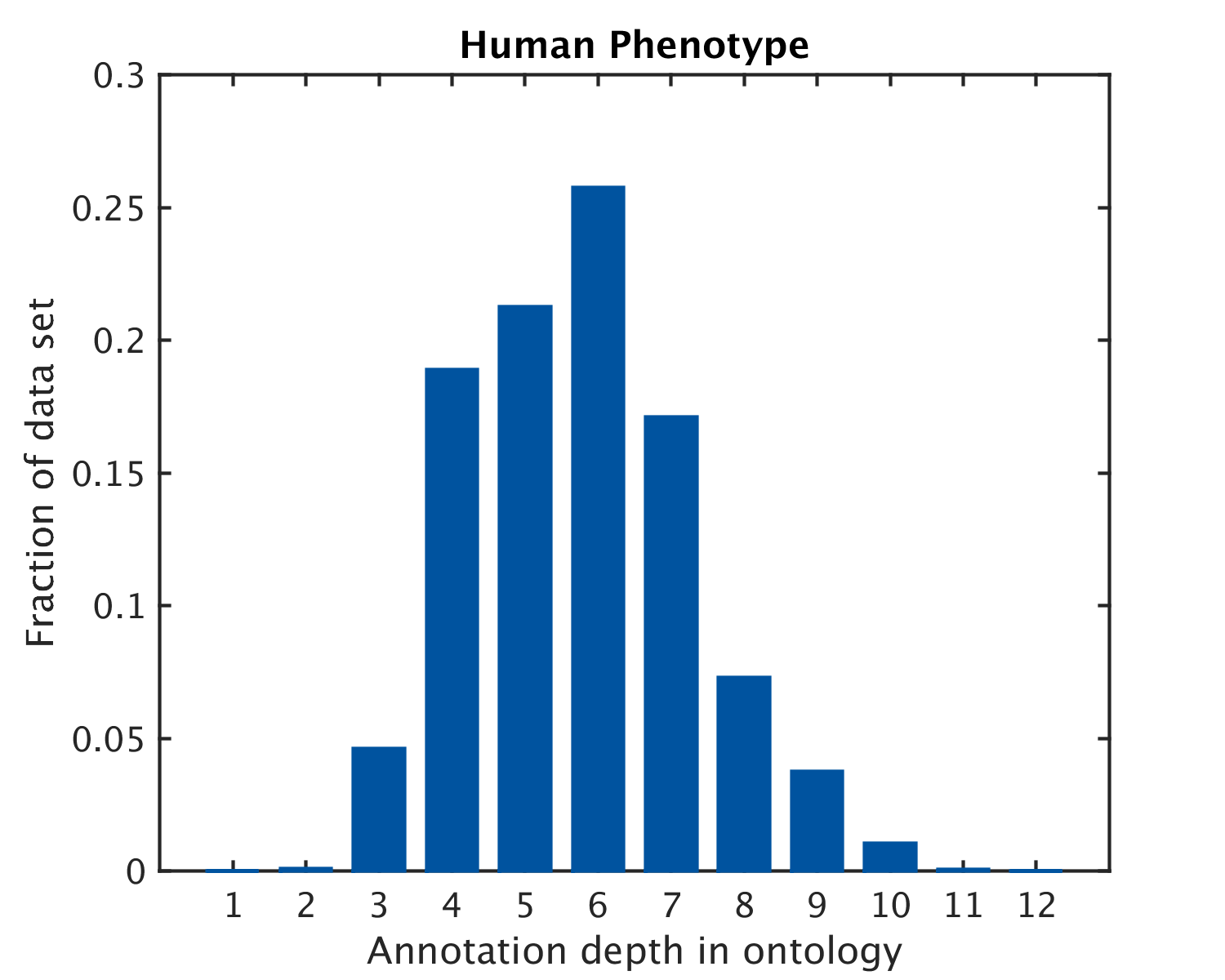}}
\end{figure}

\newpage

\paragraph{\labelstyle Supplementary Figure 2}
Precision-recall curves for the top-performing methods for (A) Molecular Function ontology, (B) Biological Process ontology, (C) Cellular Component ontology and (D) Human Phenotype ontology. All panels show the top ten participating methods in each category, as well as the Na\"ive and BLAST baseline methods. Points corresponding to the maximum F-measure are marked in circles on each curve. The legend provides the maximum F-measure ($F$) and coverage ($C$) for all methods. In cases where a Principal Investigator (PI) participated with multiple teams, only the results of the best scoring method are presented.\\

\vspace{.5in}
\noindent Supplementary Figure 2A:
\begin{center}
  \includegraphics[width=\textwidth]{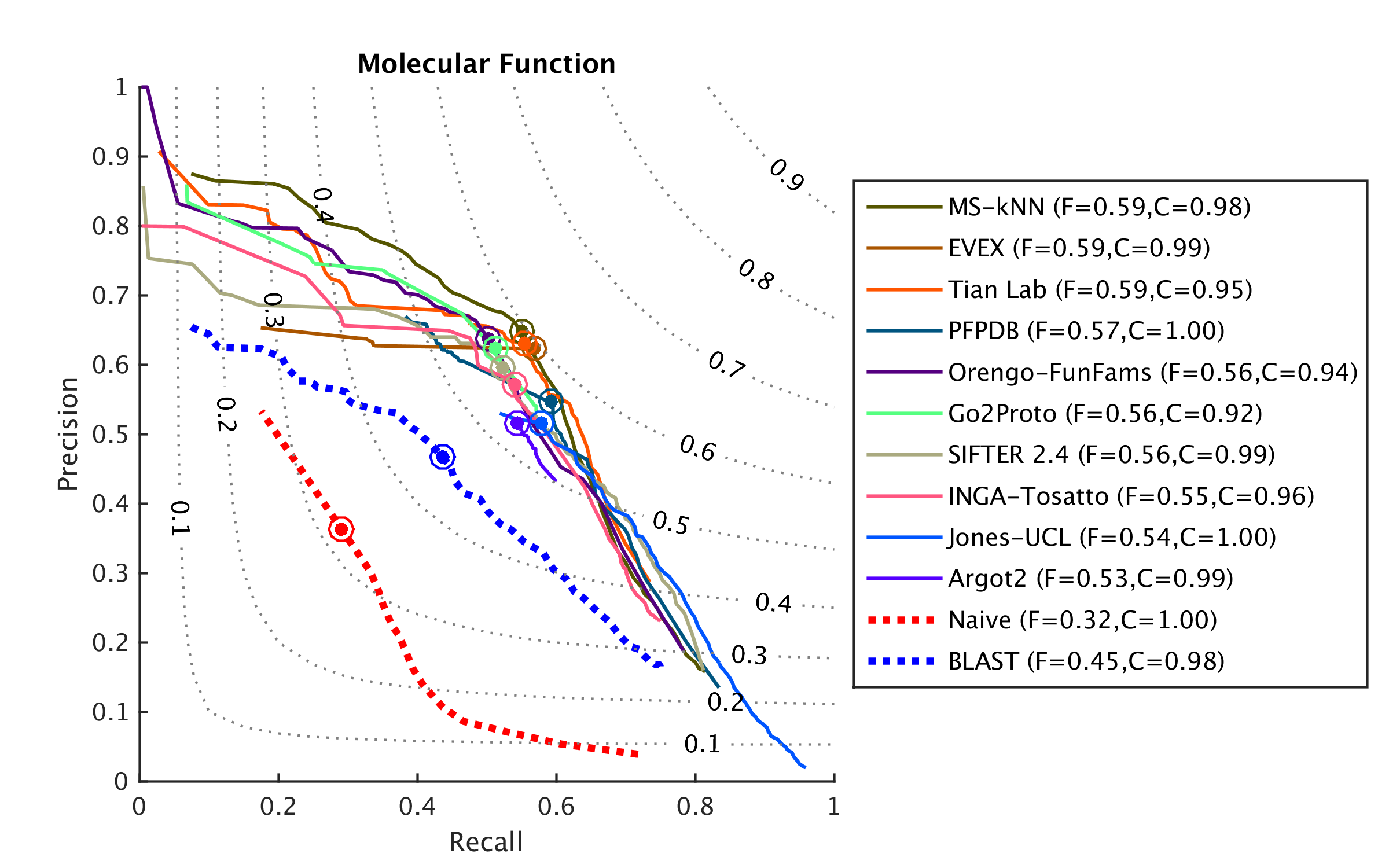}
\end{center}

\newpage

\noindent Supplementary Figure 2B:
\begin{center}
  \includegraphics[width=\textwidth]{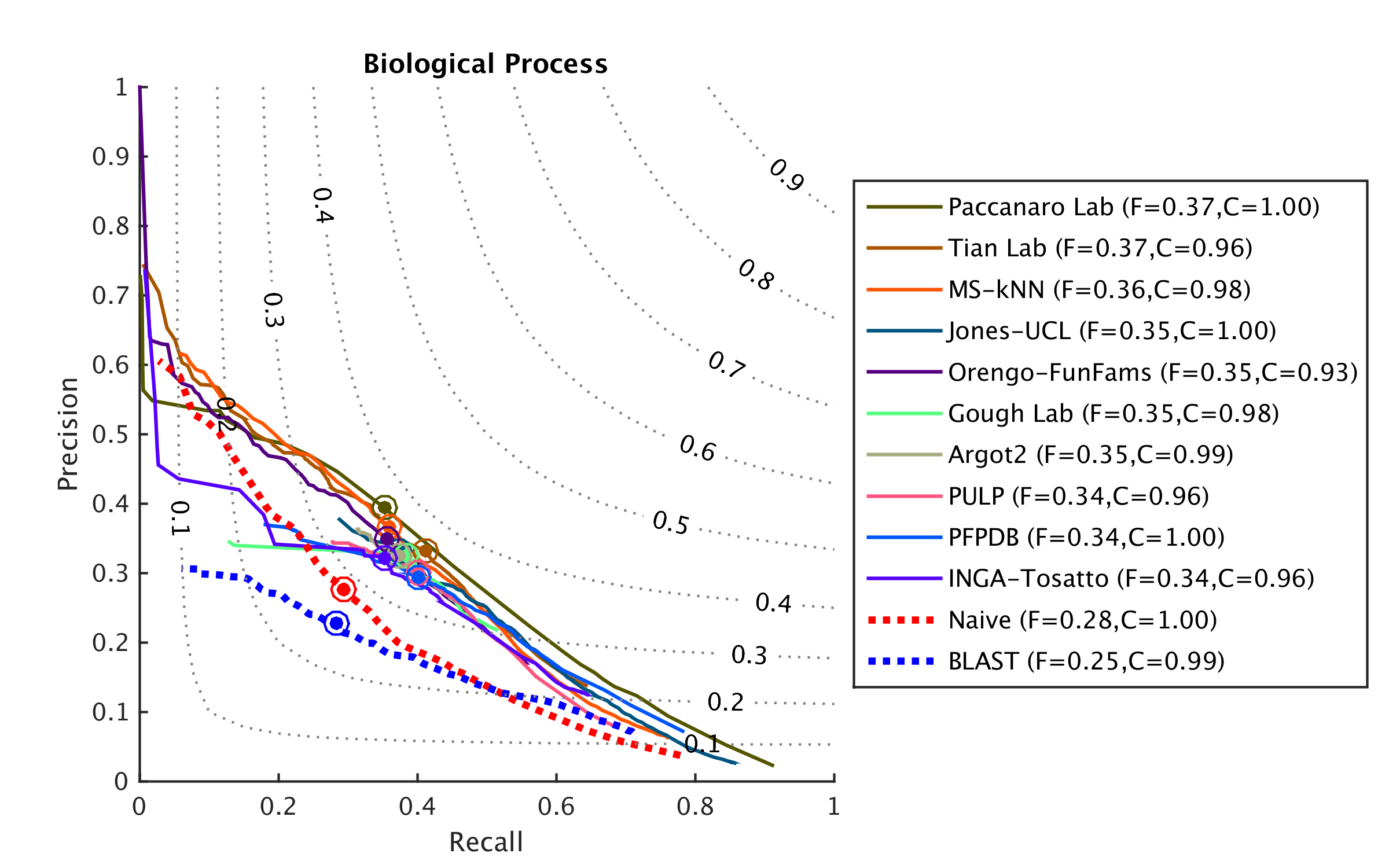}
\end{center}

\newpage

\noindent Supplementary Figure 2C:
\begin{center}
  \includegraphics[width=\textwidth]{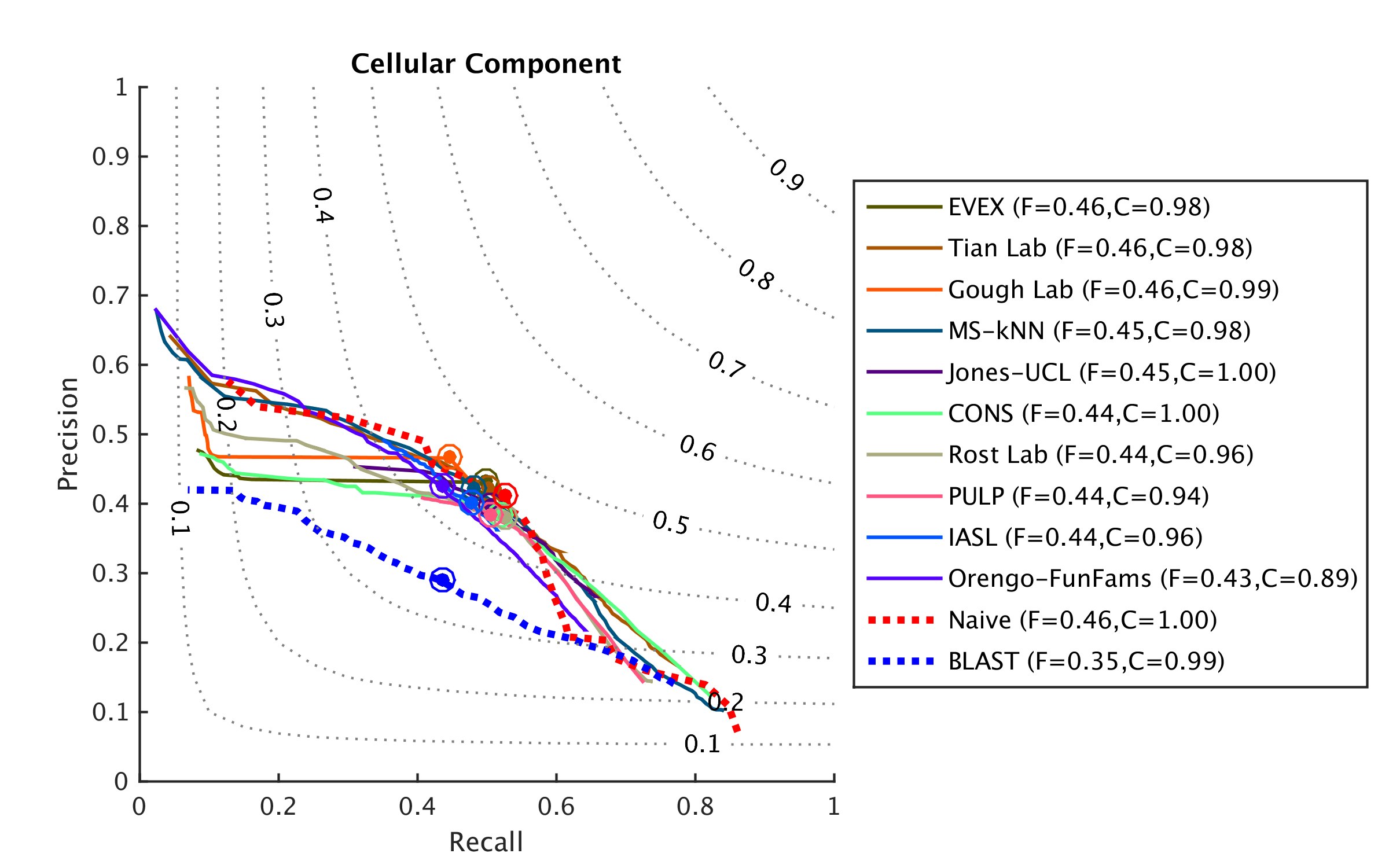}
\end{center}

\newpage

\noindent Supplementary Figure 2D:
\begin{center}
  \includegraphics[width=\textwidth]{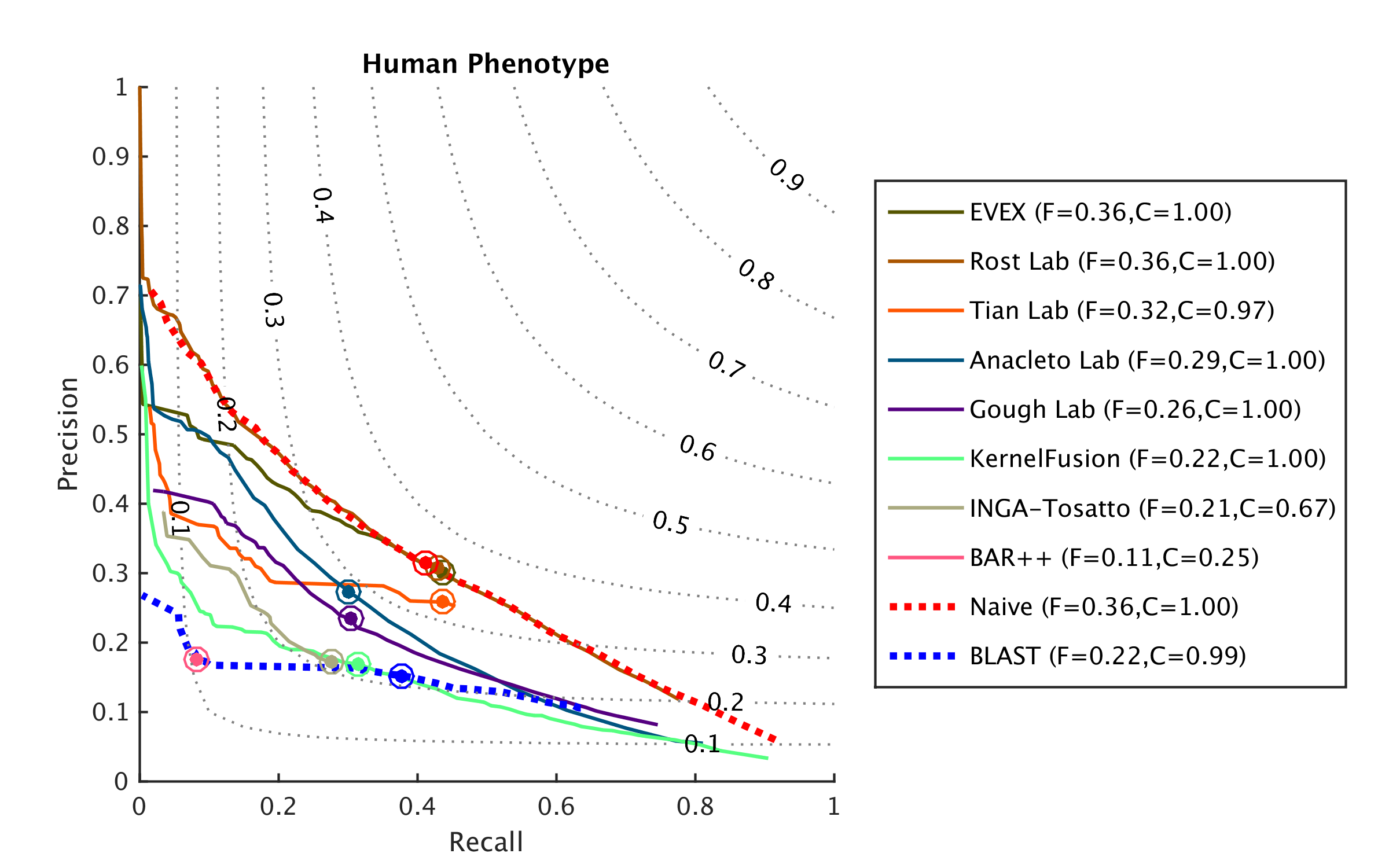}
\end{center}

\newpage

\paragraph{\labelstyle Supplementary Figure 3}
The histogram of pairwise sequence identities between each benchmark proteins and the experimentally annotated template most similar to it: (A) Molecular Function ontology, (B) Biological Process ontology, and (C) Cellular Component ontology. The histograms roughly determine two groups of benchmarks: \emph{easy} -- with maximum global sequence identity greater than or equal to $60\%$, and \emph{difficult} -- with maximum global sequence identity below $60\%$.\\

\vspace{.5in}
\begin{figure}[hp!]
  \centering
  \setcounter{subfigure}{0}
  \subfigure[]{\includegraphics[width=.48\textwidth]{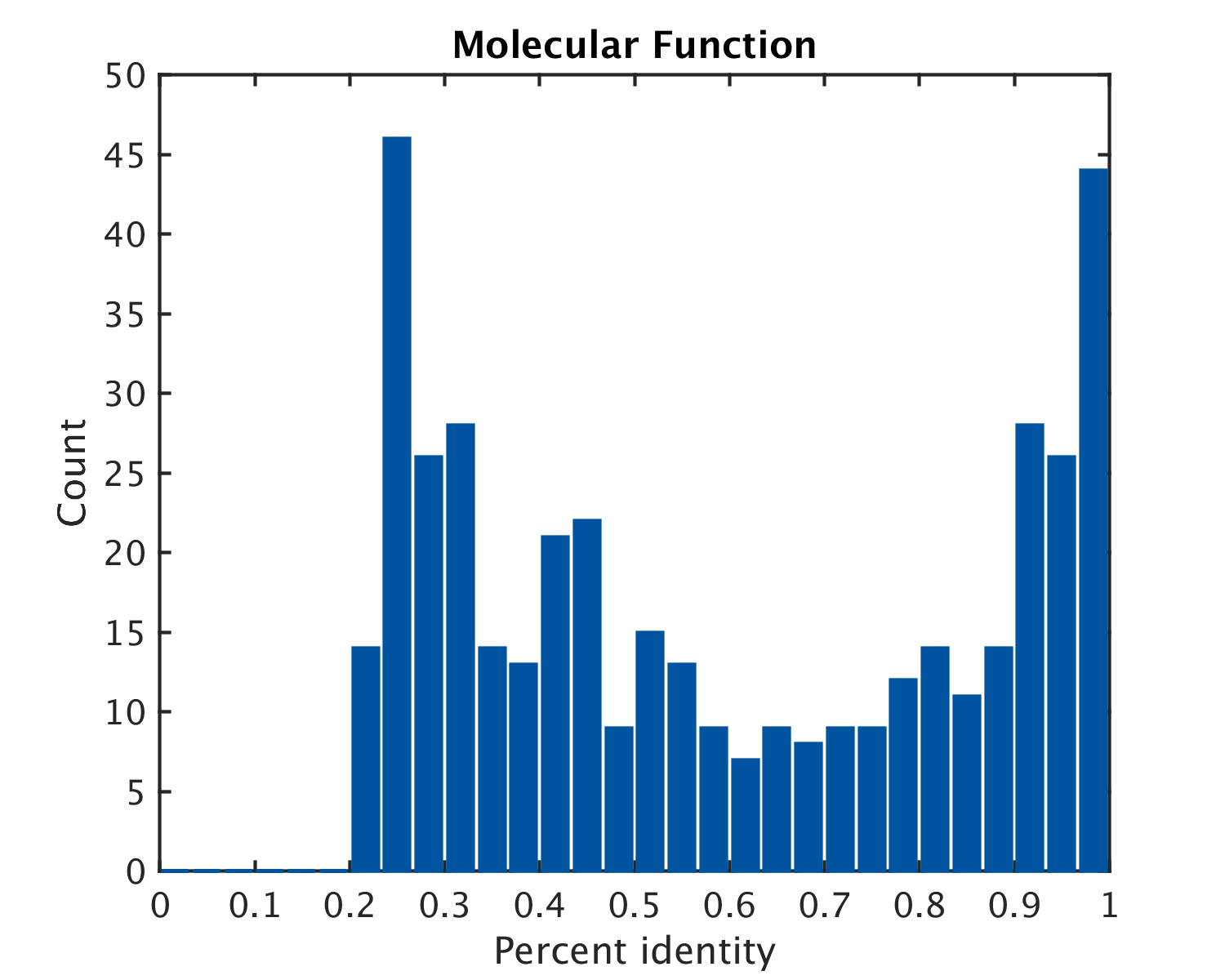}}
  \subfigure[]{\includegraphics[width=.48\textwidth]{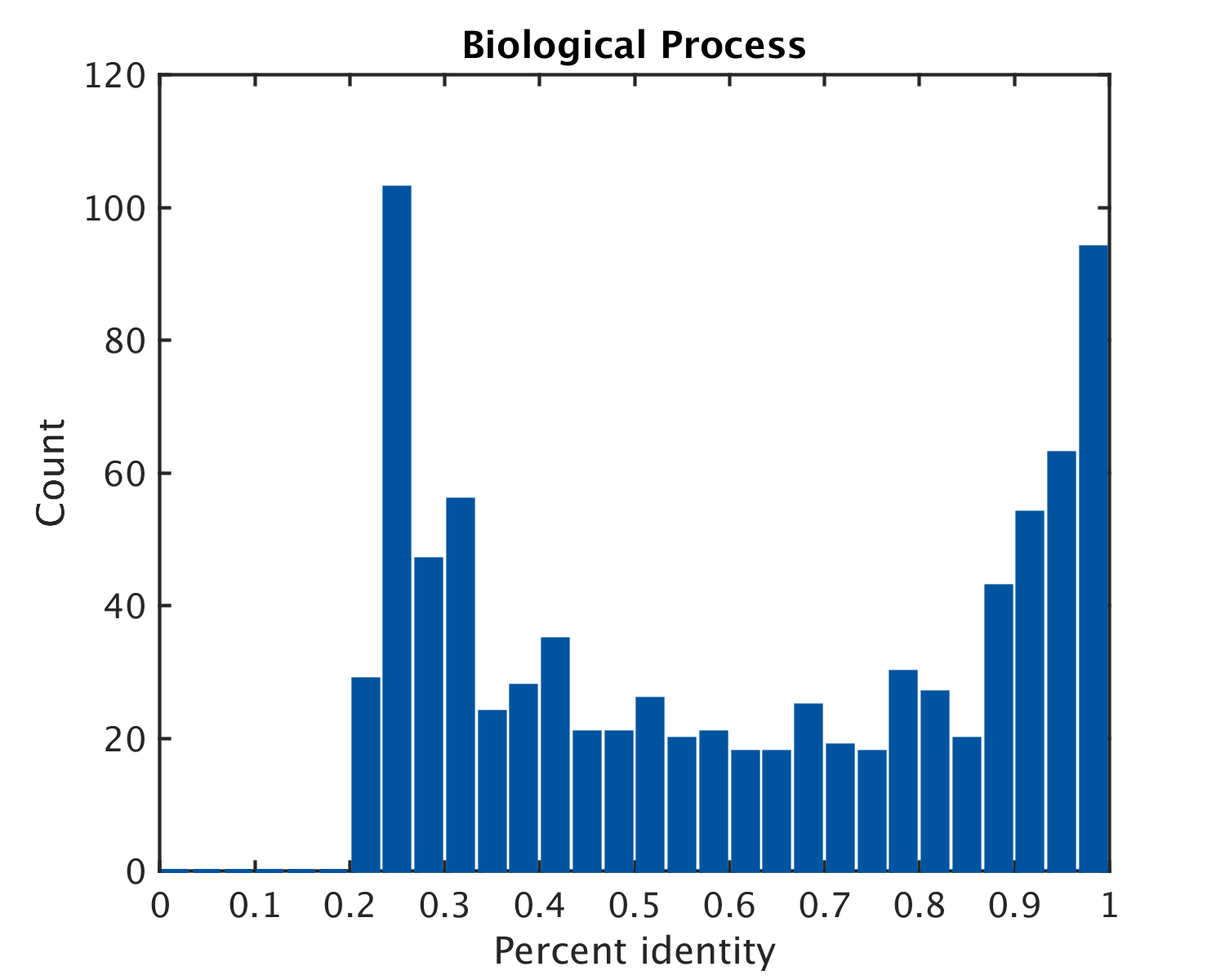}}\\
  \subfigure[]{\includegraphics[width=.48\textwidth]{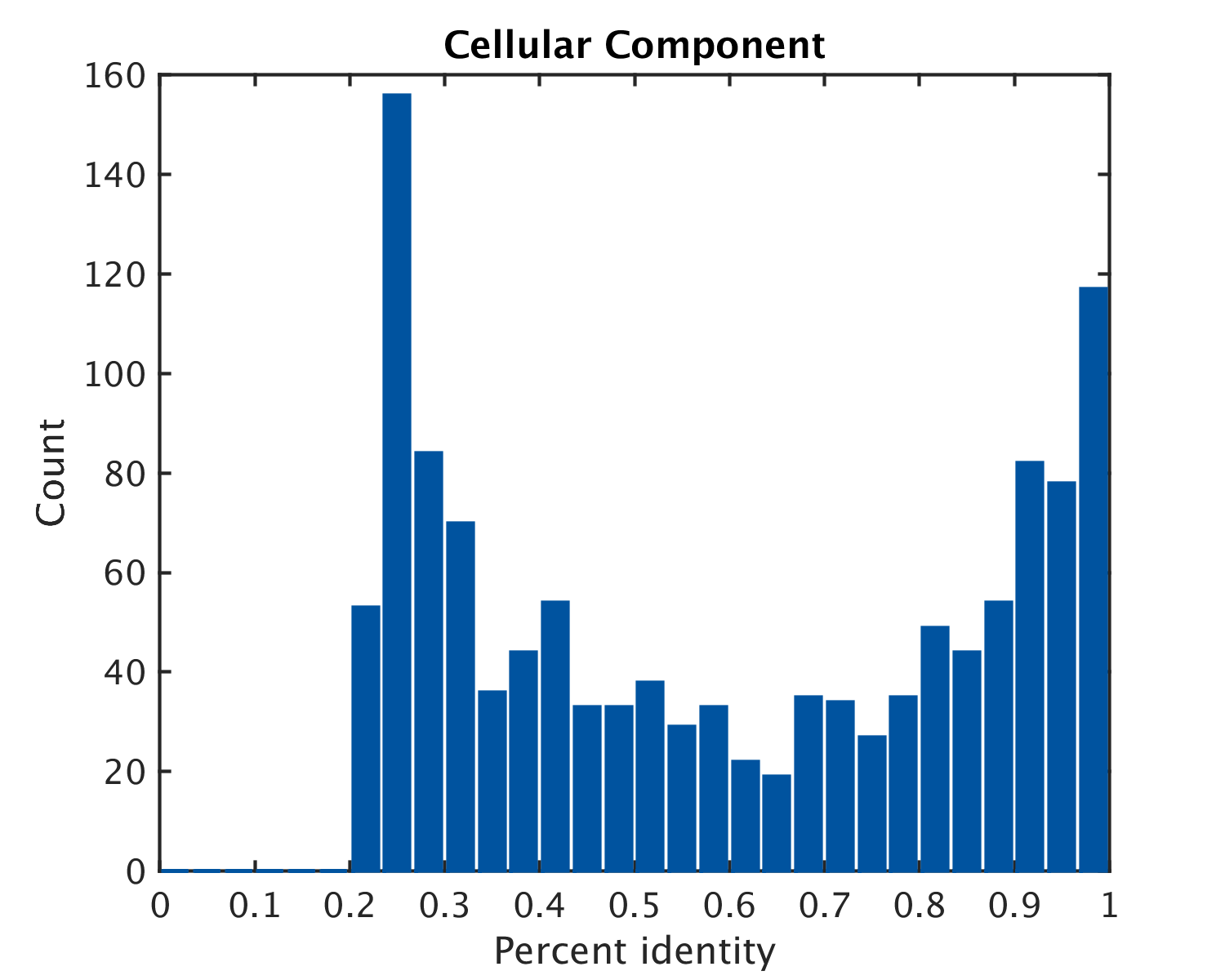}}
\end{figure}

\newpage

\paragraph{\labelstyle Supplementary Figure 4}
Precision-recall curves for the top-performing methods for (A) easy benchmark category and Molecular Function ontology, (B) difficult benchmark category and Molecular Function ontology, (C) easy benchmark category and Biological Process ontology, (D) difficult benchmark category and Biological Process ontology, (E) easy benchmark category and Cellular Component ontology and (F) difficult benchmark category and Cellular Component ontology. All panels show the top ten participating methods in each category, as well as the Na\"ive and BLAST baseline methods. Points corresponding to the maximum F-measure are marked in circles on each curve. The legend provides the maximum F-measure ($F$) and coverage ($C$) for all methods. In cases where a Principal Investigator (PI) participated with multiple teams, only the results of the best scoring method are presented.\\

\newpage 

\noindent Supplementary Figure 4A (easy):
\begin{center}
  \includegraphics[width=\textwidth]{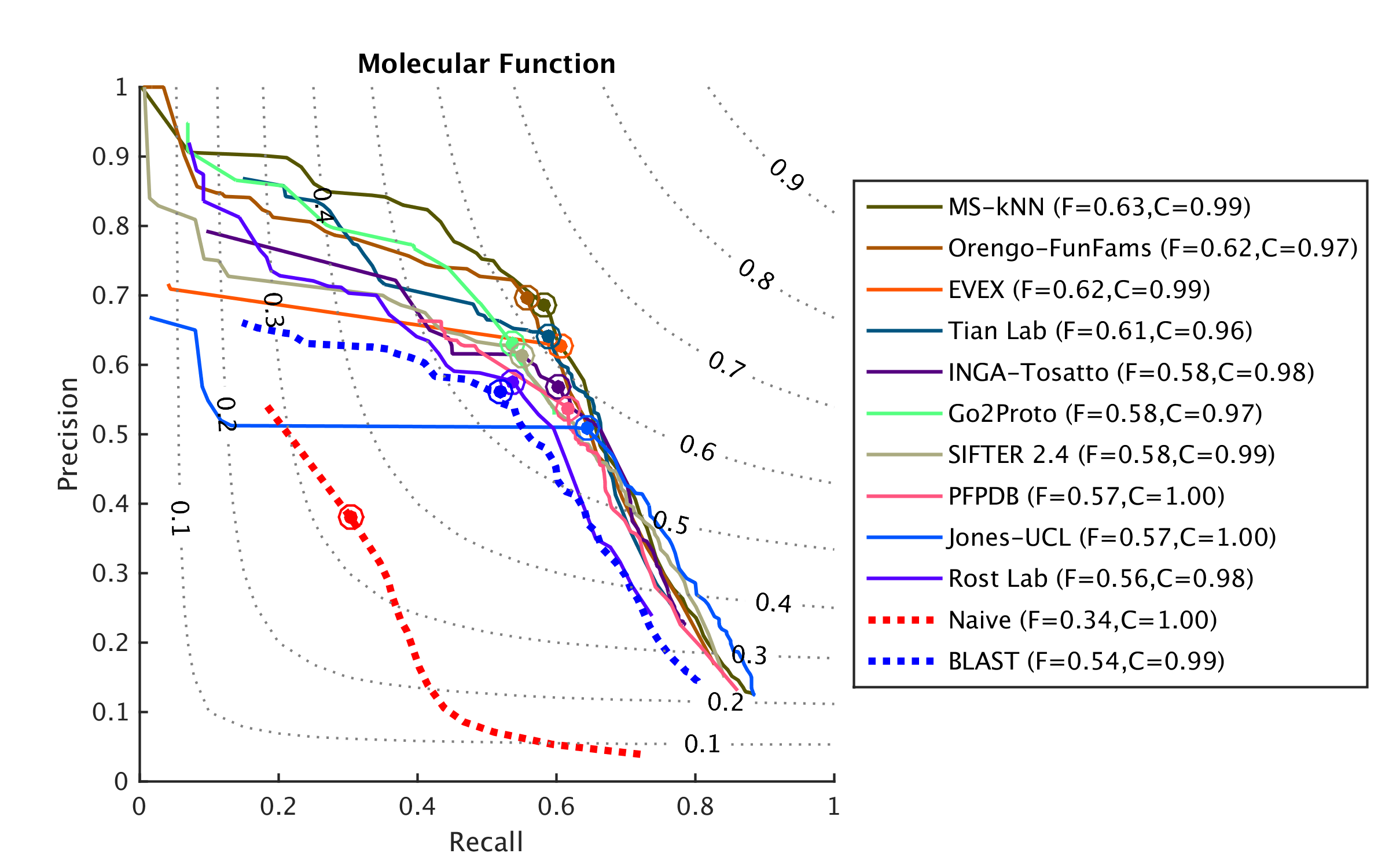}
\end{center}

\noindent Supplementary Figure 4B (difficult):
\begin{center}
  \includegraphics[width=\textwidth]{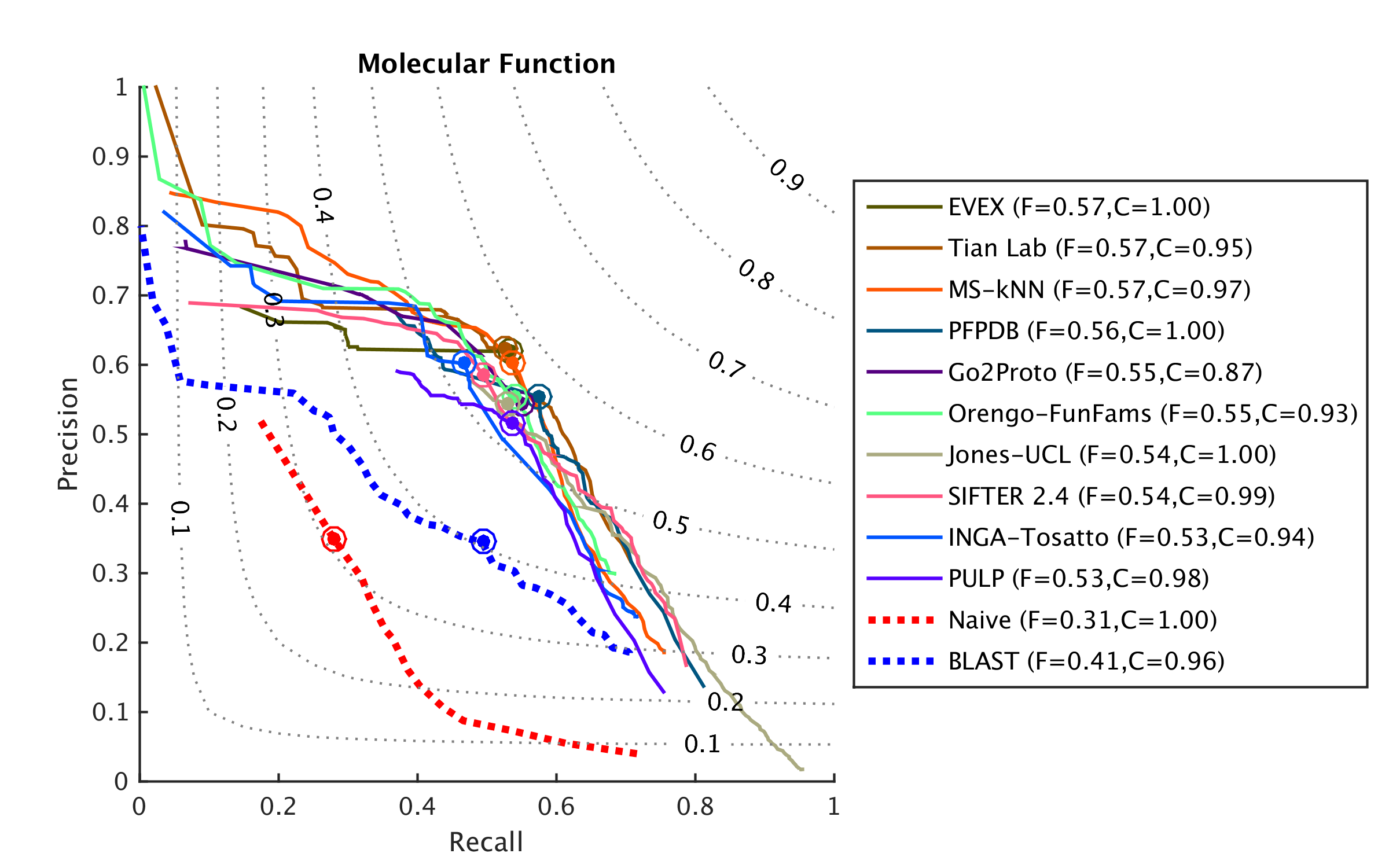}
\end{center}

\newpage

\noindent Supplementary Figure 4C (easy):
\begin{center}
  \includegraphics[width=\textwidth]{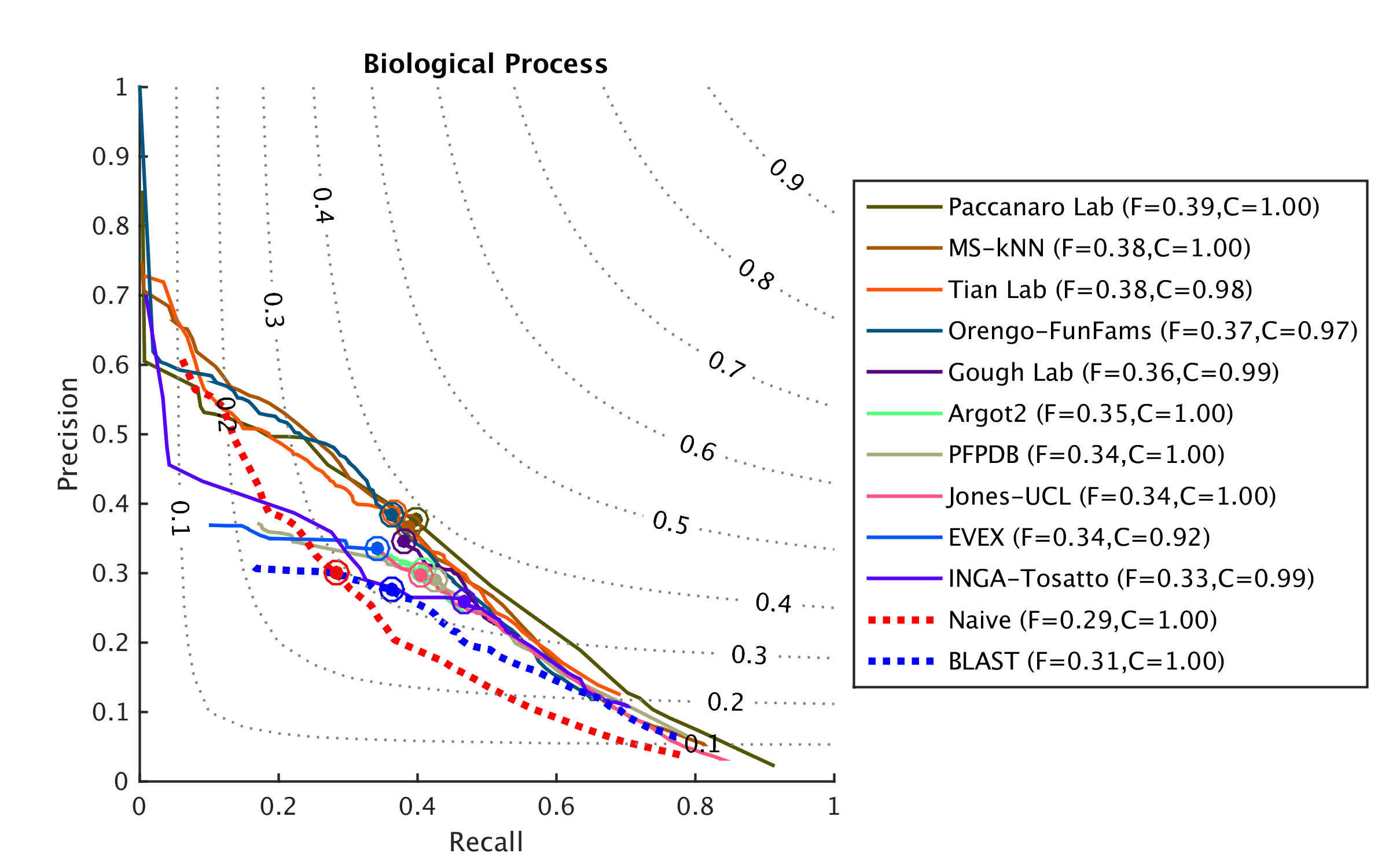}
\end{center}

\noindent Supplementary Figure 4D (difficult):
\begin{center}
  \includegraphics[width=\textwidth]{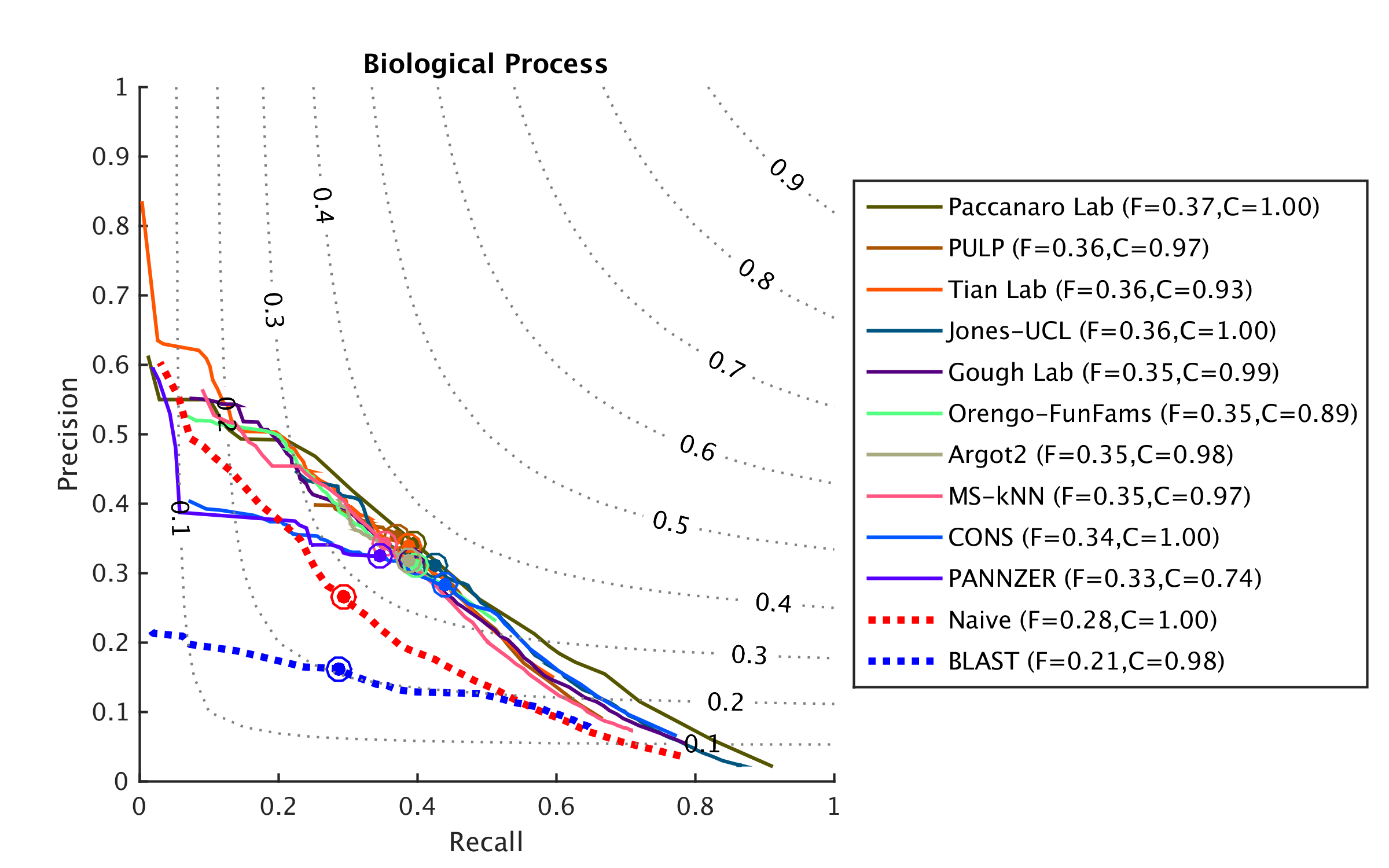}
\end{center}

\newpage

\noindent Supplementary Figure 4E (easy):
\begin{center}
  \includegraphics[width=\textwidth]{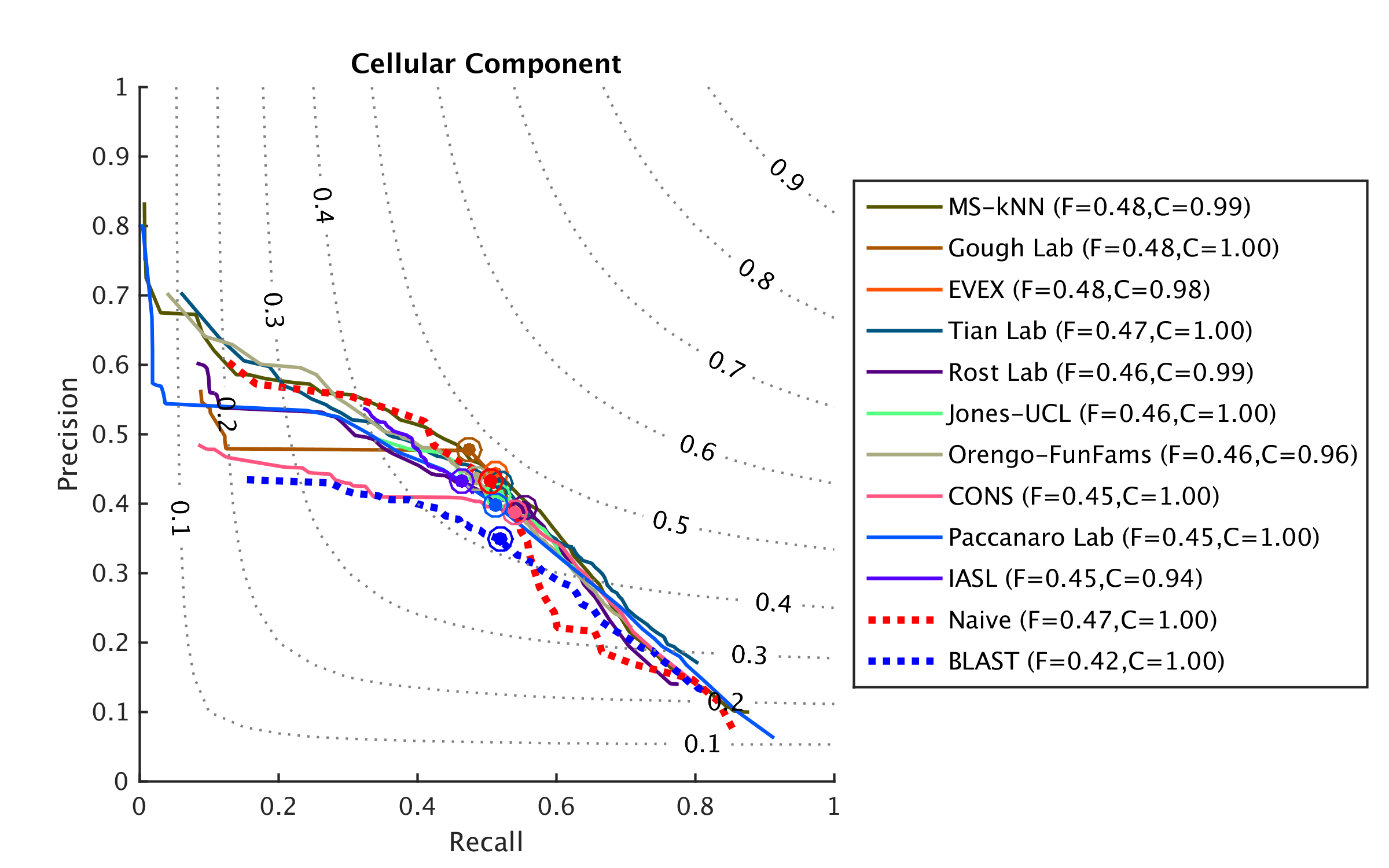}
\end{center}

\noindent Supplementary Figure 4F (difficult):
\begin{center}
  \includegraphics[width=\textwidth]{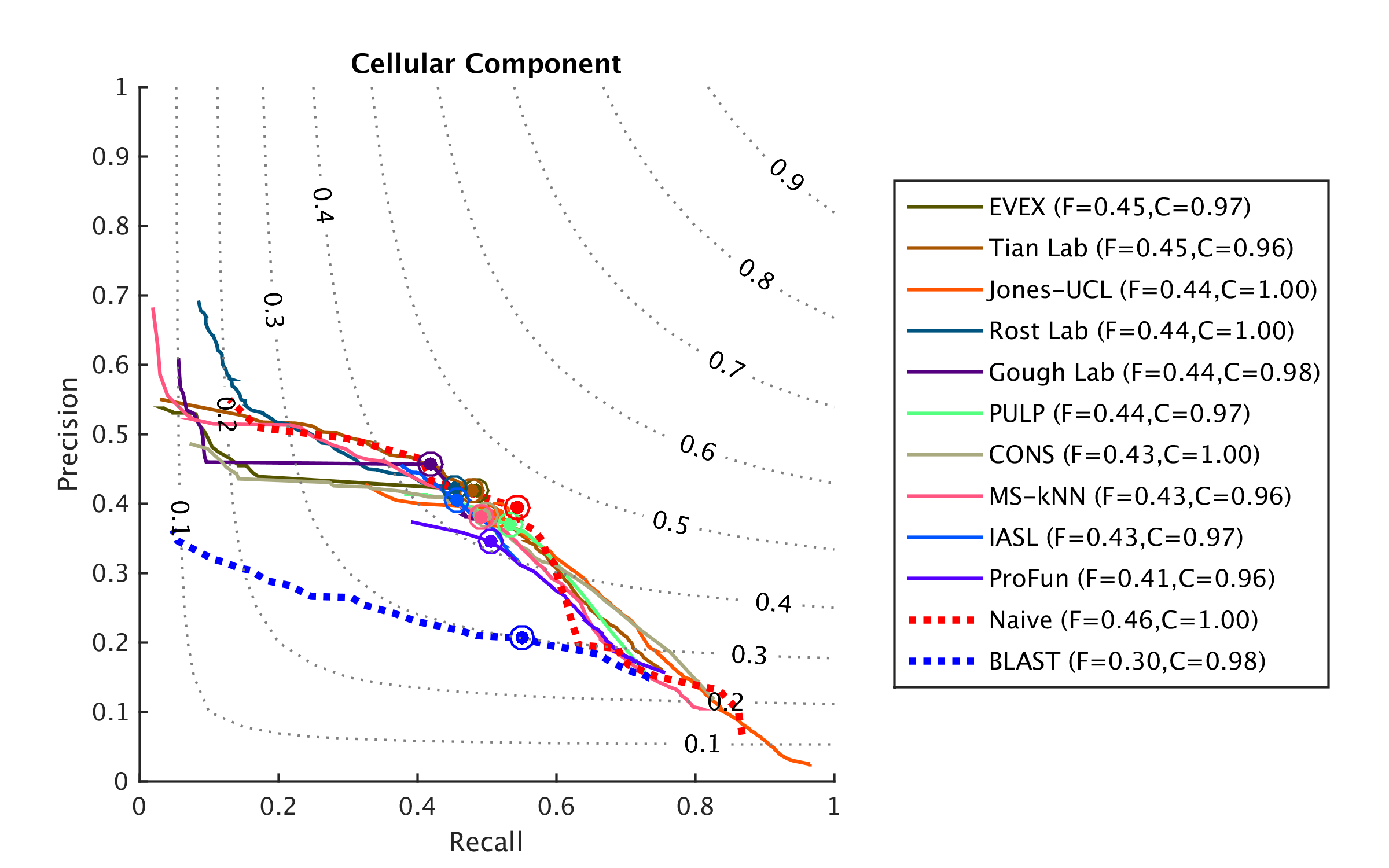}
\end{center}

\newpage

\paragraph{\labelstyle Supplementary Figure 5}
Precision-recall curves for the top-performing methods for (A) eukaryotic benchmark category and Molecular Function ontology, (B) prokaryotic benchmark category and Molecular Function ontology, (C) eukaryotic benchmark category and Biological Process ontology, (D) prokaryotic benchmark category and Biological Process ontology, (E) eukaryotic benchmark category and Cellular Component ontology and (F) prokaryotic benchmark category and Cellular Component ontology. All panels show the top ten participating methods in each category, as well as the Na\"ive and BLAST baseline methods. Points corresponding to the maximum F-measure are marked in circles on each curve. The legend provides the maximum F-measure ($F$) and coverage ($C$) for all methods. In cases where a Principal Investigator (PI) participated with multiple teams, only the results of the best scoring method are presented.\\

\newpage

\noindent Supplementary Figure 5A (eukarya):
\begin{center}
  \includegraphics[width=\textwidth]{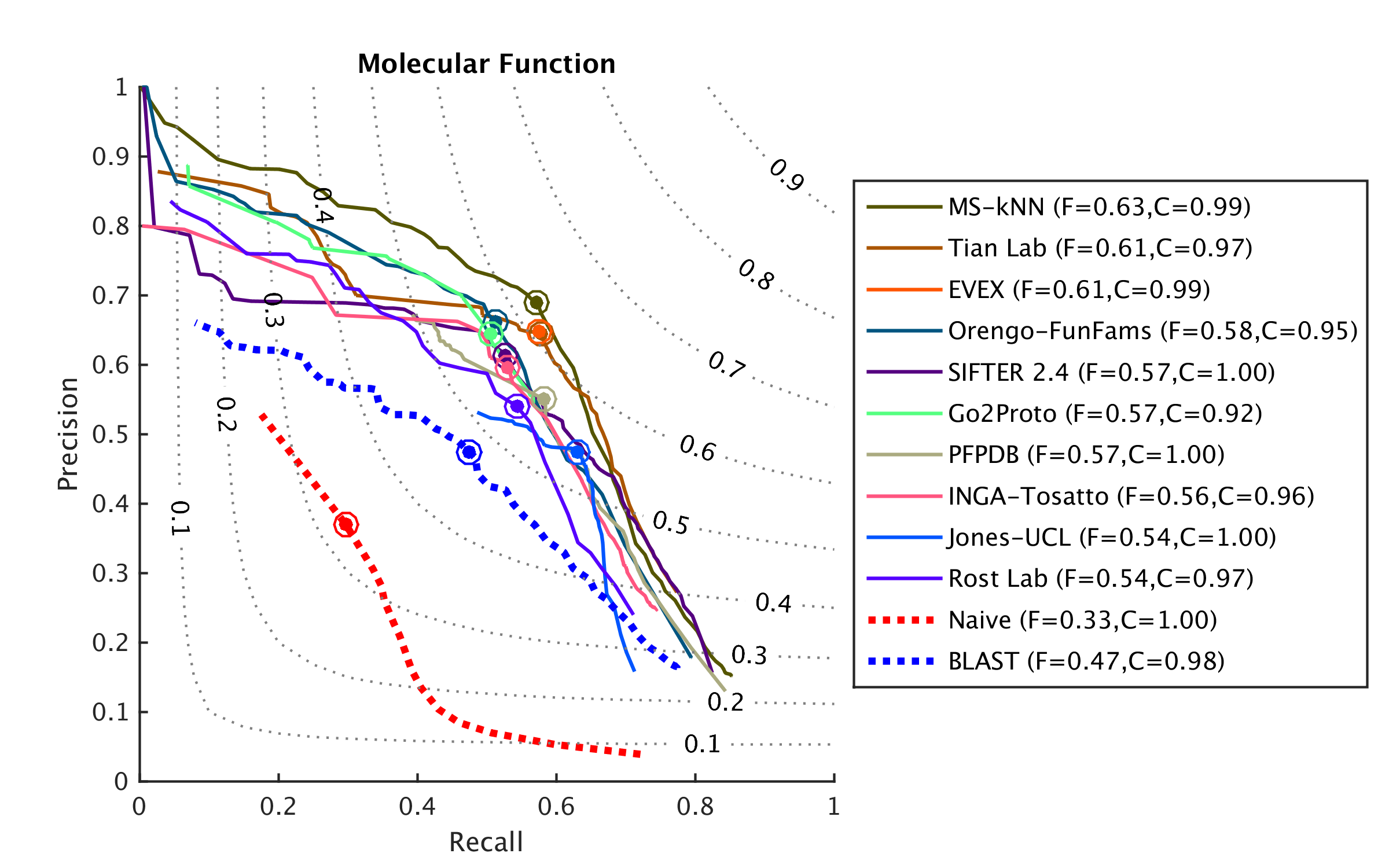}
\end{center}

\noindent Supplementary Figure 5B (prokarya):
\begin{center}
  \includegraphics[width=\textwidth]{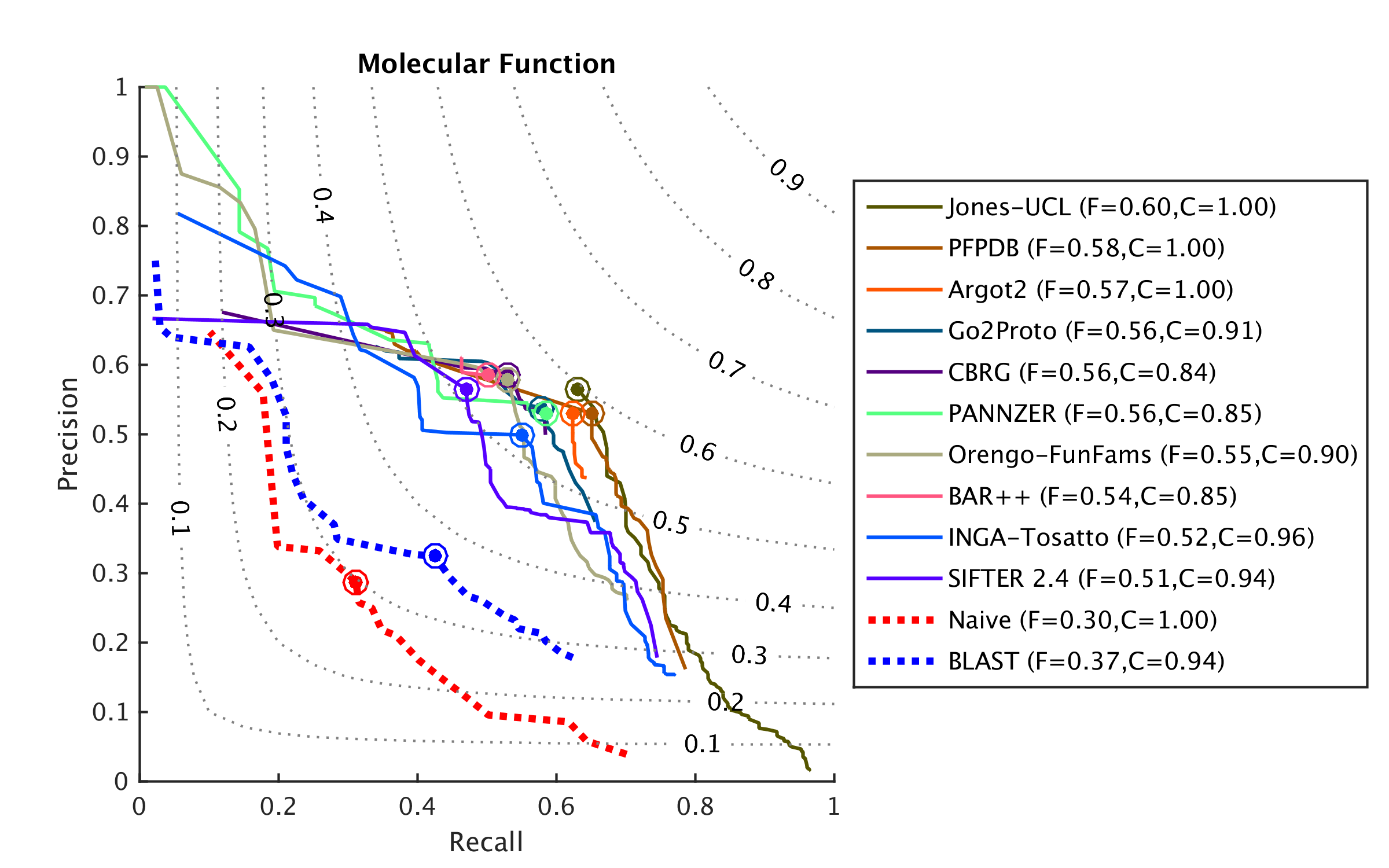}
\end{center}

\newpage

\noindent Supplementary Figure 5C (eukarya):
\begin{center}
  \includegraphics[width=\textwidth]{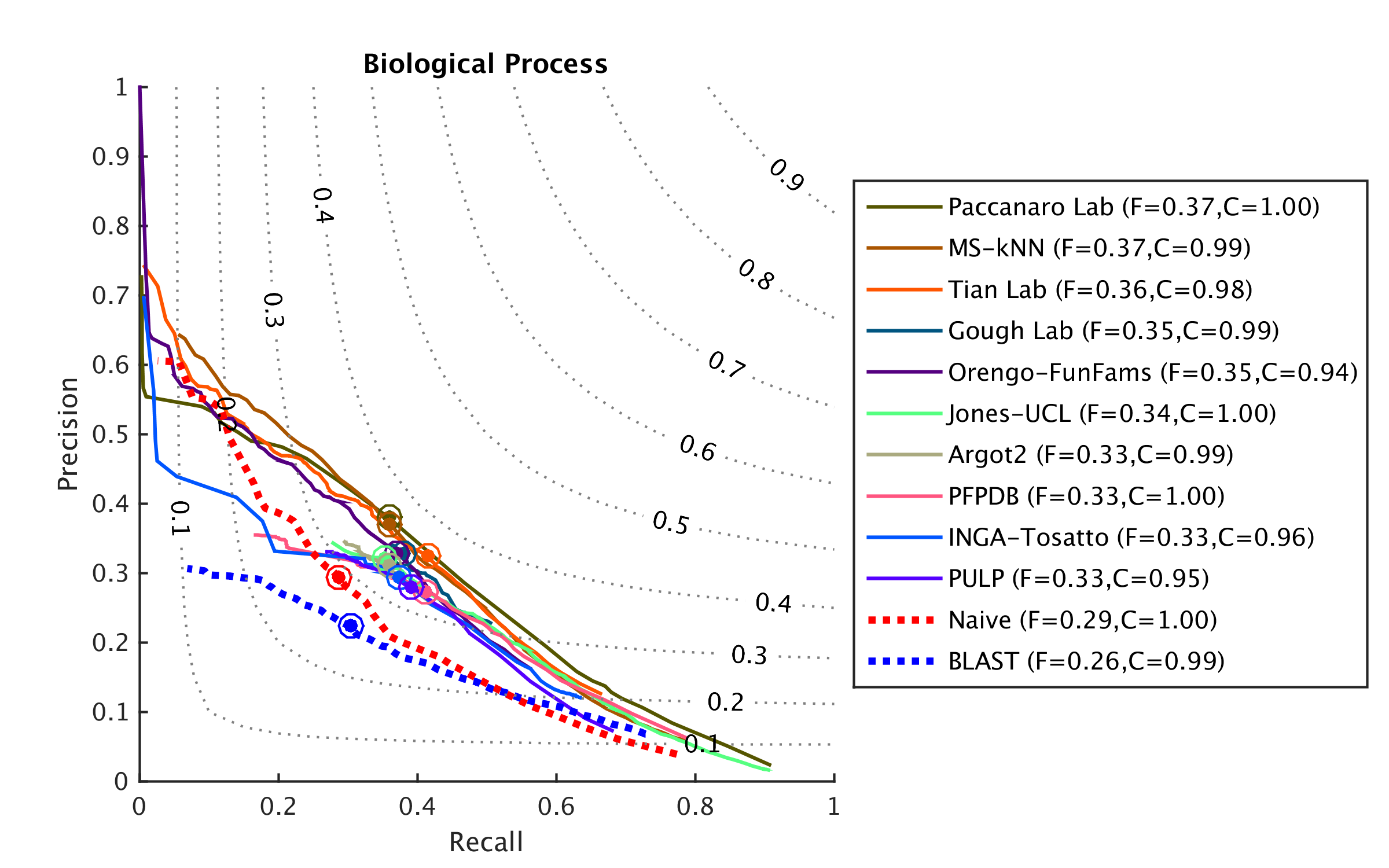}
\end{center}

\noindent Supplementary Figure 5D (prokarya):
\begin{center}
  \includegraphics[width=\textwidth]{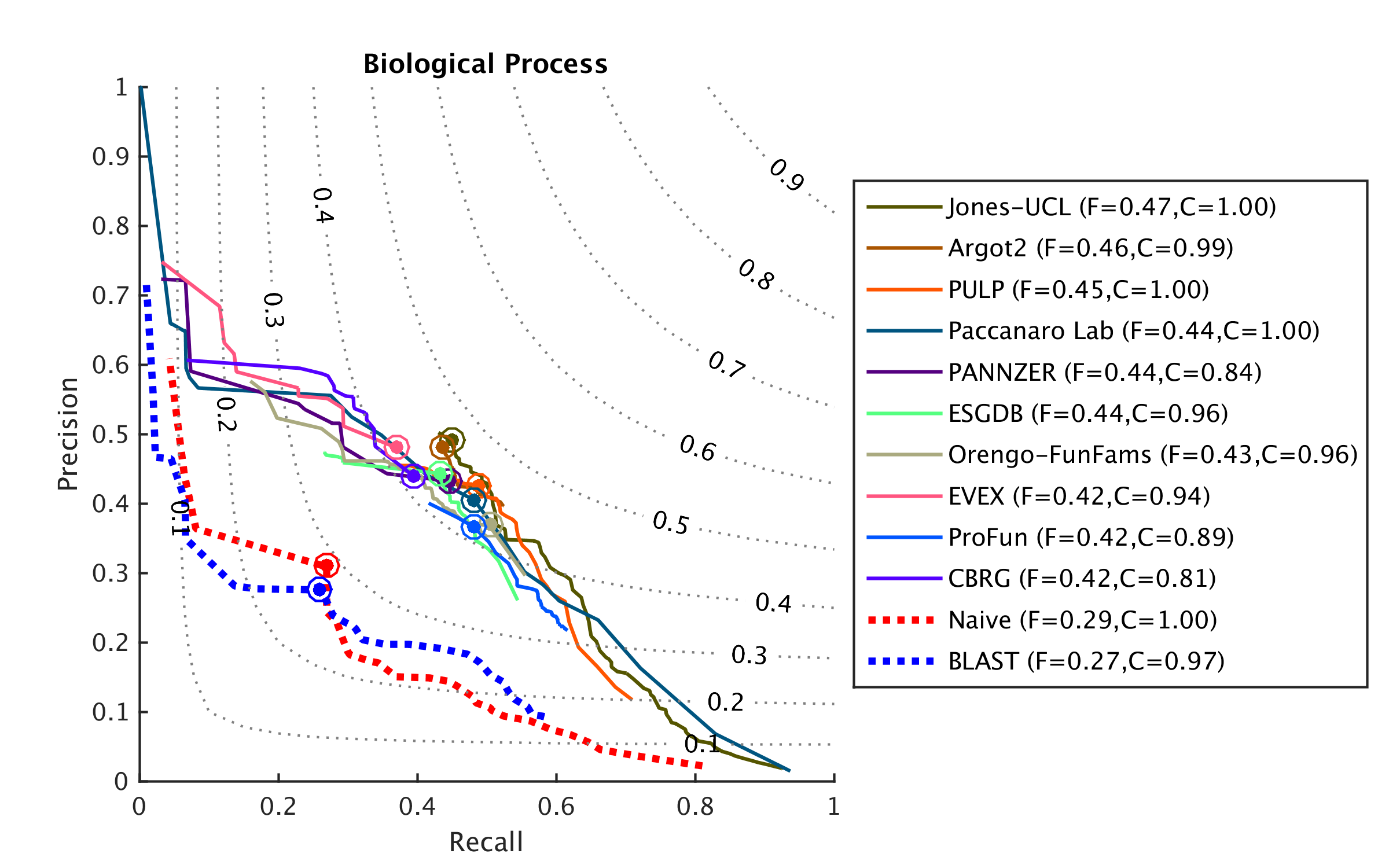}
\end{center}

\newpage

\noindent Supplementary Figure 5E (eukarya):
\begin{center}
  \includegraphics[width=\textwidth]{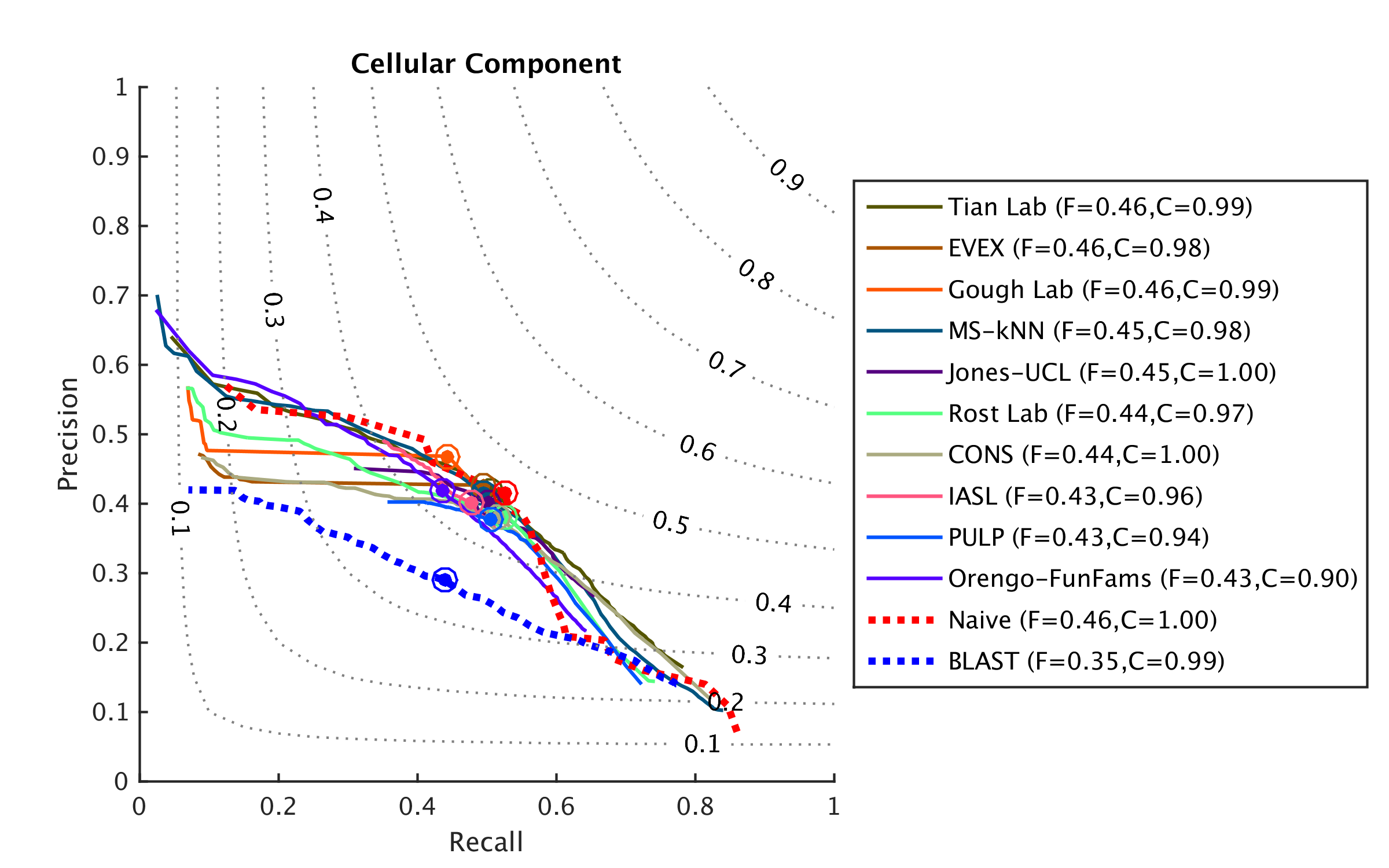}
\end{center}

\noindent Supplementary Figure 5F (prokarya):
\begin{center}
  \includegraphics[width=\textwidth]{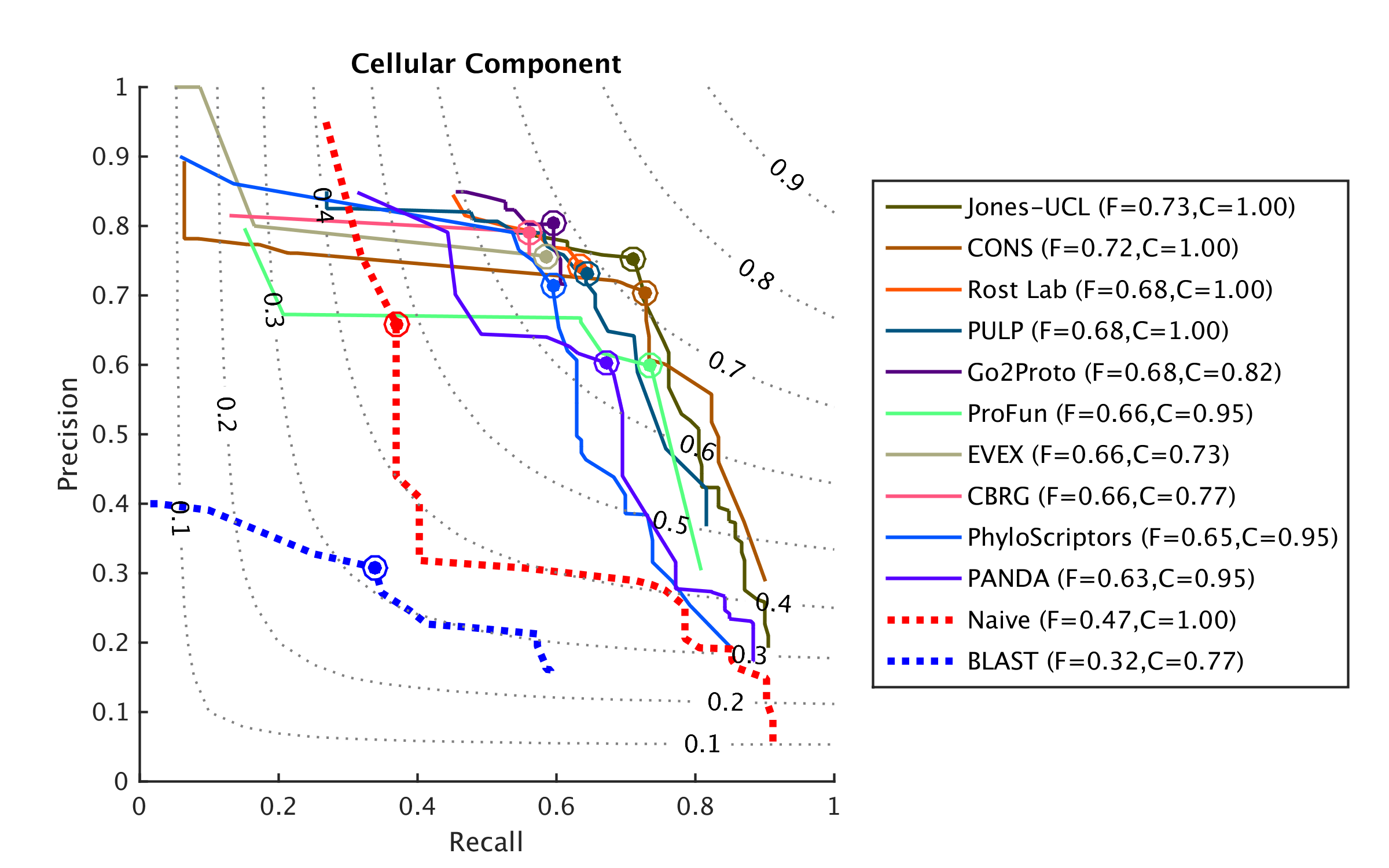}
\end{center}

\newpage

\paragraph{\labelstyle Supplementary Figure 6}
Performance evaluation based on the maximum F-measure for the top-performing methods for the Molecular Function ontology (A--F), Biological Process ontology (G--O), and Cellular Component ontology (P--V). Only the species with 15 benchmark proteins or more are included. All bars show the top ten participating methods as well as the Na\"ive and BLAST baseline methods. A perfect predictor would be characterized with $\fmax$ of $1$. Confidence interval ($95\%$) were determined using bootstrapping with 10,000 iterations on the set of target sequences.

\newpage

\noindent Supplementary Figure 6A ({\speciesstyle Arabidopsis thaliana}):
\begin{center}
  \includegraphics[height=.45\textheight]{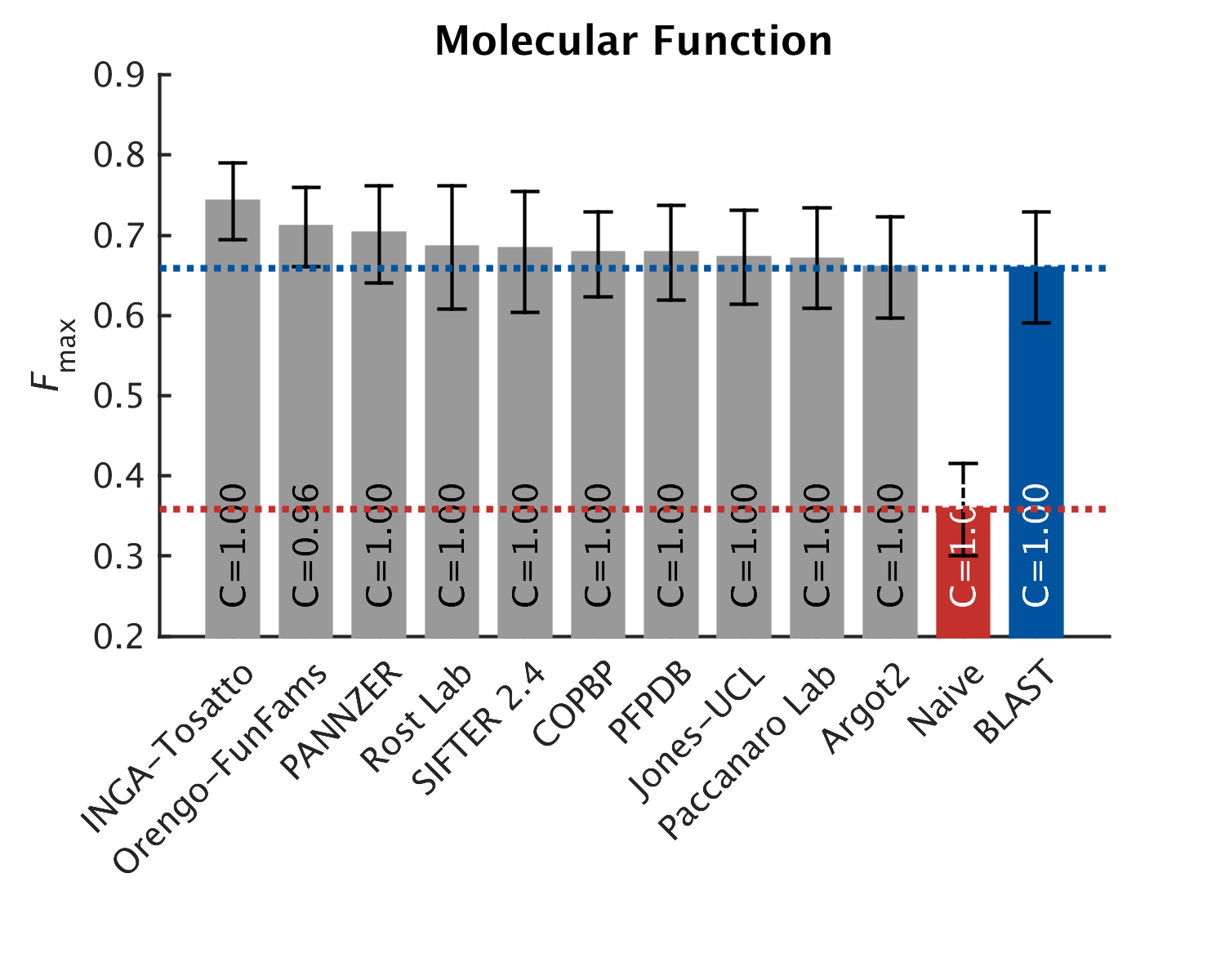}
\end{center}

\noindent Supplementary Figure 6B ({\speciesstyle Escherichia coli K12}):
\begin{center}
  \includegraphics[height=.45\textheight]{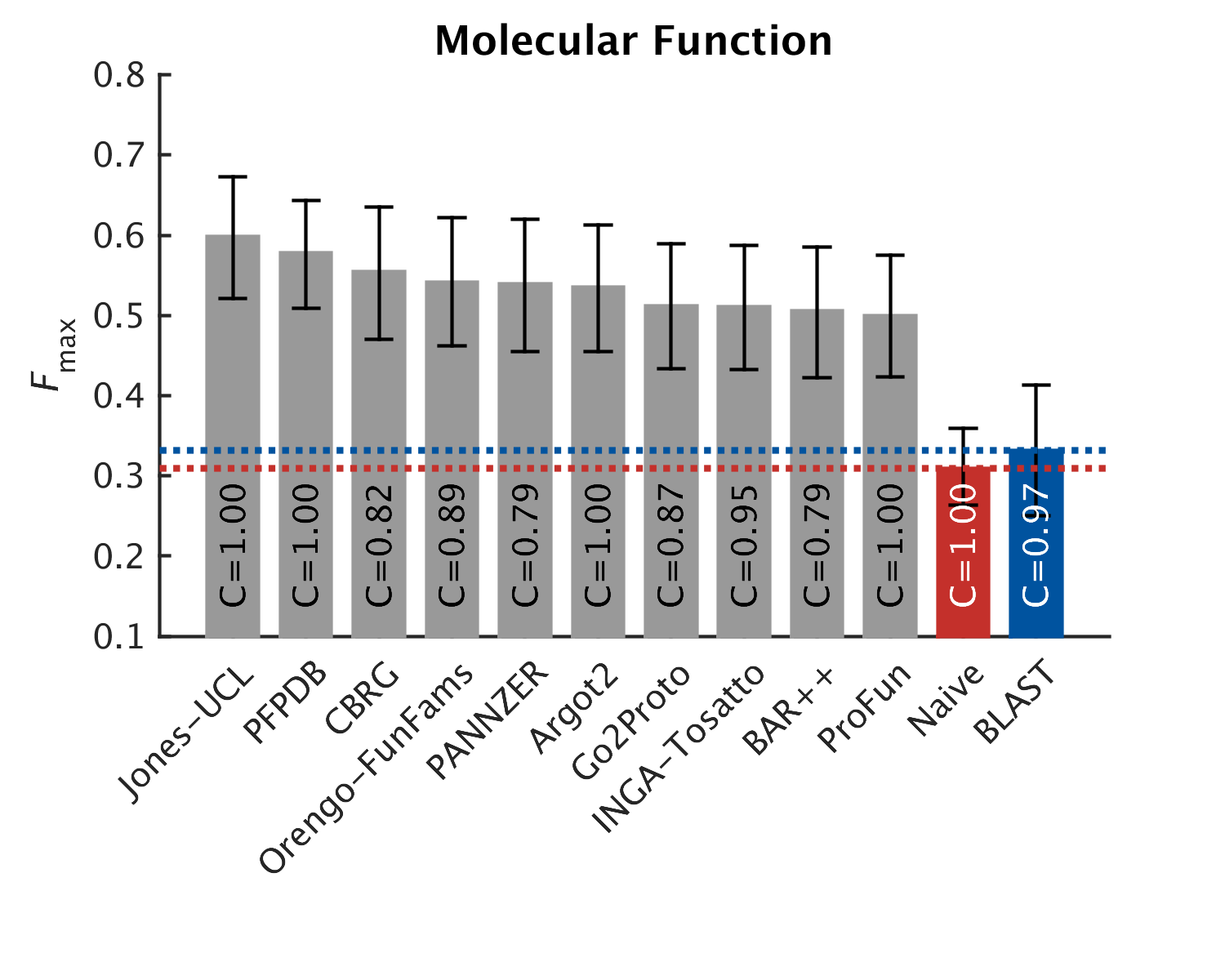}
\end{center}

\newpage

\noindent Supplementary Figure 6C ({\speciesstyle Homo sapiens}):
\begin{center}
  \includegraphics[height=.45\textheight]{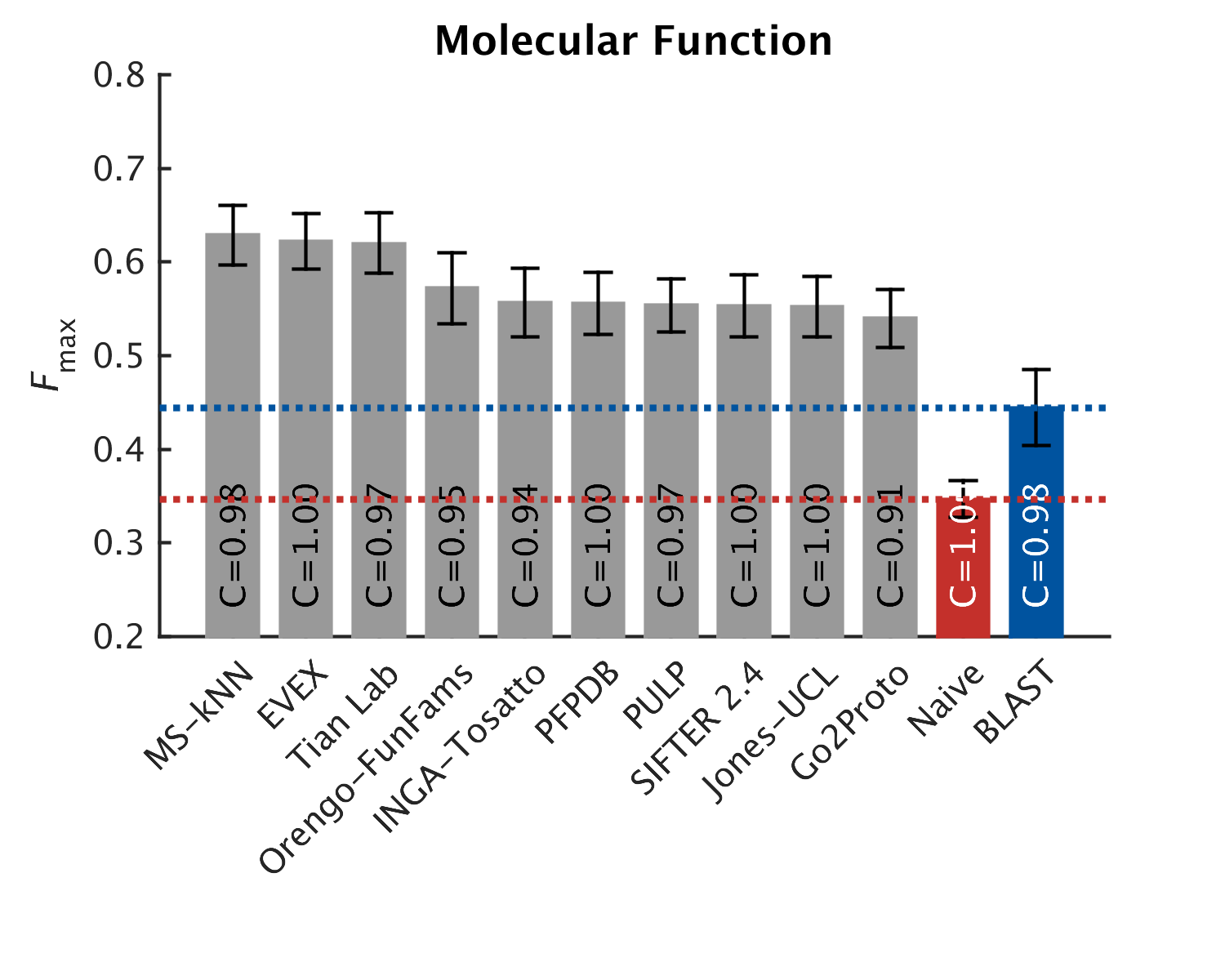}
\end{center}

\noindent Supplementary Figure 6D ({\speciesstyle Mus musculus}):
\begin{center}
  \includegraphics[height=.45\textheight]{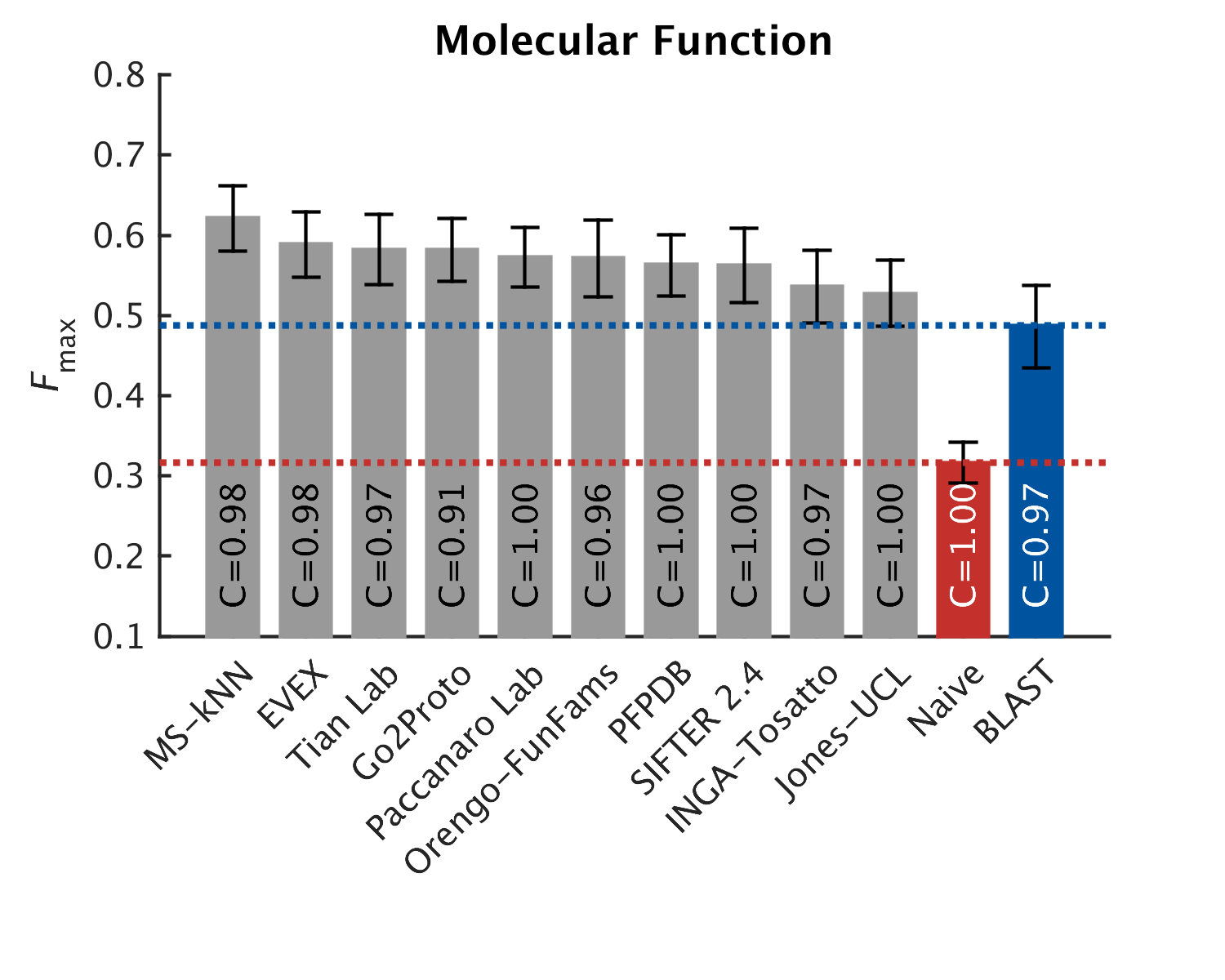}
\end{center}

\newpage

\noindent Supplementary Figure 6E ({\speciesstyle Pseudomonas aeruginosa}):
\begin{center}
  \includegraphics[height=.45\textheight]{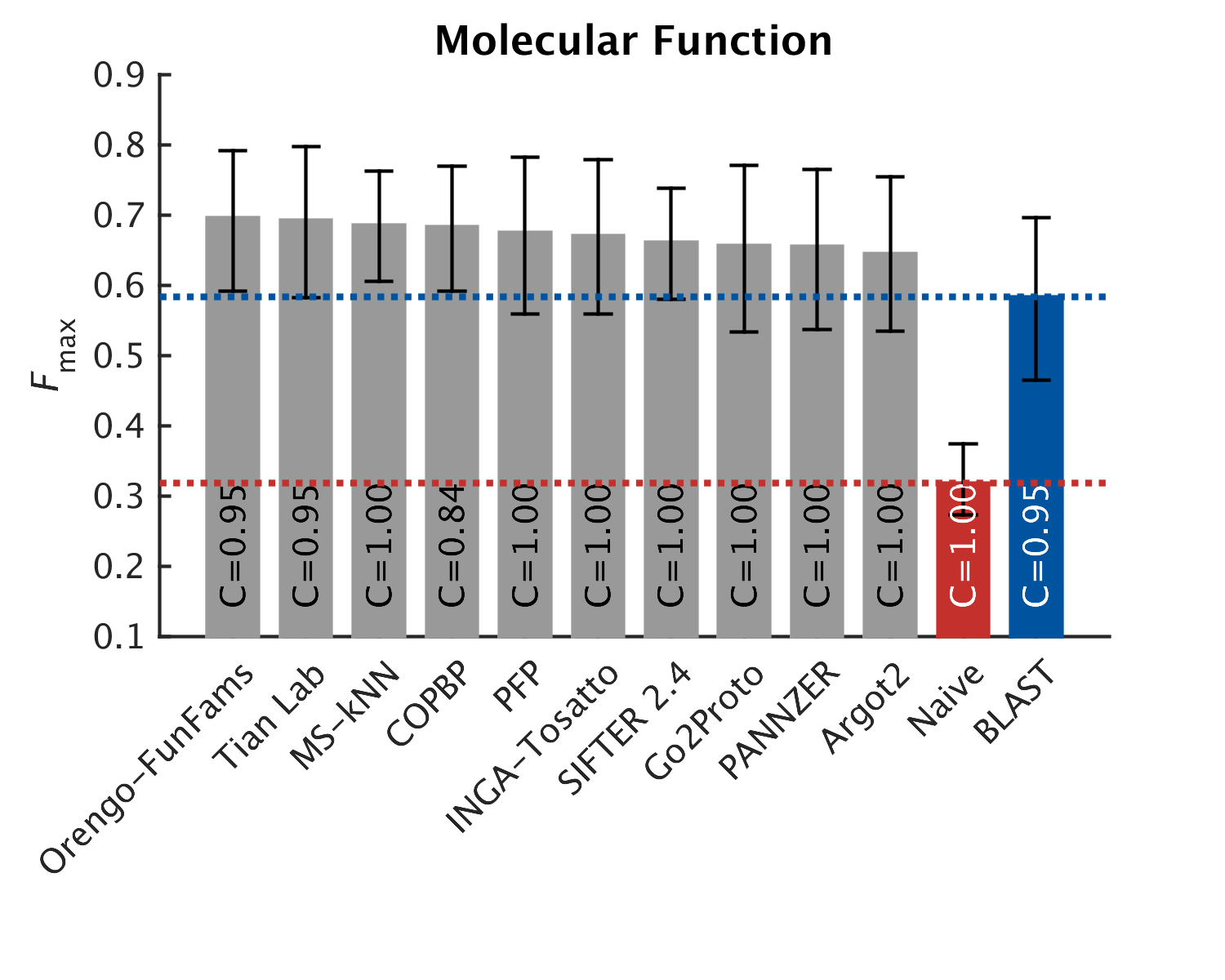}
\end{center}

\noindent Supplementary Figure 6F ({\speciesstyle Rattus norvegicus}):
\begin{center}
  \includegraphics[height=.45\textheight]{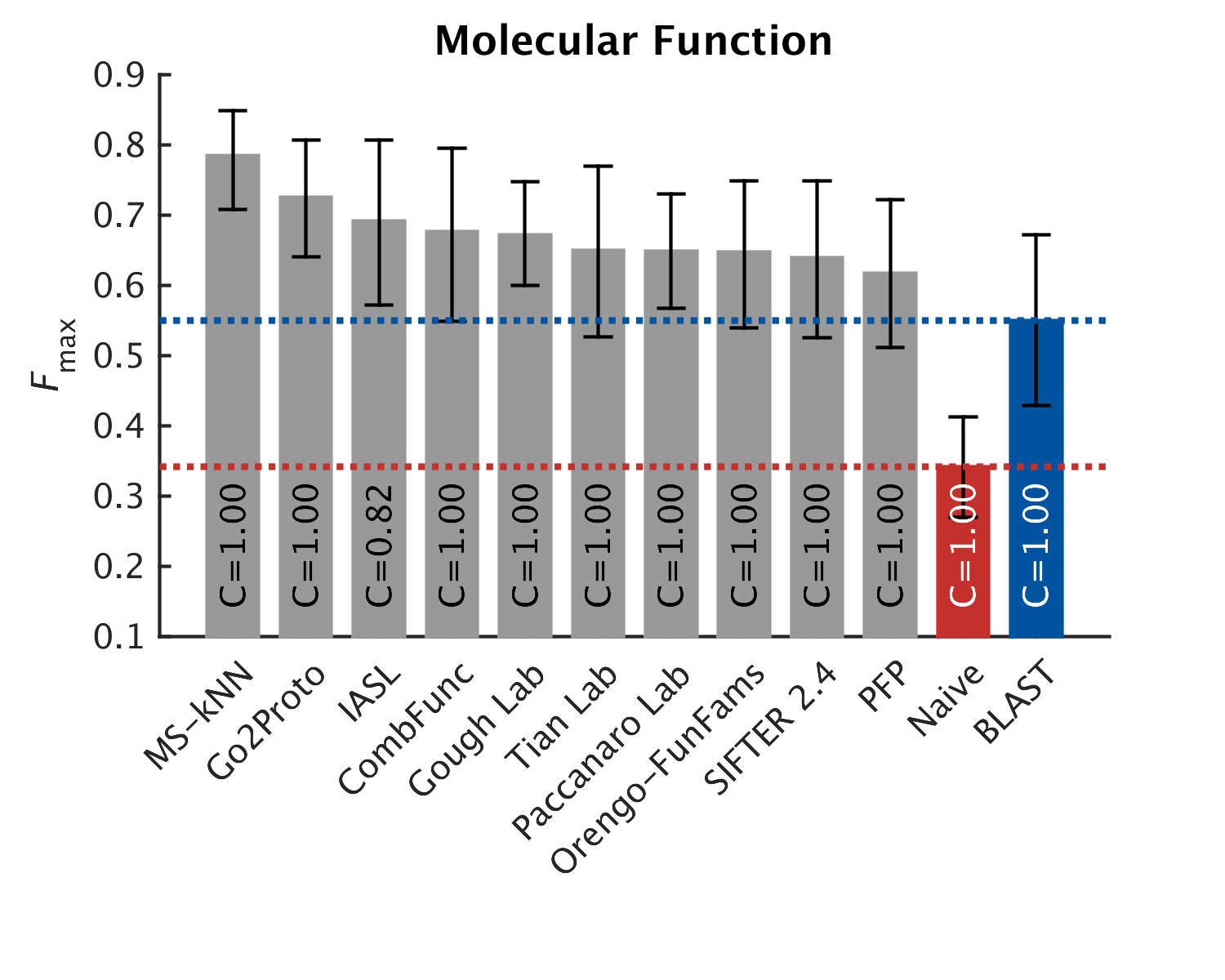}
\end{center}

\newpage

\noindent Supplementary Figure 6G ({\speciesstyle Arabidopsis thaliana}):
\begin{center}
  \includegraphics[height=.45\textheight]{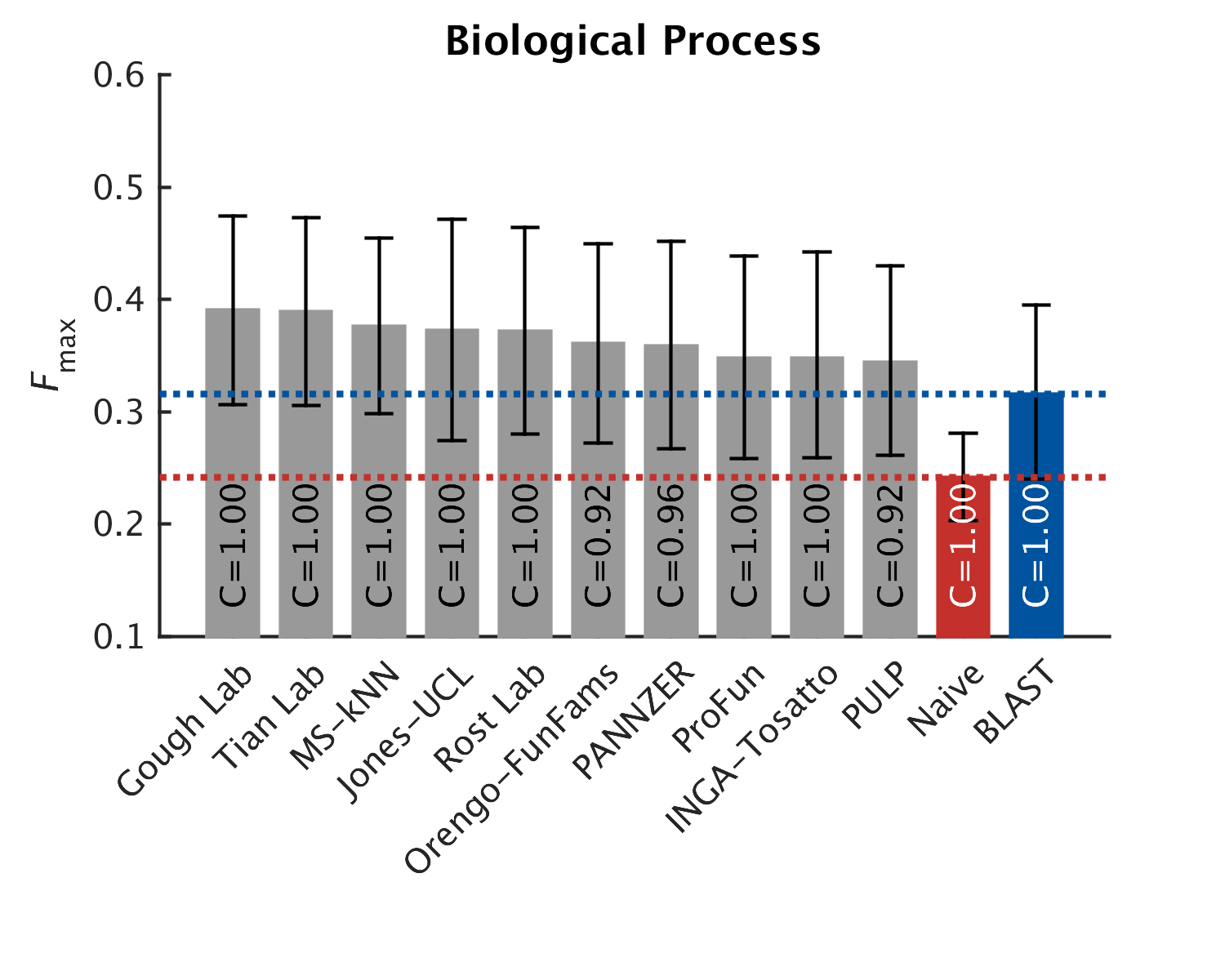}
\end{center}

\noindent Supplementary Figure 6H ({\speciesstyle Danio rerio}):
\begin{center}
  \includegraphics[height=.45\textheight]{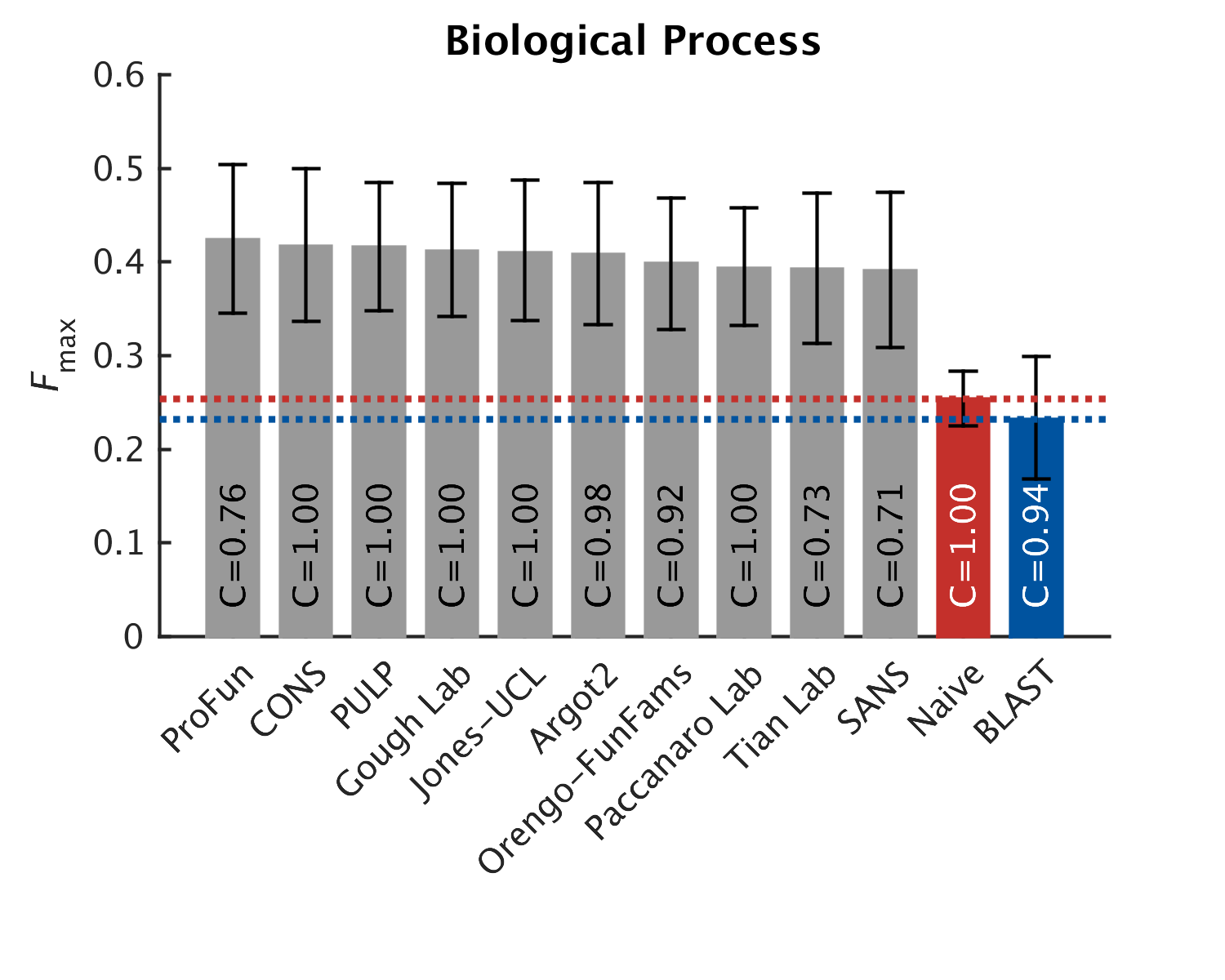}
\end{center}

\newpage

\noindent Supplementary Figure 6I ({\speciesstyle Dictyostelium discoideum}):
\begin{center}
  \includegraphics[height=.45\textheight]{./picssuppl/bpo_ARATH_type1_mode1_top10_fmax_bar.png}
\end{center}

\noindent Supplementary Figure 6J ({\speciesstyle Drosophila melanogaster}):
\begin{center}
  \includegraphics[height=.45\textheight]{./picssuppl/bpo_ECOLI_type1_mode1_top10_fmax_bar.png}
\end{center}

\newpage

\noindent Supplementary Figure 6K ({\speciesstyle Escherichia coli K12}):
\begin{center}
  \includegraphics[height=.45\textheight]{./picssuppl/bpo_ECOLI_type1_mode1_top10_fmax_bar.png}
\end{center}

\noindent Supplementary Figure 6L ({\speciesstyle Homo sapiens}):
\begin{center}
  \includegraphics[height=.45\textheight]{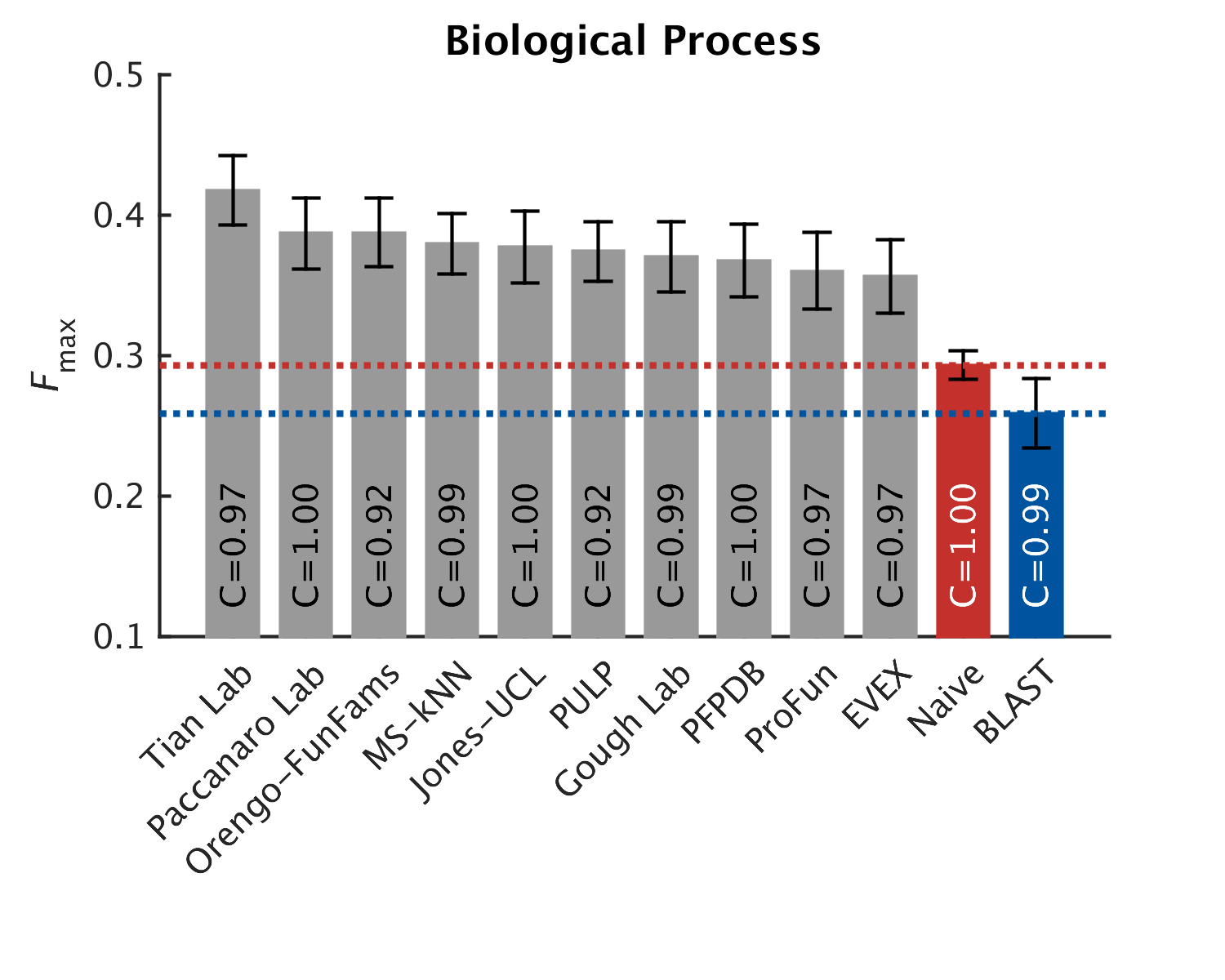}
\end{center}

\newpage

\noindent Supplementary Figure 6M ({\speciesstyle Mus musculus}):
\begin{center}
  \includegraphics[height=.45\textheight]{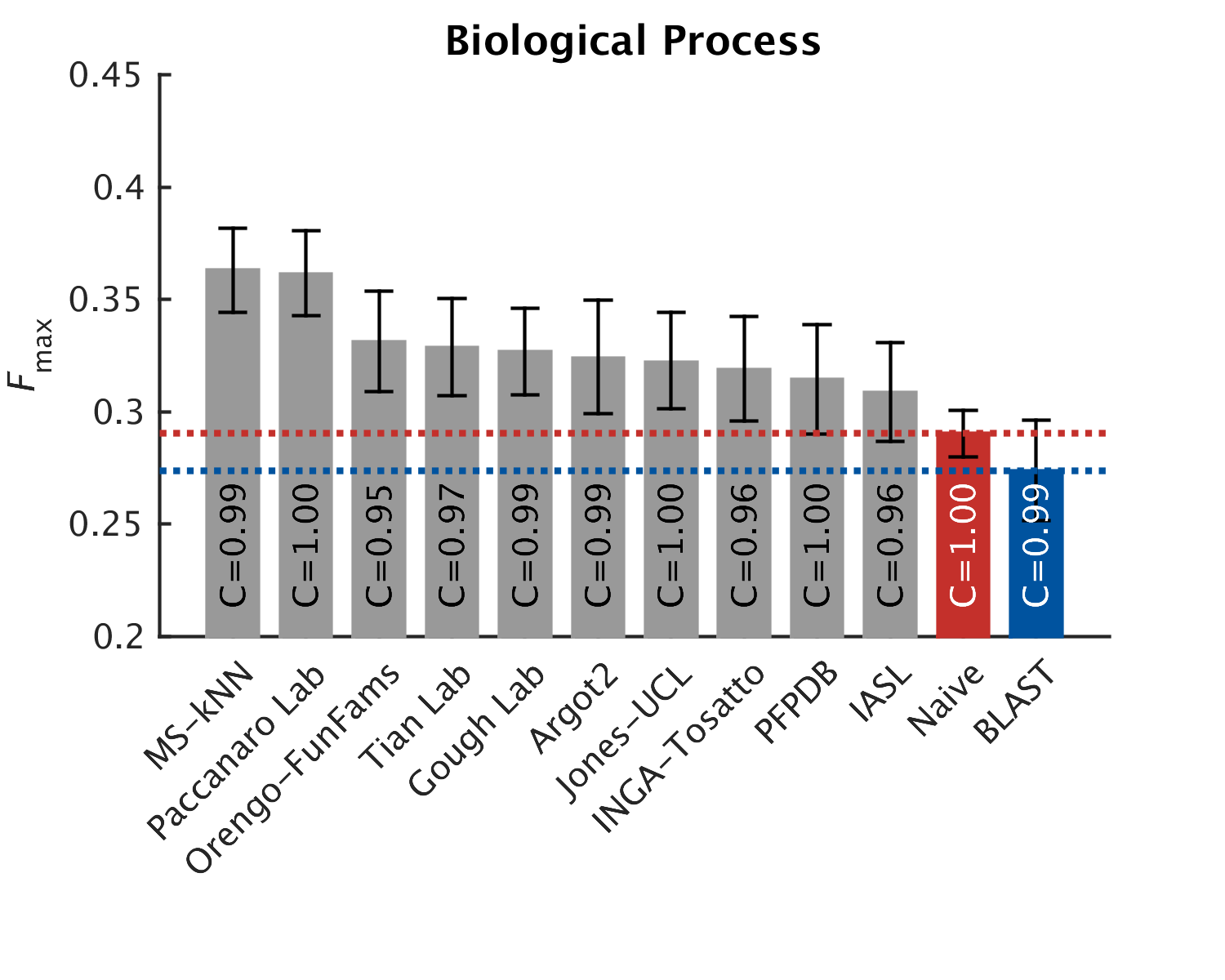}
\end{center}

\noindent Supplementary Figure 6N ({\speciesstyle Pseudomonas aeruginosa}):
\begin{center}
  \includegraphics[height=.45\textheight]{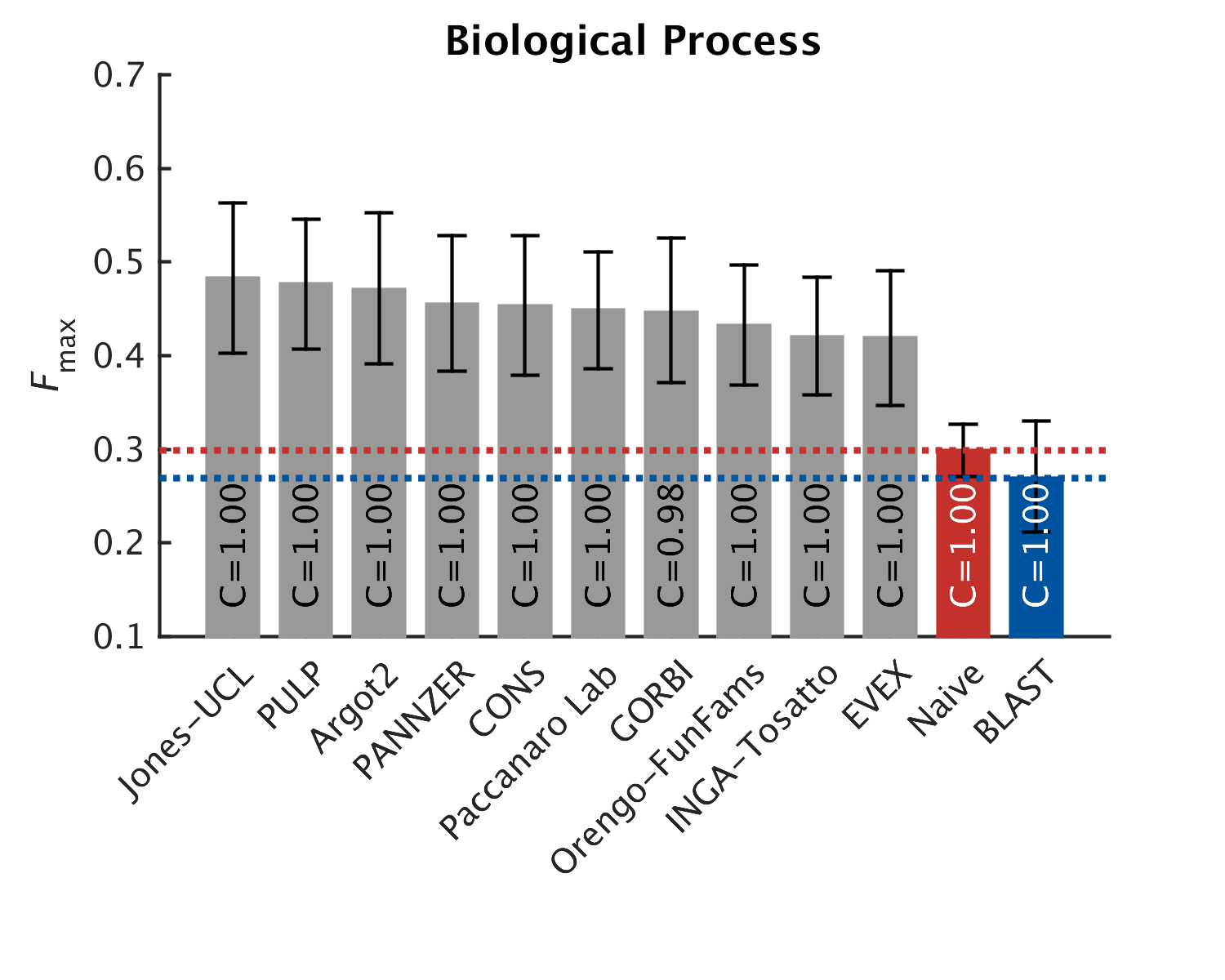}
\end{center}

\newpage

\noindent Supplementary Figure 6O ({\speciesstyle Rattus norvegicus}):
\begin{center}
  \includegraphics[height=.45\textheight]{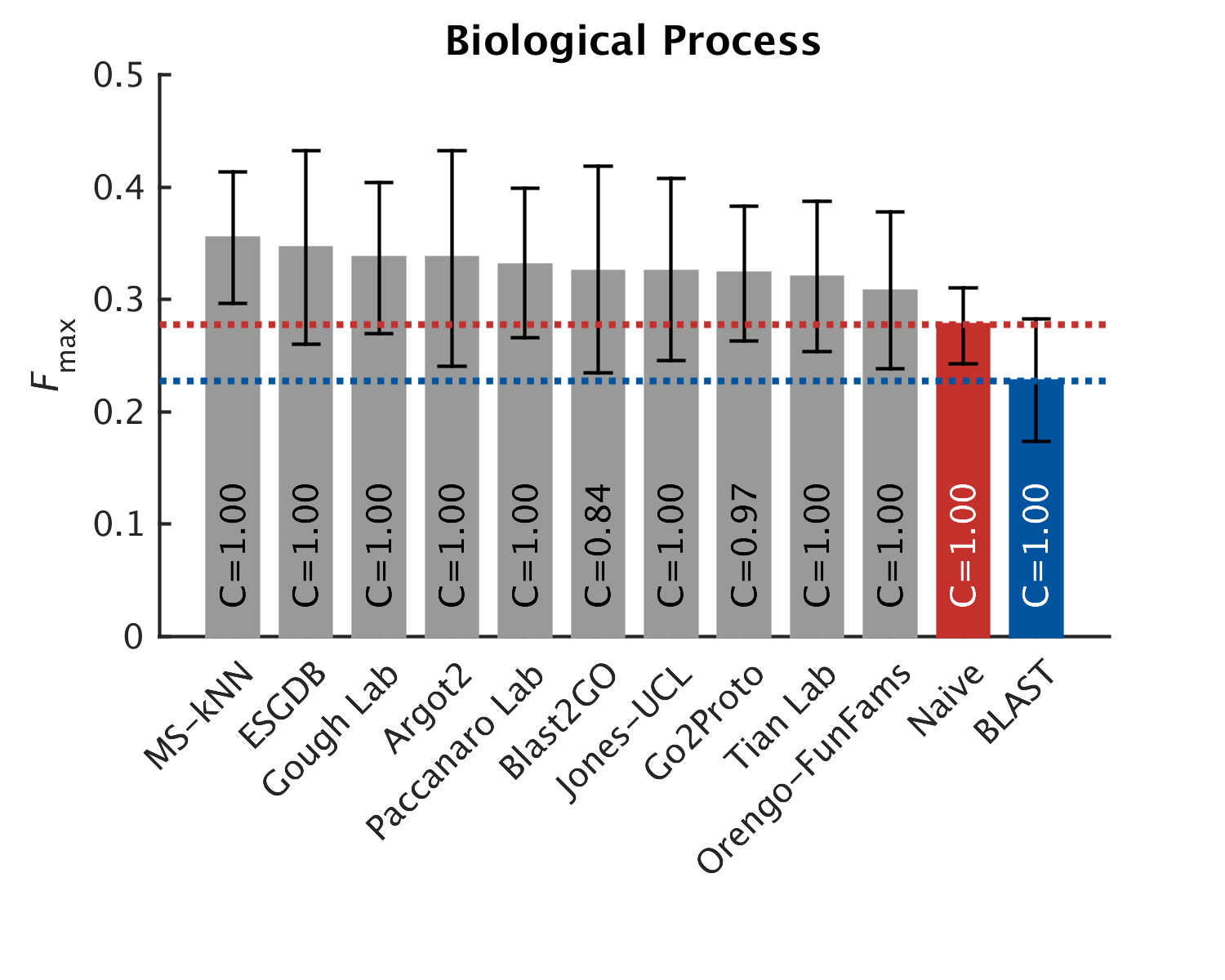}
\end{center}

\newpage

\noindent Supplementary Figure 6P ({\speciesstyle Arabidopsis thaliana}):
\begin{center}
  \includegraphics[height=.45\textheight]{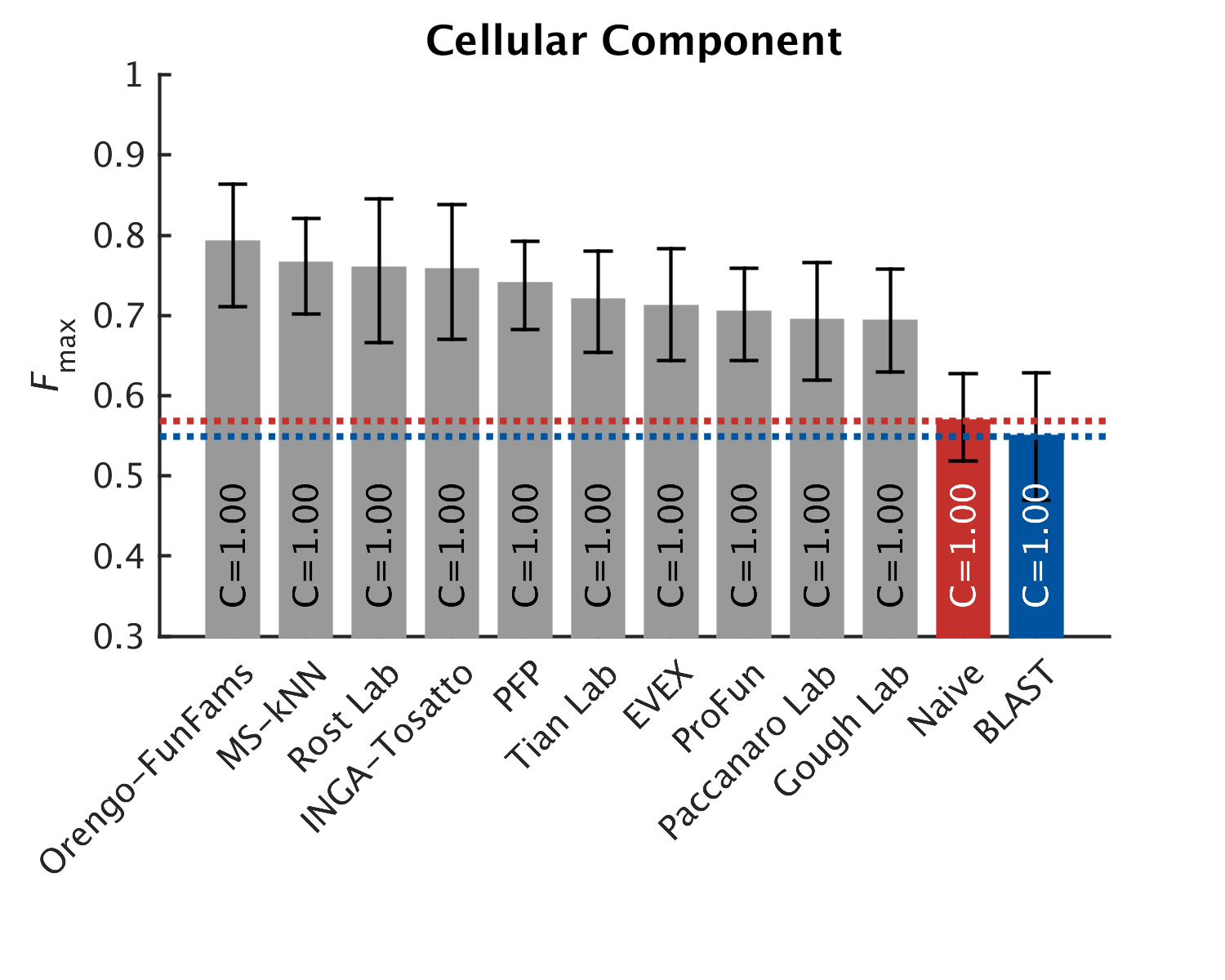}
\end{center}

\noindent Supplementary Figure 6Q ({\speciesstyle Drosophila melanogaster}):
\begin{center}
  \includegraphics[height=.45\textheight]{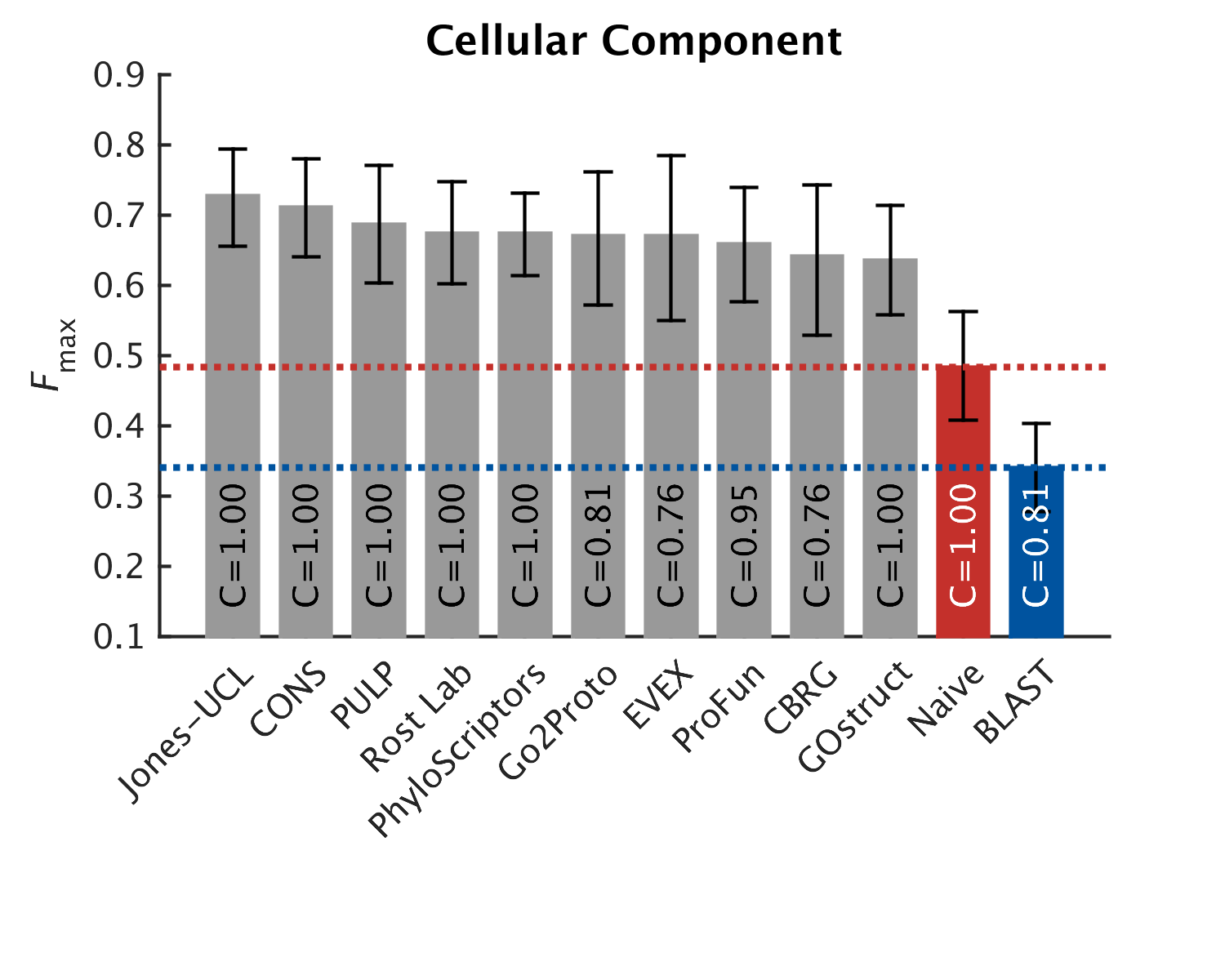}
\end{center}

\newpage

\noindent Supplementary Figure 6R ({\speciesstyle Escherichia coli K12}):
\begin{center}
  \includegraphics[height=.45\textheight]{./picssuppl/cco_ECOLI_type1_mode1_top10_fmax_bar.png}
\end{center}

\noindent Supplementary Figure 6S ({\speciesstyle Homo sapiens}):
\begin{center}
  \includegraphics[height=.45\textheight]{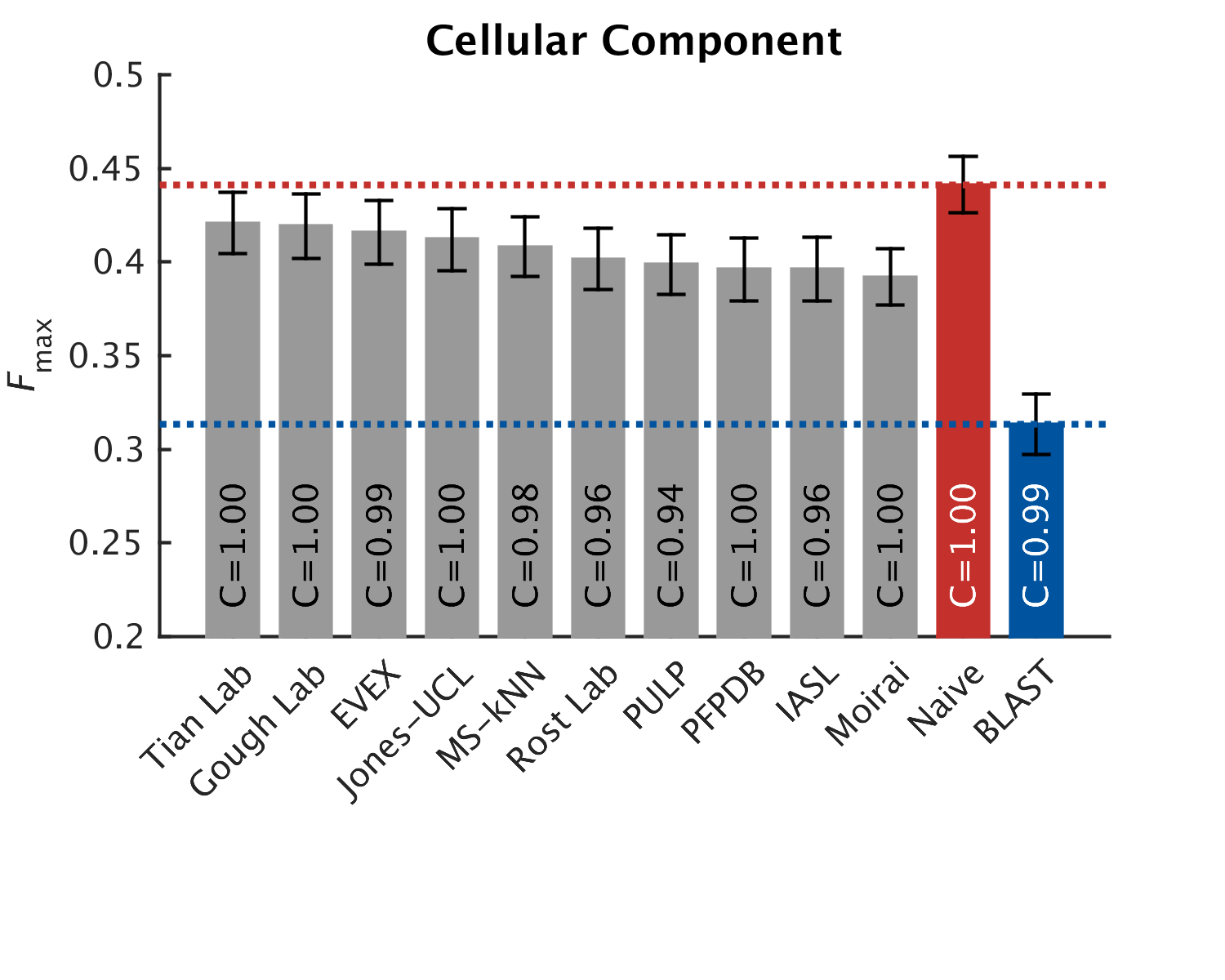}
\end{center}

\newpage

\noindent Supplementary Figure 6T ({\speciesstyle Mus musculus}):
\begin{center}
  \includegraphics[height=.45\textheight]{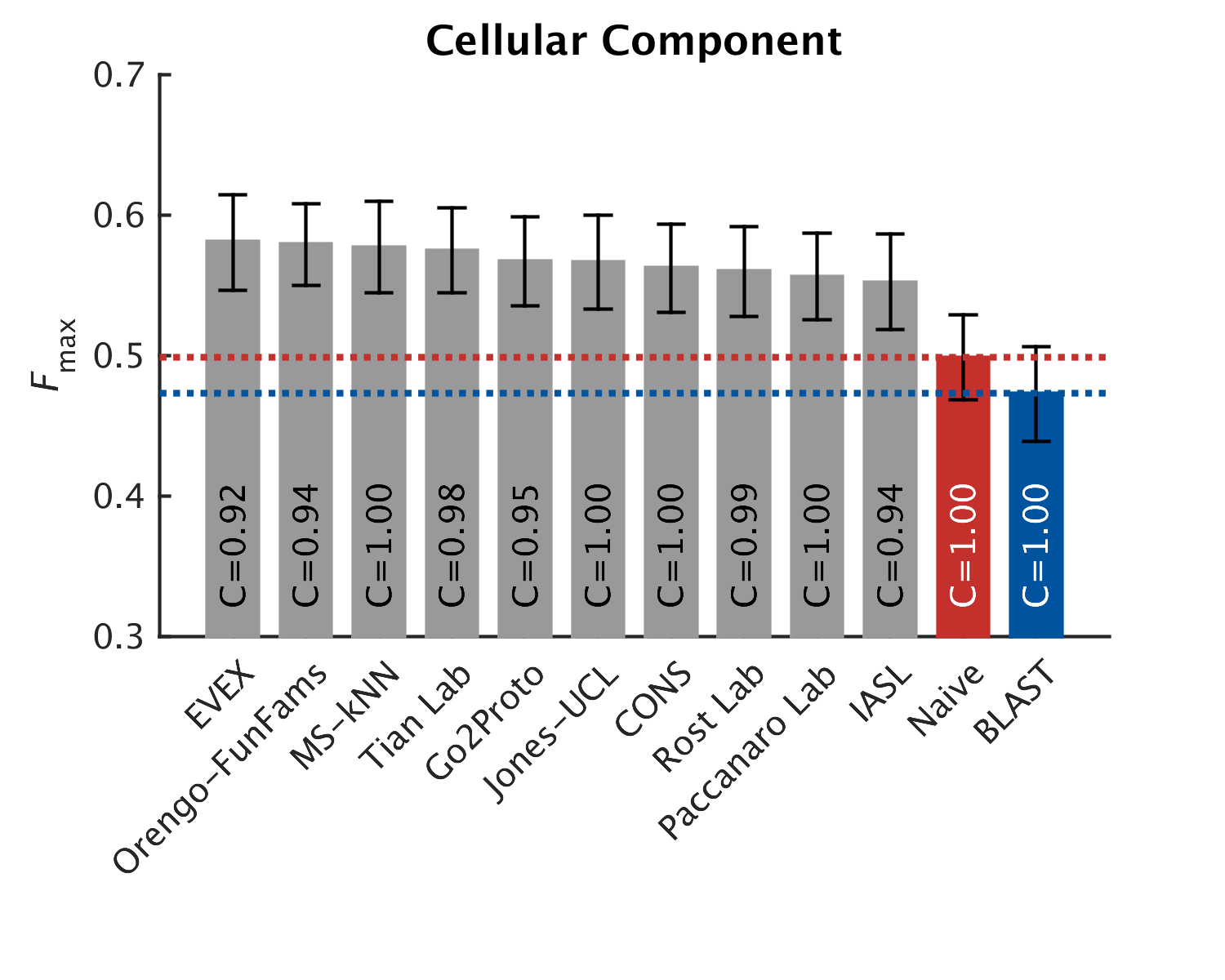}
\end{center}

\noindent Supplementary Figure 6U ({\speciesstyle Rattus norvegicus}):
\begin{center}
  \includegraphics[height=.45\textheight]{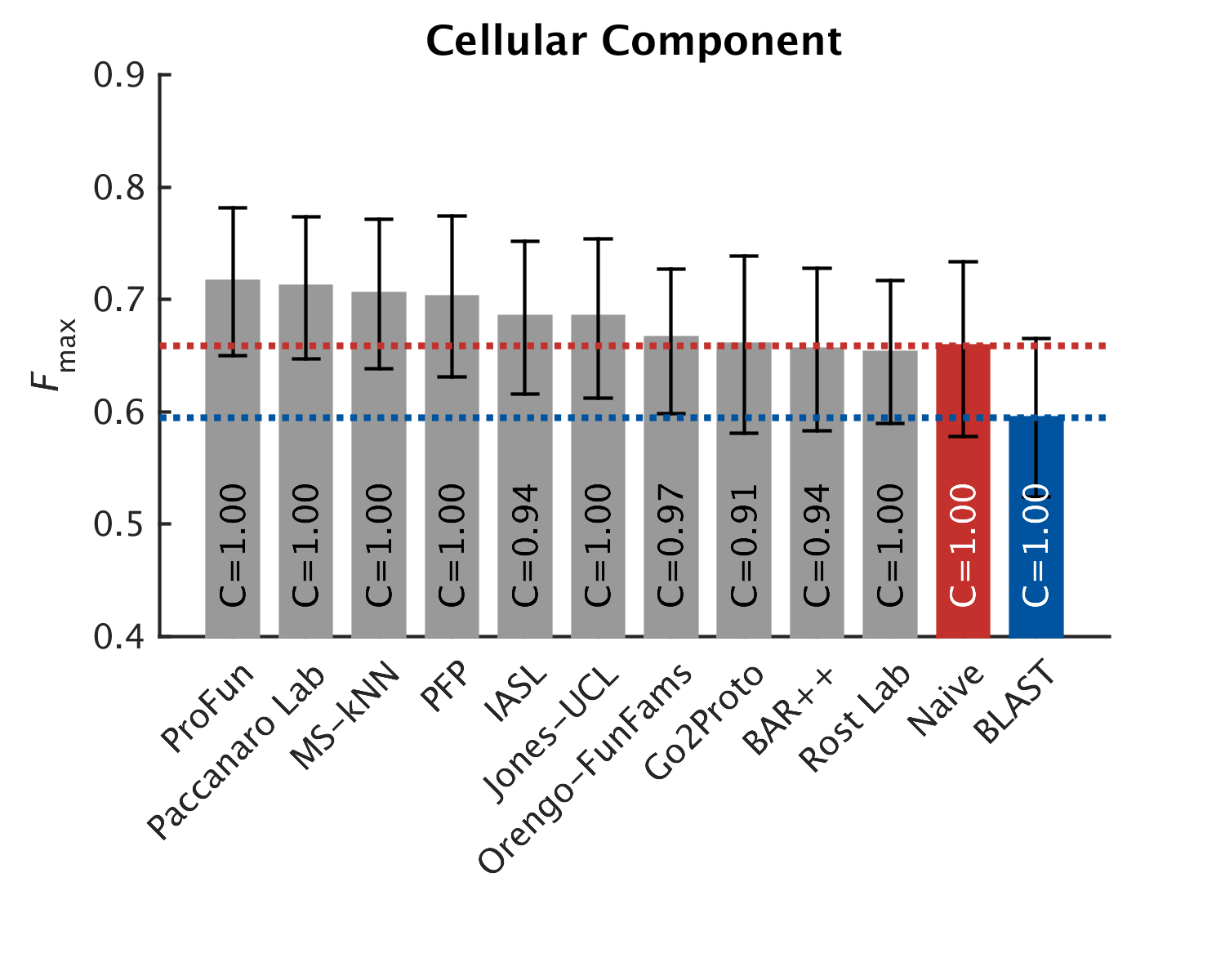}
\end{center}

\newpage

\noindent Supplementary Figure 6V ({\speciesstyle Saccharomyces cerevisiae}):
\begin{center}
  \includegraphics[height=.45\textheight]{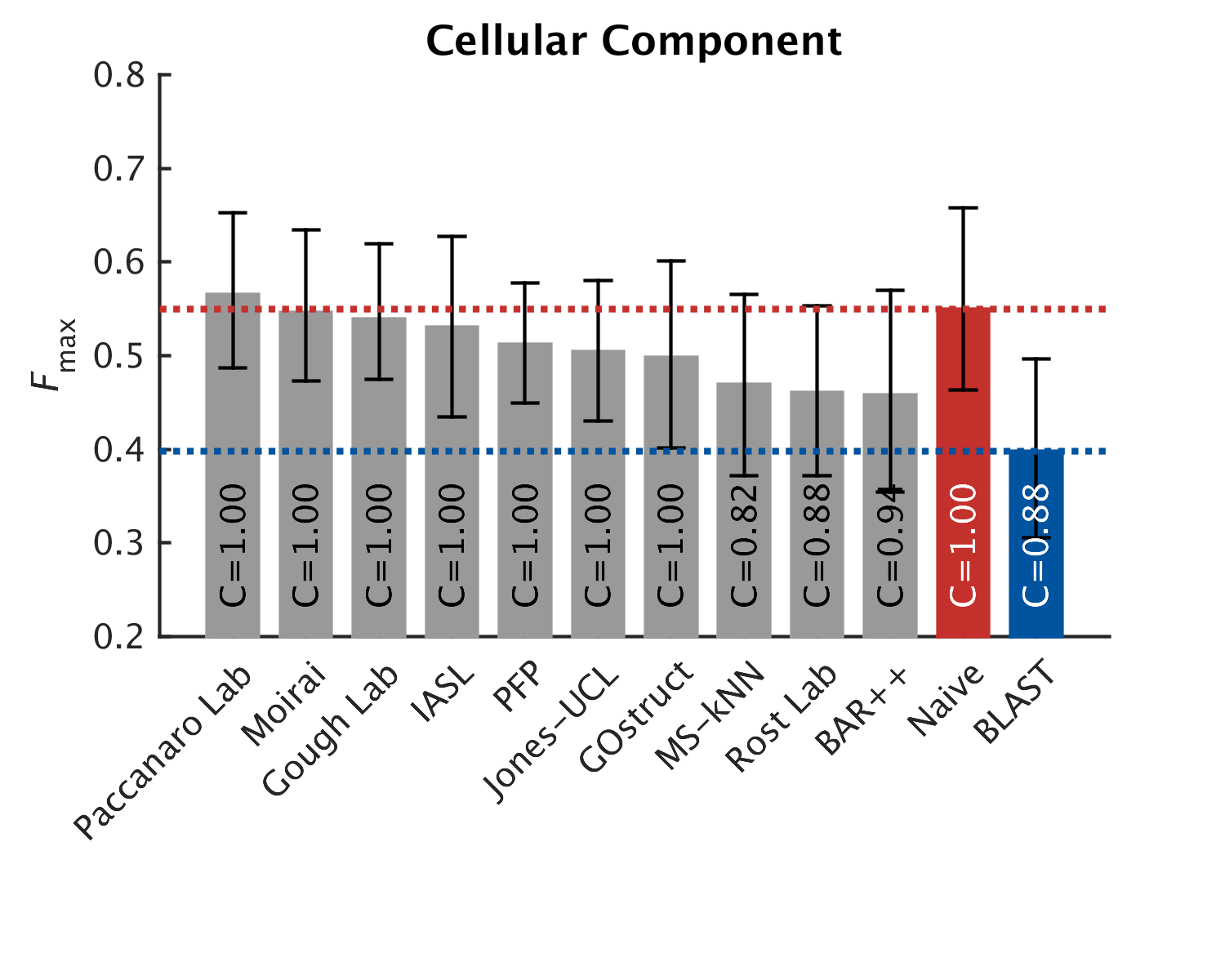}
\end{center}

\newpage 

\paragraph{\labelstyle Supplementary Figure 7}
Weighted precision-recall curves for the top-performing methods for (A) Molecular Function ontology, (B) Biological Process ontology, (C) Cellular Component ontology and (D) Human Phenotype ontology. All panels show the top ten participating methods in each category, as well as the Na\"ive and BLAST baseline methods. Points corresponding to the maximum weighted F-measure are marked in circles on each curve. The legend provides the maximum weighted F-measure ($F$) and coverage ($C$) for all methods. In cases where a Principal Investigator (PI) participated with multiple teams, only the results of the best scoring method are presented.\\

\paragraph{\labelstyle Calculation of the weighted precision-recall curve.} Each term $f$ in the ontology was weighted according to the information content of that term. The information content of the term $f$ was calculated as
\begin{eqnarray*}
  ic(f) = \log_{2}\frac{1}{\mathrm{Pr}\left(f | \mathcal{P}(f) \right)},
\end{eqnarray*}
where $\mathrm{Pr}\left(f | \mathcal{P}(f) \right)$ is the probability that the term $f$ in the ontology is associated to a protein given that all of its parents are associated. (probabilities were determined based on the union of Swiss-Prot, UniProt-GOA and GO Consortium databases). Weighted precisions and recalls are calculated as
\begin{eqnarray*}
  wpr(\tau) &=& \frac{1}{m(\tau)}\sum_{i=1}^{m(\tau)}
  \frac{\sum_{f} ic(f) \cdot \mathbbm{1}\left( f \in P_{i}(\tau) \wedge T_{i}(\tau)\right)}{\sum_{f} ic(f) \cdot \mathbbm{1}\left( f \in P_{i}(\tau) \right)}, \quad \mbox{and}\\
  wrc(\tau) &=& \frac{1}{n_{e}}\sum_{i=1}^{n_{e}}
  \frac{\sum_{f} ic(f) \cdot \mathbbm{1}\left( f \in P_{i}(\tau) \wedge T_{i}(\tau)\right)}{\sum_{f} ic(f) \cdot \mathbbm{1}\left( f \in T_{i}(\tau) \right)},
\end{eqnarray*}
where $P_{i}(\tau)$ is the set of predicted terms for protein $i$ with score no less than threshold $\tau$ and $T_{i}$ is the set of true terms for protein $i$, $m(\tau)$ is the number of sequences with at least one predicted score greater than or equal to $\tau$, and $n_{e}$ is the number of proteins used in a particular mode of evaluation. In the full evaluation mode $n_{e} = n$, the number of benchmark proteins, whereas in the partial evaluation mode $n_{e} = m(0)$.

\newpage

\noindent Supplementary Figure 7A:
\begin{center}
  \includegraphics[width=\textwidth]{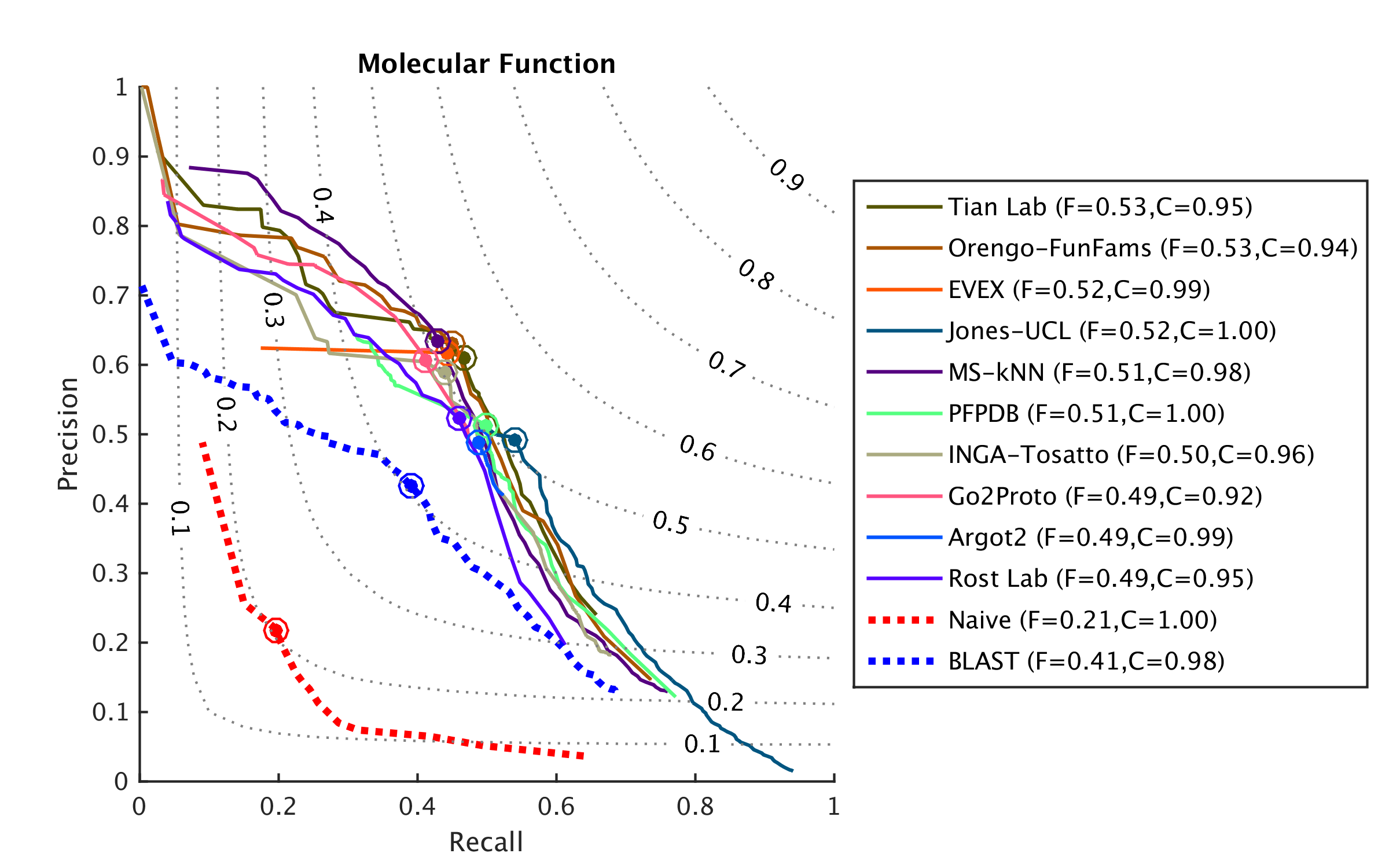}
\end{center}

\noindent Supplementary Figure 7B:
\begin{center}
  \includegraphics[width=\textwidth]{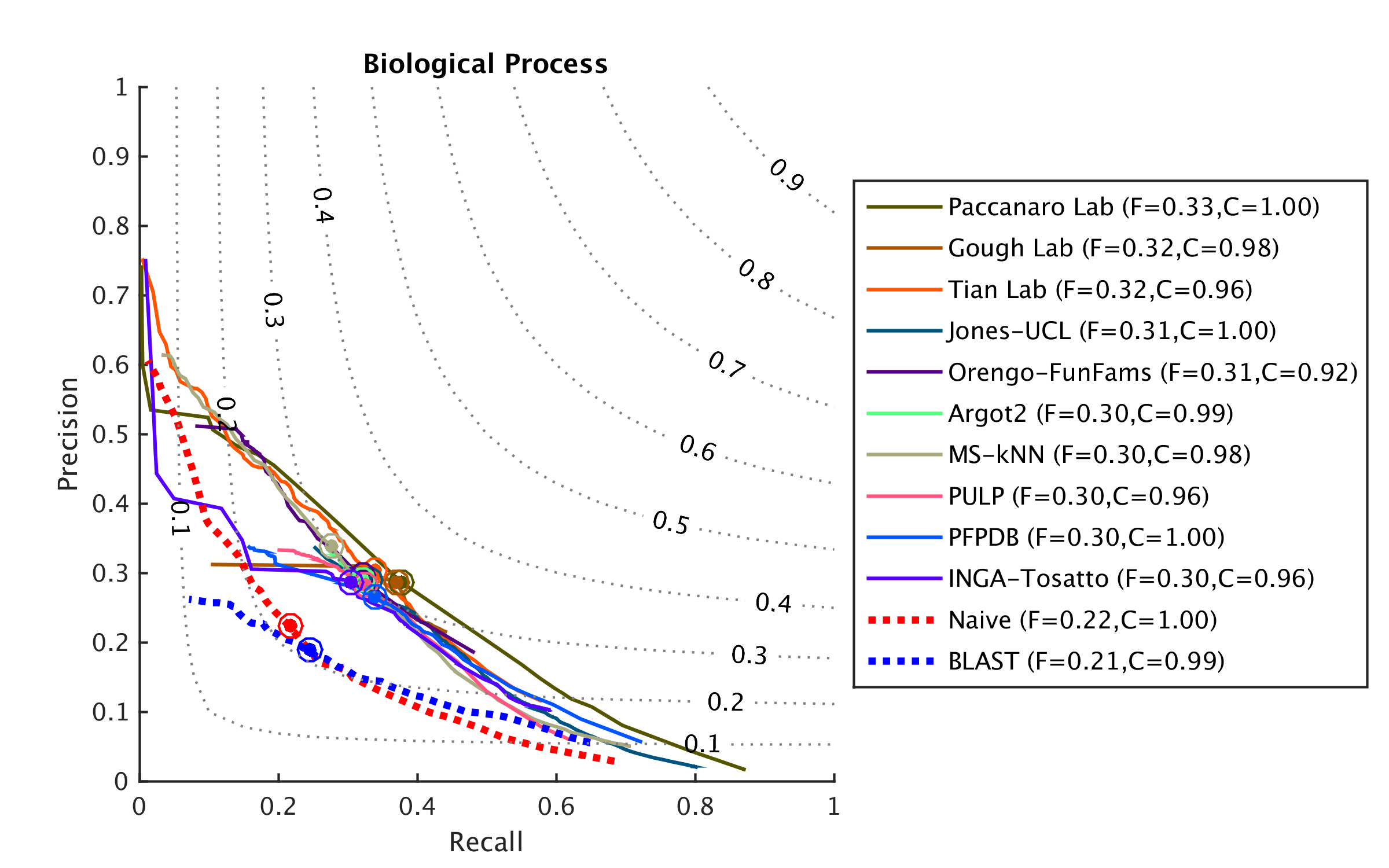}
\end{center}

\newpage

\noindent Supplementary Figure 7C:
\begin{center}
  \includegraphics[width=\textwidth]{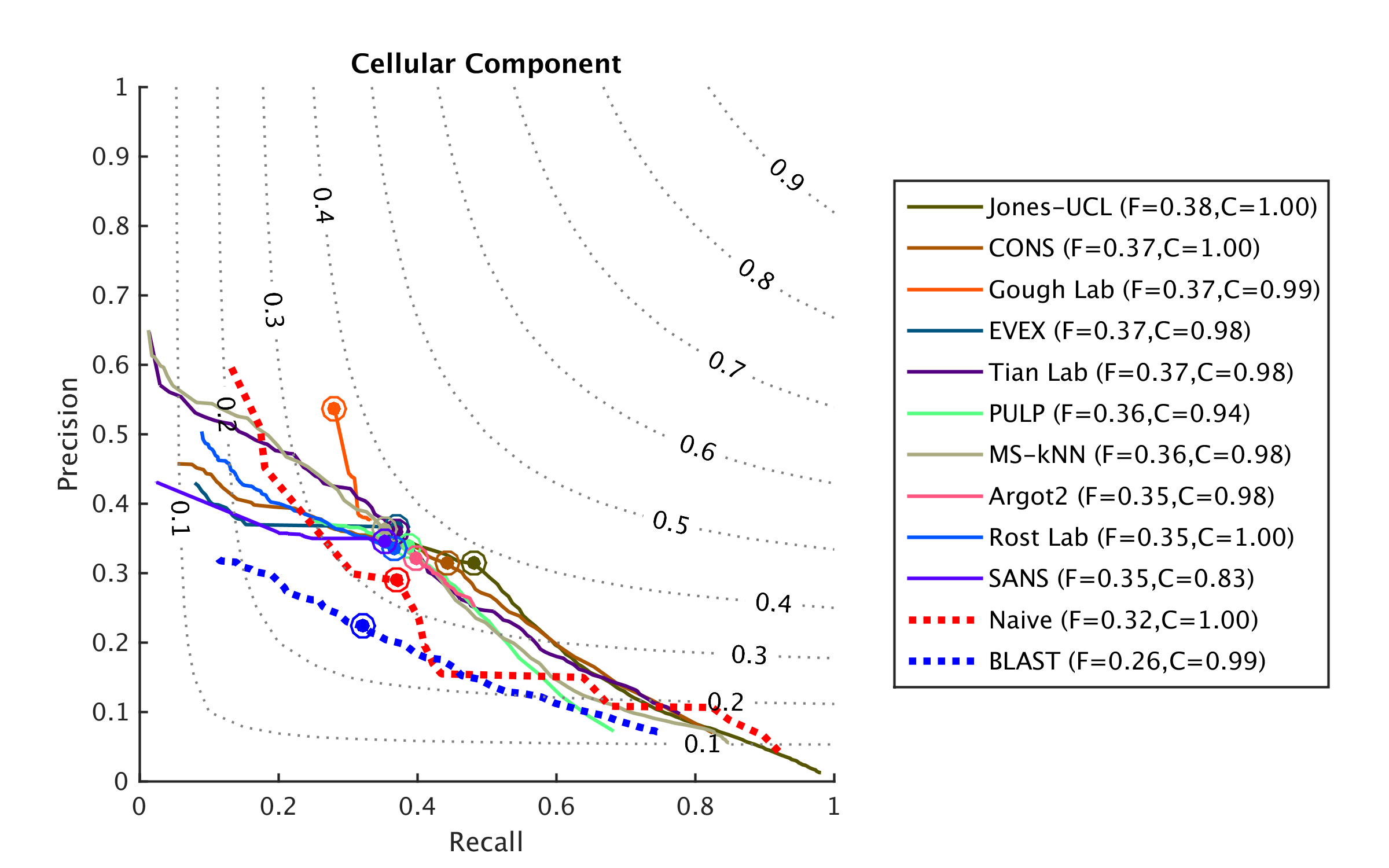}
\end{center}

\noindent Supplementary Figure 7D:
\begin{center}
  \includegraphics[width=\textwidth]{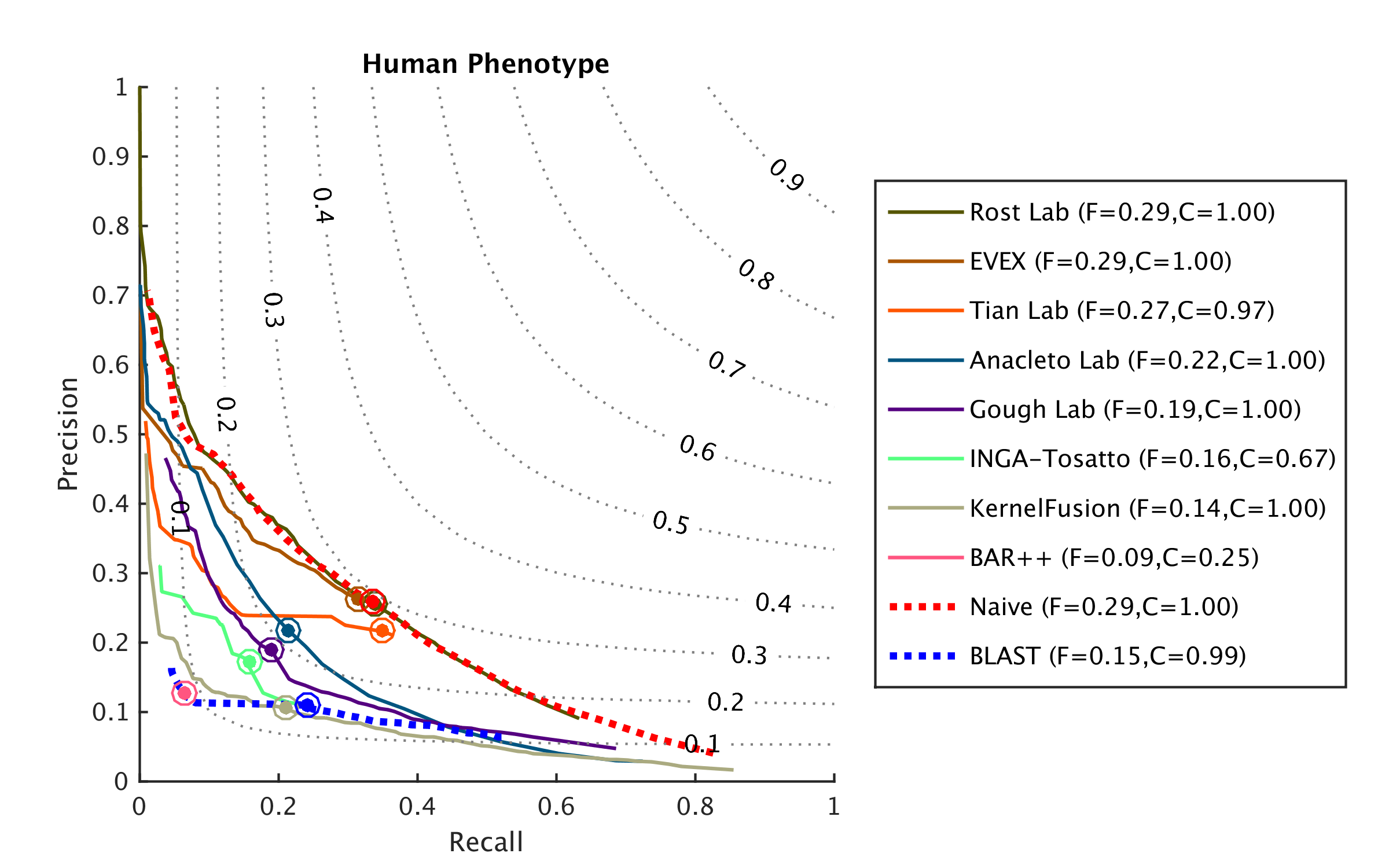}
\end{center}

\newpage

\paragraph{\labelstyle Supplementary Figure 8}
Normalized remaining uncertainty-misinformation curves for the top-performing methods for (A) Molecular Function ontology, (B) Biological Process ontology, (C) Cellular Component ontology and (D) Human Phenotype ontology. All panels show the top ten participating methods in each category, as well as the Na\"ive and BLAST baseline methods. Points corresponding to the minimum normalized semantic distance are marked in circles on each curve. The legend provides the minimum normalized semantic distance ($S$) and coverage ($C$) for all methods. In cases where a Principal Investigator (PI) participated with multiple teams, only the results of the best scoring method are presented.\\

\paragraph{\labelstyle Calculation of the normalized remaining uncertainty-misinformation curve.} 
\begin{eqnarray*}
  nru(\tau) &=& \frac{1}{n_e}\sum_{i=1}^{n_e} 
  \frac{\sum_{f} ic(f) \cdot \mathbbm{1}\left( f \notin P_{i}(\tau) \wedge f \in T_{i} \right)}{\sum_{f} ic(f) \cdot \mathbbm{1} \left( f \in P_{i}(\tau) \vee f \in T_{i} \right)}, \quad \mbox{and}\\
  nmi(\tau) &=& \frac{1}{n_e}\sum_{i=1}^{n_e}
  \frac{\sum_{f} ic(f) \cdot \mathbbm{1}\left( f \in P_{i}(\tau) \wedge f \notin T_{i} \right)}{\sum_{f} ic(f) \cdot \mathbbm{1} \left( f \in P_{i}(\tau) \vee f \in T_{i} \right)},
\end{eqnarray*}
where $P_{i}(\tau)$ is the set of predicted terms for protein $i$ with score no less than threshold $\tau$ and $T_{i}$ is the set of true terms for protein $i$, and $n_{e}$ is the number of proteins used in a particular mode of evaluation. In the full evaluation mode $n_{e} = n$, the number of benchmark proteins, whereas in the partial evaluation mode $n_{e}$ is the number of proteins that have at least one positive predicted score.

\newpage

\noindent Supplementary Figure 8A:
\begin{center}
  \includegraphics[width=\textwidth]{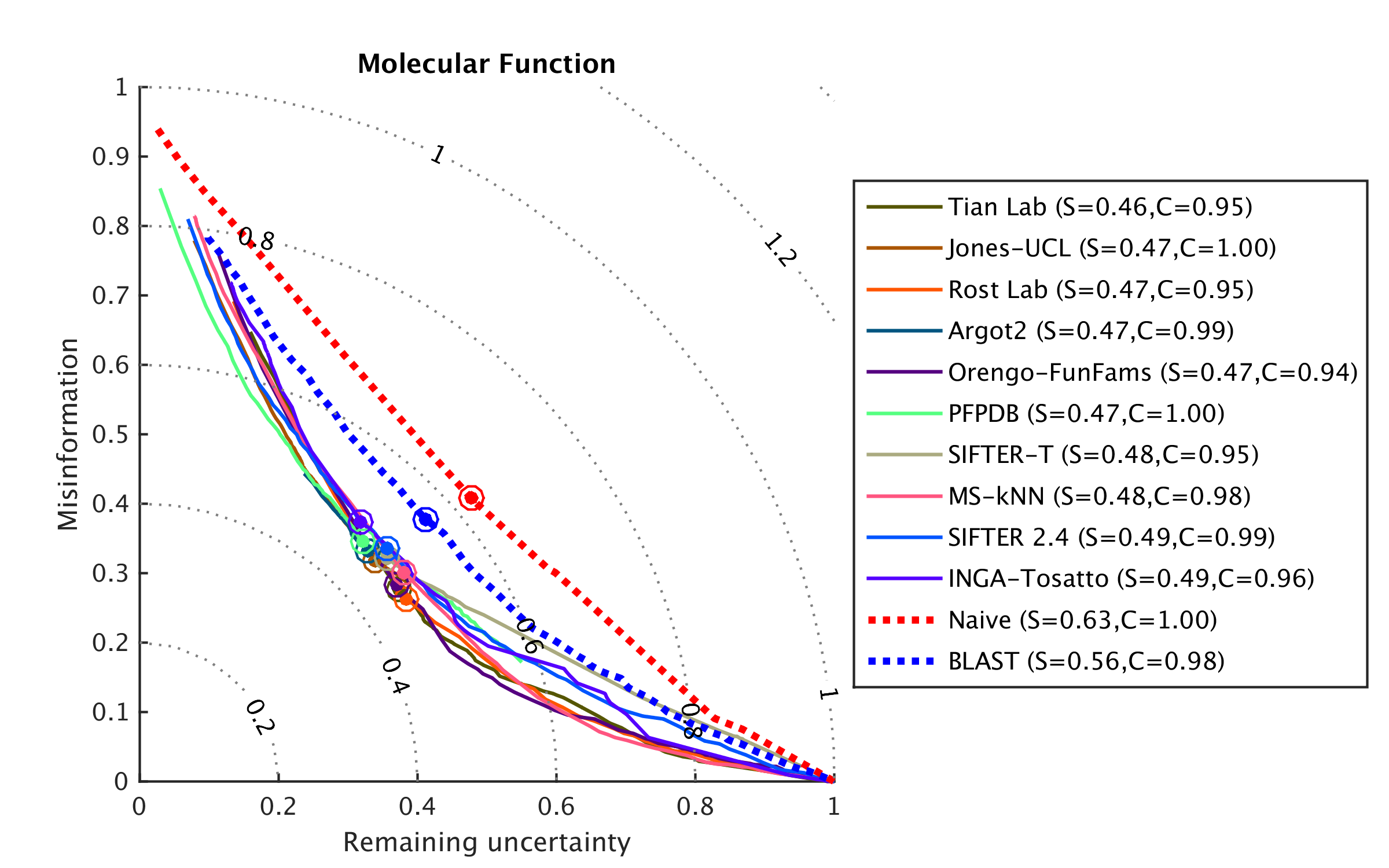}
\end{center}

\noindent Supplementary Figure 8B:
\begin{center}
  \includegraphics[width=\textwidth]{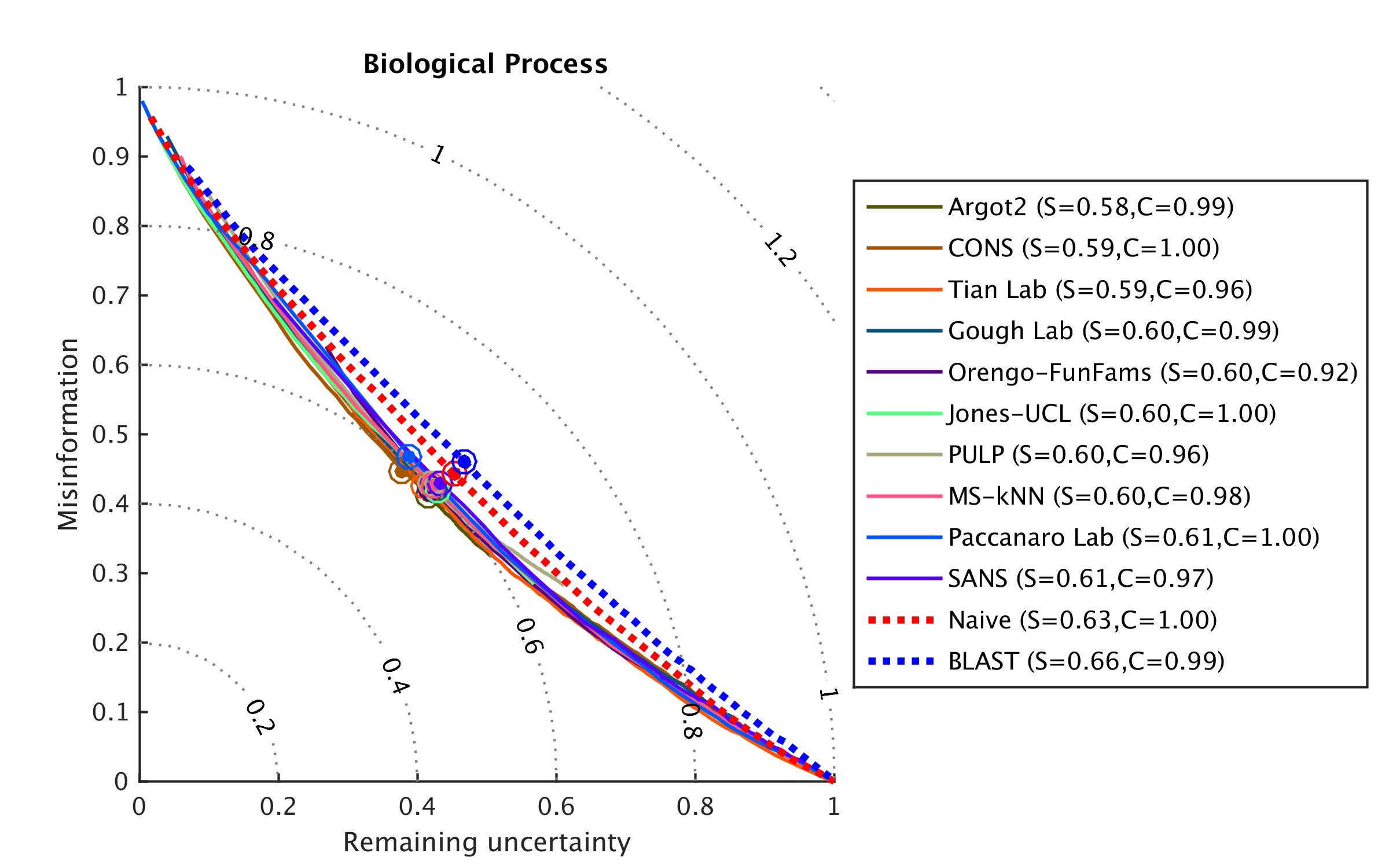}
\end{center}

\newpage

\noindent Supplementary Figure 8C:
\begin{center}
  \includegraphics[width=\textwidth]{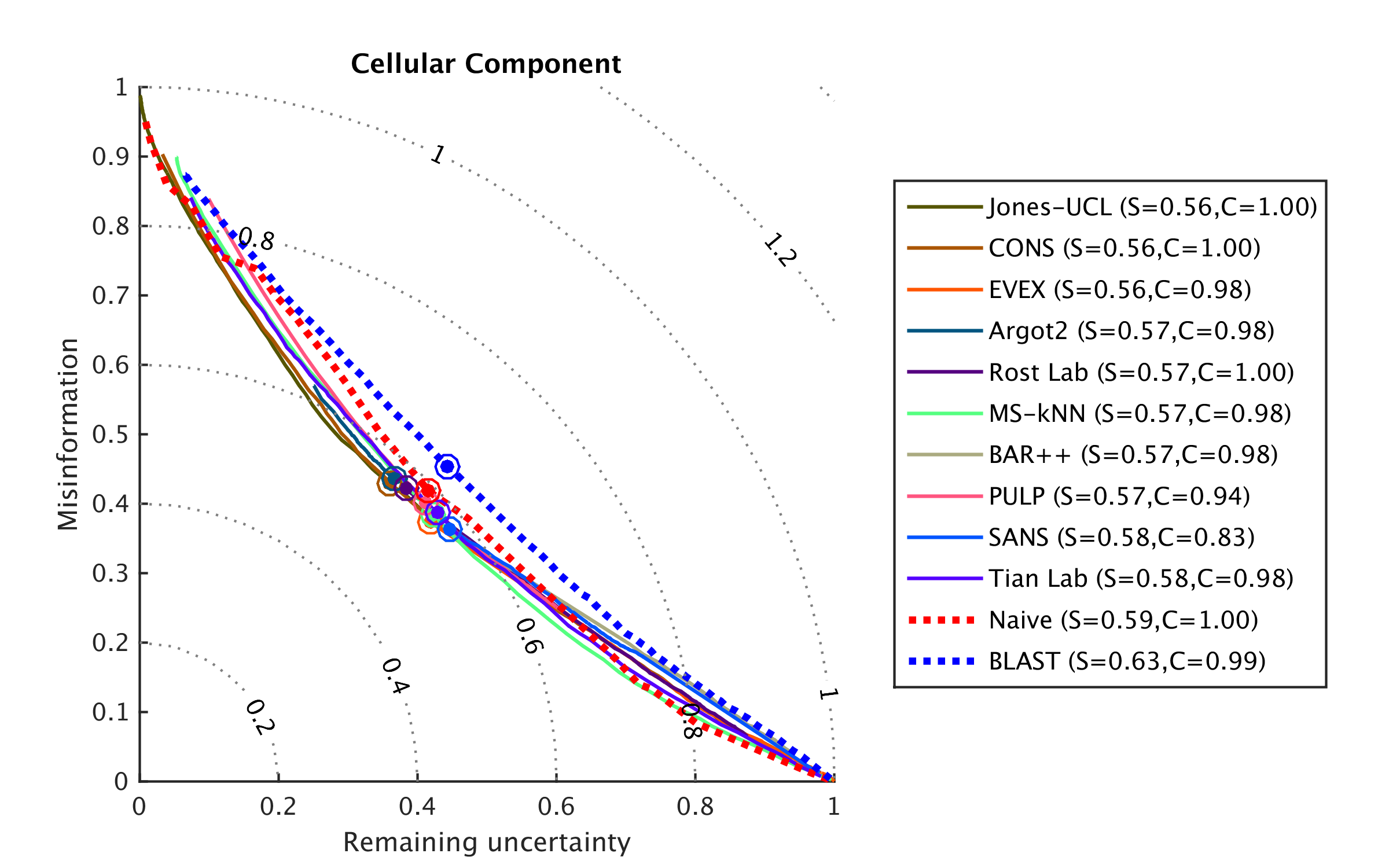}
\end{center}

\noindent Supplementary Figure 8D:
\begin{center}
  \includegraphics[width=\textwidth]{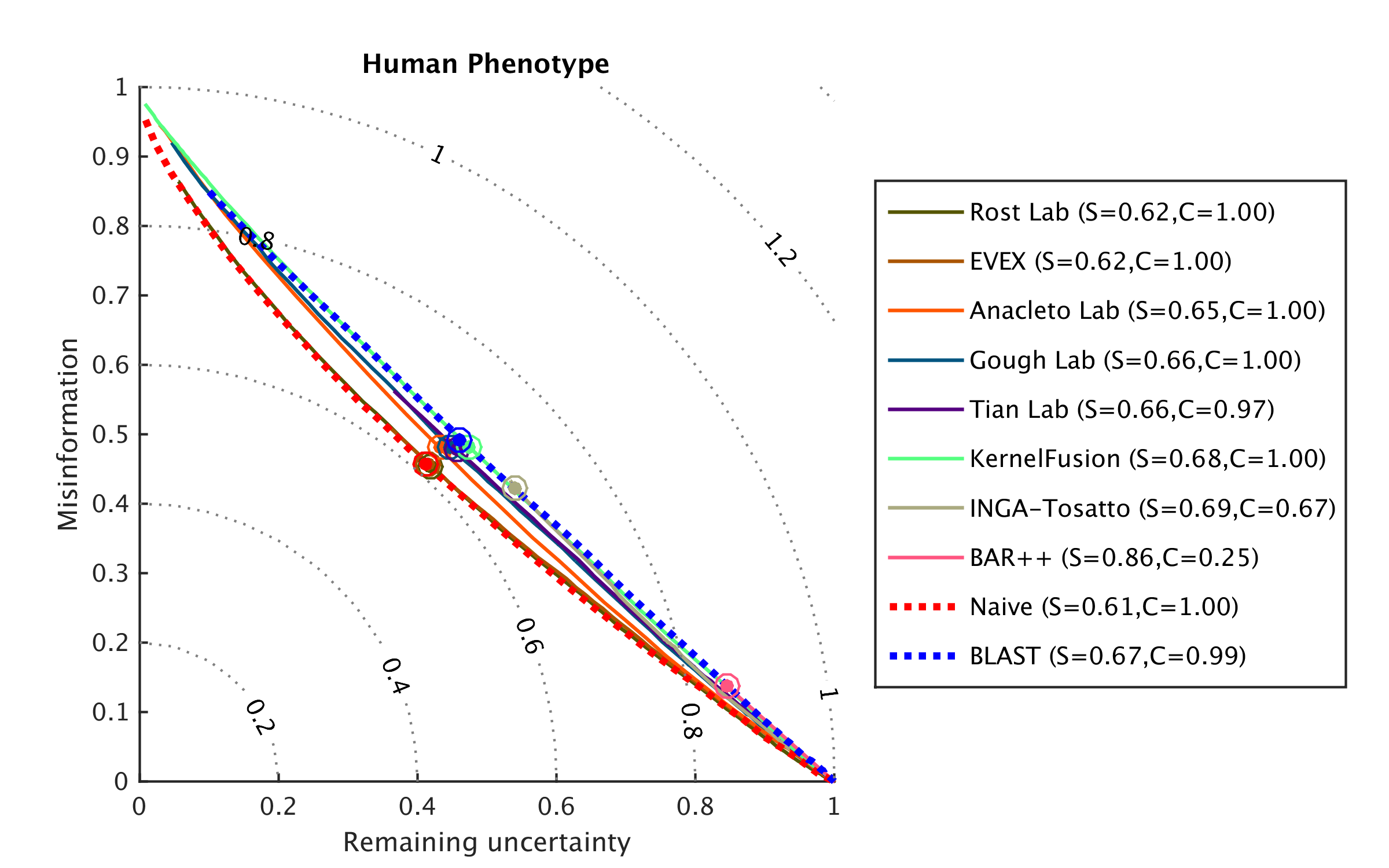}
\end{center}

\newpage

\paragraph{\labelstyle Supplementary Figure 9}
Similarity network of participated methods for (A) Molecular Function ontology, (B) Biological Process ontology, (C) Cellular Component ontology and (D) Human Phenotype ontology. For all panels, similarities are computed as the Pearson's correlation coefficient between methods with a $0.75$ cutoff for illustration purposes. A unique color is assigned to all methods submitted under the same principal investigator. Not evaluated (organizer's) methods are shown in triangles, while benchmark methods (Na\"ive and BLAST) are shown in squares. Top~10 methods are highlighted with enlarged nodes and circled in red. Edge width indicates the strength of similarity. Nodes are labelled with the name of methods followed by ``team-model'' if multiple teams/models are submitted.

\newpage

\noindent Supplementary Figure 9A:
\begin{center}
  \includegraphics[width=\textwidth]{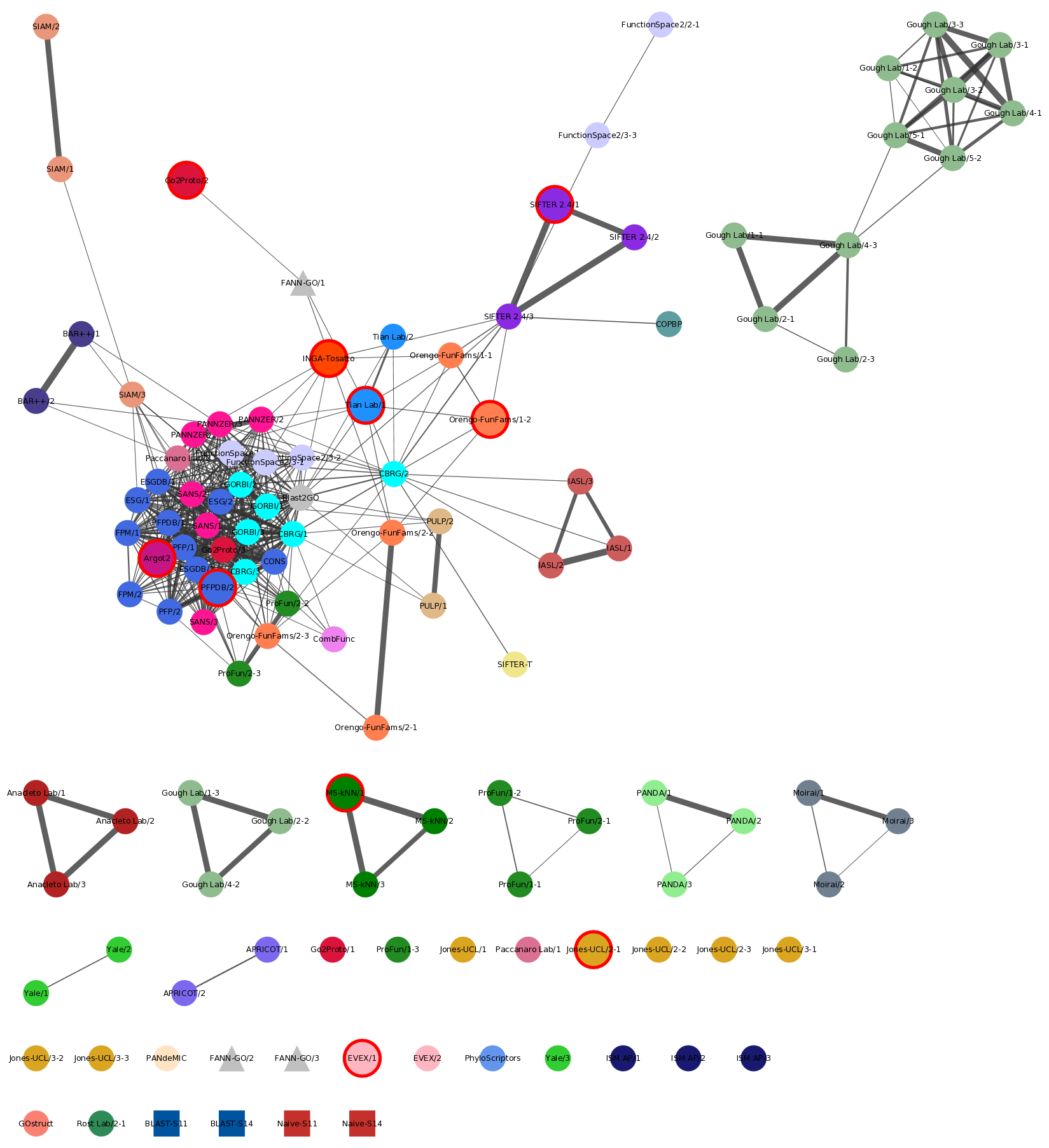}
\end{center}

\newpage

\noindent Supplementary Figure 9B:
\begin{center}
  \includegraphics[width=\textwidth]{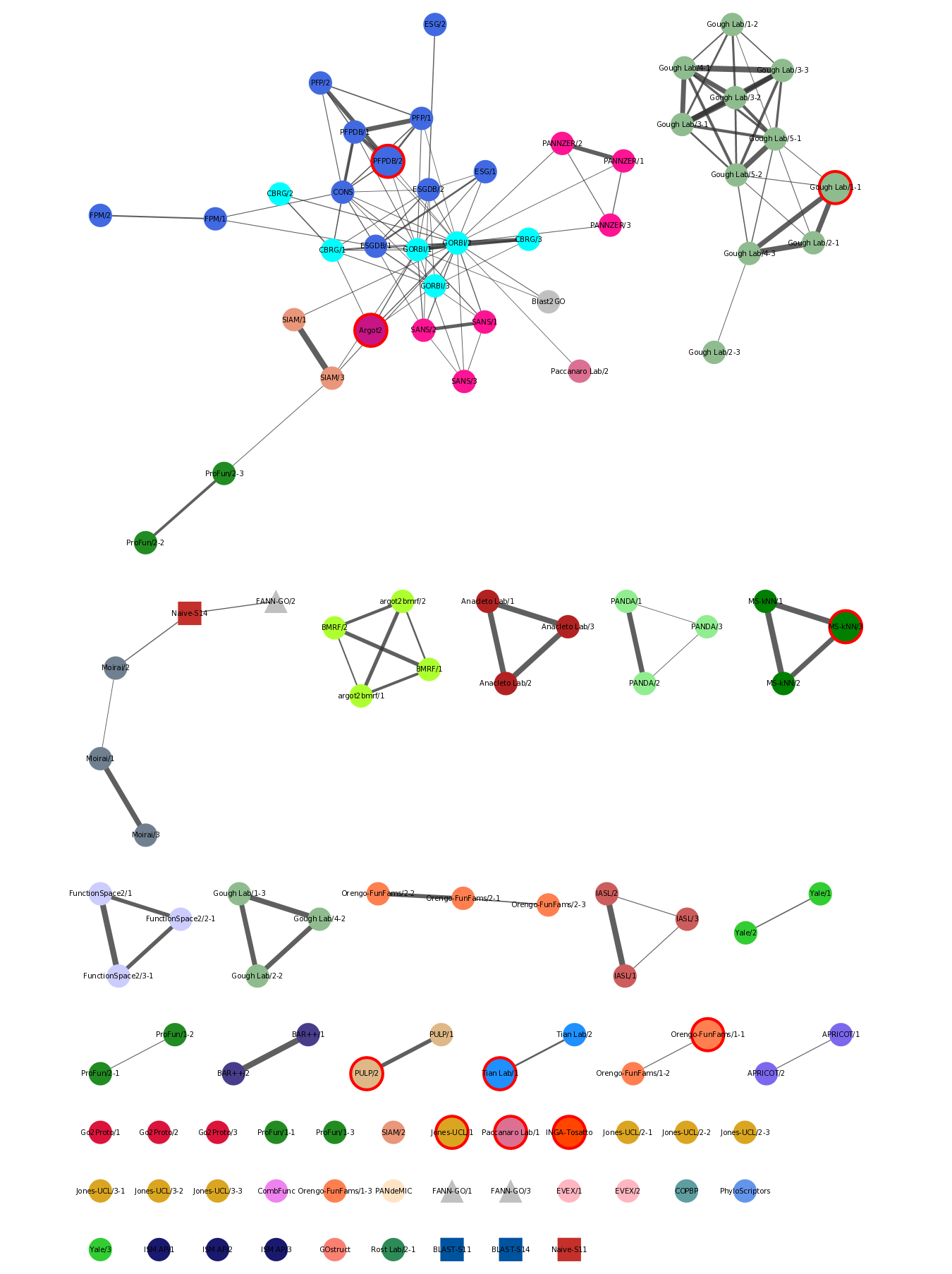}
\end{center}

\newpage

\noindent Supplementary Figure 9C:
\begin{center}
  \includegraphics[width=\textwidth]{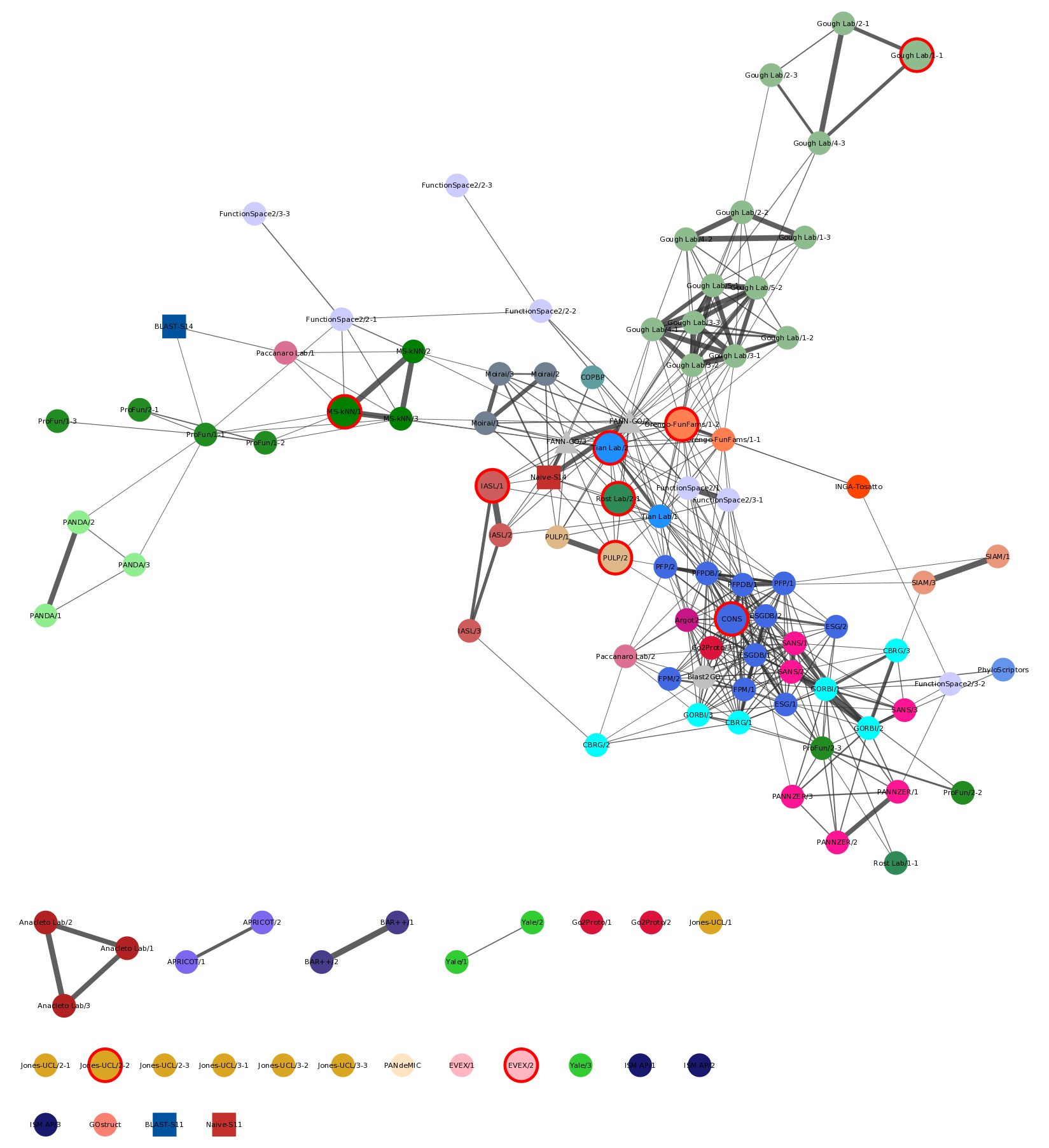}
\end{center}

\newpage

\noindent Supplementary Figure 9D:
\begin{center}
  \includegraphics[width=\textwidth]{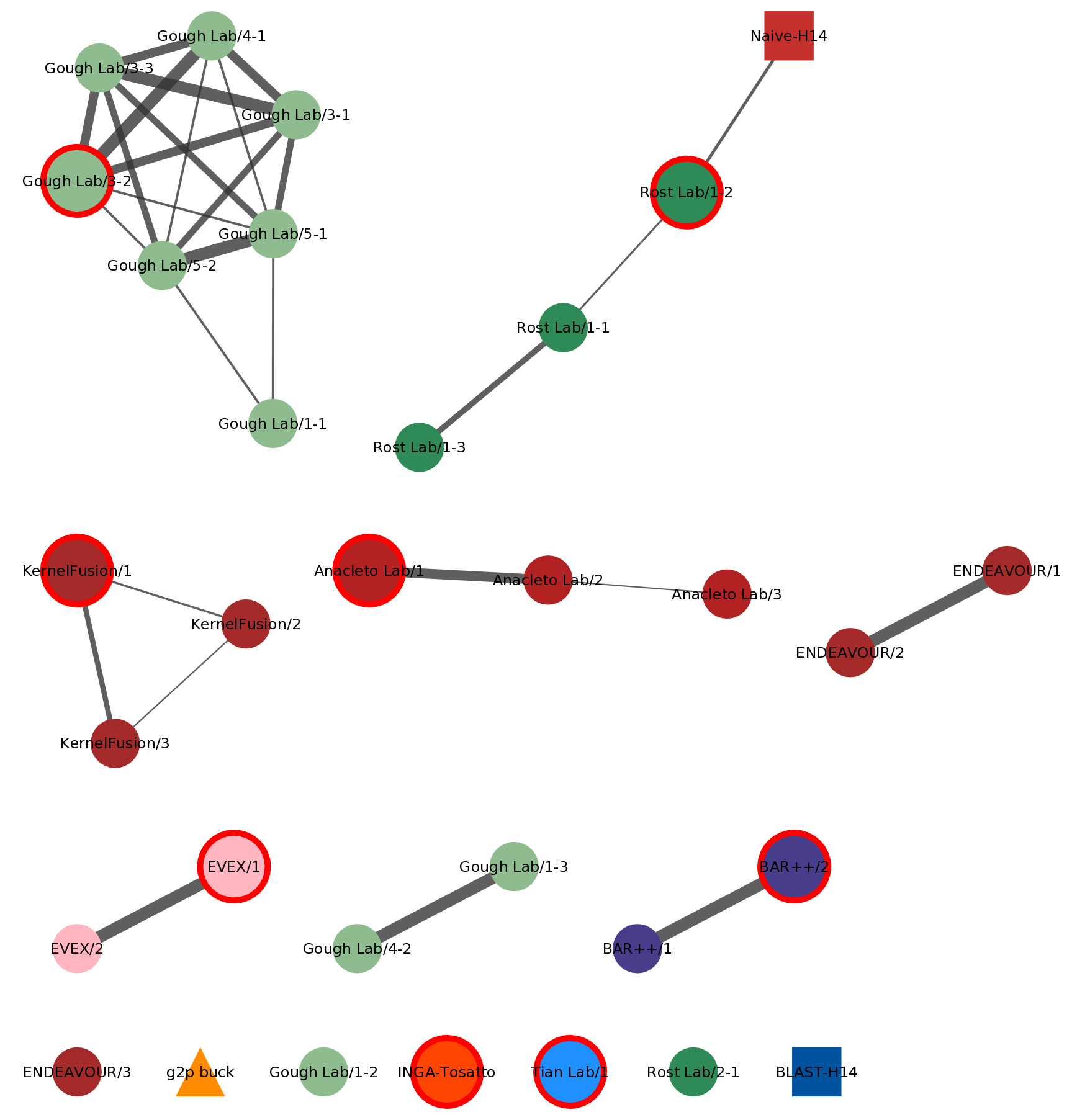}
\end{center}

\newpage

{\footnotesize
\begin{center}
\begin{longtable}{|l|l|l|l|}
    \caption{Participating methods grouped by Principal Investigators }  \\
    \hline
    \textbf{Principal Investigator} & \textbf{Method Name} & \textbf{Model (keywords)} & \textbf{Citation} \\ \hline\hline
    \endfirsthead
    \hline
    \textbf{Principal Investigator} & \textbf{Method Name} & \textbf{Model (keywords)} & \textbf{Citation} \\ \hline\hline 
    \endhead
    Asa Ben-Hur                          & GOstruct                                  & Model 1 (sa,sp,pp,pi,ge,gi,lt,gc,ml,nlp) & \cite{Sokolov2010}    \\ \hline
    Richard Bonneau     & PULP                     & Model 1 (ph,sp,pp,pi,ge,ps,pps,dp,ml,or) & {\cite{Youngs2013,Youngs2014a,Youngs2014b}}\\
                                         &                                           & Model 2 (ph,sp,pp,pi,ge,ps,pps,dp,ml)    &                       \\ \hline
    Steven Brenner      & SIFTER~2.4~*          & Model 1 (ph,ml,or,pa,ho)                 & {\cite{Sahraeian2015,Engelhardt2011}} \\
                                         &                                           & Model 2 (ph,ml,or,pa,ho)                 &                       \\
                                         &                                           & Model 3 (ph,ml,or,pa,ho)                 &                       \\ \hline
    Rita Casadio        & BAR++                    & Model 1 (sa,spa,pp,pps,ml,ho,hmm)        & {\cite{Bartoli2009, Piovesan2011}} \\
                                         &                                           & Model 2 (sa,spa,pp,pps,ml,ho,hmm)        &                       \\ \hline
    Jianlin Cheng       & ProFun                   & Model 1 (spa,sp,gi,gc,dp,gd)             & {\cite{Cao2015}} \\
                                         &                                           & Model 2 (spa,dp)                         &                       \\
                                         &                                           & Model 3 (spa,gi,gc,dp,gd)                &                       \\ \cline{2-4}
                                         & {ProFun/donet}             & Model 1 (ppa,spa)                        & {\cite{Wang2013}} \\
                                         &                                           & Model 2 (ppa,spa)                        &                       \\
                                         &                                           & Model 3 (ppa,spa)                        &                       \\ \hline
    Wyatt Clark         & {Yale}                     & Model 1 (pi)                             &                       \\
                                         &                                           & Model 2 (pi)                             &                       \\
                                         &                                           & Model 3 (pi)                             &                       \\ \hline
    {Christophe Dessimoz} & {GORBI}                    & Model 1 (ml,or,pa,ho,gc)                 & {\cite{Skunca2013}} \\
                                         &                                           & Model 2 (ml,or,pa,ho,gc)                 &                       \\
                                         &                                           & Model 3 (or,pa,ho,sa,spa,ppa,ph,hmm)     &                       \\ \cline{2-4}
                                         & {CBRG}                     & Model 1 (or,pa,ho)                       & {\cite{Altenhoff2015}} \\
                                         &                                           & Model 2 (or,pa,ho)                       &                       \\
                                         &                                           & Model 3 (or)                             &                       \\ \hline
    Tunca Dogan                          & PANdeMIC                                  & Model 1 (sa,ml,ho)                       &                       \\ \hline
    {Filip Ginter}        & {EVEX}                     & Model 1 (sa,ml,sp)                       & {\cite{Vanlandeghem2012}} \\
                                         &                                           & Model 2 (sa,ml,sp)                       &                       \\ \hline
    {Julian Gough}       & {Gough Lab/GoughGroup}     & Model 1 (sa,spa,hmm)                     &                       \\
                                         &                                           & Model 2 (pps,hmm)                        &                       \\
                                         &                                           & Model 3 (pi)                             &                       \\ \cline{2-4}
                                         & {Gough Lab/D2P2}           & Model 1 (pp,sa,spa,hmm)                  & {\cite{Oates2013}} \\
                                         &                                           & Model 2 (pp,pi)                          &                       \\
                                         &                                           & Model 3 (pp)                             &                       \\ \cline{2-4}
                                         & {Gough Lab/dcGO}           & Model 1 (pps,pp,sa,spa,hmm,pi)           & {\cite{Fang2013}} \\
                                         &                                           & Model 2 (pps,pp,sa,spa,hmm,pi)           &                       \\
                                         &                                           & Model 3 (pps,pp,sa,spa,hmm,pi)           &                       \\ \cline{2-4}
                                         & {Gough Lab/SUPERFAMILY}    & Model 1 (pps,pp,sa,spa,hmm,pi)           & {\cite{LimaMorais2011}} \\
                                         &                                           & Model 2 (pi)                             &                       \\
                                         &                                           & Model 3 (pp,sa,spa,hmm)                  &                       \\ \cline{2-4}
                                         & {Gough Lab/dcGOpredictor}  & Model 1 (pps,sa,spa,hmm,pi)              &                       \\
                                         &                                           & Model 2 (pps,sa,spa,hmm,pi)              &                       \\ \hline
    {Liisa Holm}          & {SANS}                     & Model 1 (sa)                             & {\cite{Koskinen2012}} \\
                                         &                                           & Model 2 (sa)                             &                       \\
                                         &                                           & Model 3 (sa)                             &                       \\ \cline{2-4}
                                         & {PANNZER}                  & Model 1 (sa,ph,or,pa,ho,nlp,ofi)         & {\cite{Koskinen2015}} \\
                                         &                                           & Model 2 (sa,ph,or,pa,ho,nlp,ofi)         &                       \\
                                         &                                           & Model 3 (sa,ph,or,pa,ho,nlp,ofi)         &                       \\ \hline
    {Wen-Lian Hsu}        & {IASL}                     & Model 1 (sa,spa,sp)                      &                       \\
                                         &                                           & Model 2 (sa,spa,sp)                      &                       \\
                                         &                                           & Model 3 (sa,spa,sp)                      &                       \\ \hline
    {David Jones}         & Jones-UCL/jfpred-RF                       & Model 1 (hmm,ppa,sp,pi,or,lt,ml)         & {\cite{Cozzetto2013}} \\ \cline{2-3}
                                         & {Jones-UCL/jfpred-FP}      & Model 1 (hmm,ppa,sp,pi,or,lt,ml)         &                       \\
                                         &                                           & Model 2 (sp,pp,pps,ml)                   &                       \\
                                         &                                           & Model 3 (sp,pp,pps,ml)                   &                       \\ \cline{2-3}
                                         & {Jones-UCL/jfpred-PB}      & Model 1 (hmm,ppa,sp,pi,or,lt,ml)         &                       \\
                                         &                                           & Model 2 (sa,spa)                         &                       \\
                                         &                                           & Model 3 (hmm,ppa)                        &                       \\ \hline
    {Daisuke Kihara}   & {ESG}                       & Model 1 (sa)                              & {\cite{Chitale2009}} \\
                                       &                                            & Model 2 (sa)                              &                       \\ \cline{2-4}
                                       & CONS                                       & Model 1 (sa)                              & {\cite{Khan2015}} \\ \cline{2-3}
                                       & {FPM}                       & Model 1 (sa)                              &                       \\
                                       &                                            & Model 2 (sa)                              &                       \\ \cline{2-3}
                                       & {PFPDB}                     & Model 1 (sa)                              &                       \\
                                       &                                            & Model 2 (sa)                              &                       \\ \cline{2-3}
                                       & {ESGDB}                     & Model 1 (sa)                              &                       \\
                                       &                                            & Model 2 (sa)                              &                       \\ \cline{2-4}
                                       & {PFP}                       & Model 1 (sa)                              & {\cite{Hawkins2006,Hawkins2009}} \\
                                       &                                            & Model 2 (sa)                              &                       \\ \hline
    Sean Mooney                        & g2p buck (not evaluated)                   & Model 1 (N/A)                             &                       \\ \hline
    {Michal Linial}     & {Go2Proto}                  & Model 1 (sa,sp,php,pp,cm,ml,or,pa,ho,ofi) &                       \\
                                       &                                            & Model 2 (sa,sp,php,pp,cm,ml,or,pa,ho,ofi) &                       \\
                                       &                                            & Model 3 (sa,sp,php,pp,cm,ml,or,pa,ho,ofi) &                       \\ \hline
    {Yves Moreau}       & {ENDEAVOUR}                 & Model 1 (sa,ph,pi,ge,lt,ml,ofi)           & {\cite{Aerts2006}} \\
                                       &                                            & Model 2 (sa,ph,pi,ge,lt,ml,ofi)           &                       \\
                                       &                                            & Model 3 (sa,ph,pi,ge,lt,ml,ofi)           &                       \\ \cline{2-4}
                                       & {KernelFusion}              & Model 1 (sa,pi,ge,lt,ml,ofi)              & {\cite{Zakeri2011,DeBie2007}}\\
                                       &                                            & Model 2 (sa,pi,ge,lt,ml,ofi)              &                       \\
                                       &                                            & Model 3 (sa,pi,ge,lt,ml,ofi)              &                       \\ \hline
    {Christine Orengo}  & {Orengo-FunFams/MDA}        & Model 1 (ml)                              & {\cite{Das2015b}} \\
                                       &                                            & Model 2 (sp)                              &                       \\
                                       &                                            & Model 3 (pi)                              &                       \\ \cline{2-3}
                                       & {Orengo-FunFams}            & Model 1 (spa,ppa,ho,hmm)                  &                       \\
                                       &                                            & Model 2 (spa,ppa,ho,hmm)                  &                       \\
                                       &                                            & Model 3 (spa,ppa,ho,hmm)                  &                       \\ \hline
    {Alberto Paccanaro} & {Paccanaro Lab}             & Model 1 (sa,spa,pi,ge,lt,gc,ml,or.ho)     &                       \\
                                       &                                            & Model 2 (spa,hmm,ml)                      &                       \\ \hline
    {Paul Pavlidis}     & {Moirai}                    & Model 1 (ofi)                             &                       \\
                                       &                                            & Model 2 (ofi)                             &                       \\
                                       &                                            & Model 3 (ofi)                             &                       \\ \hline
    {Predrag Radivojac} & {FANN-GO (not evaluated)}   & Model 1 (sa,ml)                           & {\cite{Clark2011}} \\
                                       &                                            & Model 2 (sa,ml)                           &                       \\
                                       &                                            & Model 3 (sa,ml)                           &                       \\ \hline
    {Burkhard Rost}     & {Rost Lab}                  & Model 1 (sa,spa,ppa,sp,dp,ml)             & {\cite{Goldberg2014}} \\
                                       &                                            & Model 2 (sa,spa,ppa,sp,dp,ml)             &                       \\
                                       &                                            & Model 3 (sa,spa,ppa,sp,dp,ml)             &                       \\ \cline{2-4}
                                       & Rost Lab/metastudent2                      & Model 1 (sa,ml,or,pa,ho)                  & \cite{Hamp2013}       \\ \hline
    Asaf Salamov                       & COPBP                                      & Model 1 (N/A)                             &                       \\ \hline
    Fran Supek                         & PhyloScriptors                             & Model 1 (ph,gc,ml,pa,or)                  &                       \\ \hline
    {Weidong Tian}      & {Tian Lab}                  & Model 1 (sa)                              & {\cite{Gong2015}} \\
                                       &                                            & Model 2 (sa)                              &                       \\ \hline
    Stefano Toppo                      & Argot2                                     & Model 1 (sa,spa)                          & \cite{Falda2012}      \\ \hline
    {Toppo/van Dijk~\dag}  & {argot2bmrf}                & Model 1 (sp,pi,ge,gi,ml,sa,spa)           &                       \\
                                       &                                            & Model 2 (sp,pi,ge,gi,ml,sa,spa)           &                       \\ \hline
    Silvio Tosatto                     & INGA-Tosatto                               & Model 1 (hmm,ppa,sa,pi)                   & \cite{Piovesan2015}   \\ \hline
    {Michael Tress}     & {SIAM}                      & Model 1 (sa,ho,sp,ps,php,spa,ppa,sta,cm)  & {\cite{Maietta2014}} \\
                                       &                                            & Model 2 (ps,php,spa,ppa,sta,cm)           &                       \\
                                       &                                            & Model 3 (sa,ho,sp)                        &                       \\ \hline
    Hafeez Ur Rehman                   & PFPPipeLine                                & Model 1 (sa,pi,ml,ho,ofi)                 & \cite{Benso2013}      \\ \hline
    {Giorgio Valentini} & {Anacleto Lab}              & Model 1 (ml,sa)                           & {\cite{Re2012}} \\
                                       &                                            & Model 2 (ml,sa)                           &                       \\
                                       &                                            & Model 3 (ml,sa)                           &                       \\ \hline
    {Aalt-Jan van Dijk} & {BMRF}                      & Model 1 (sp,pi,ge,gi,ml)                  & {\cite{Kourmpetis2010, Kourmpetis2011}} \\
                                       &                                            & Model 2 (sp,pi,ge,gi,ml)                  &                       \\ \hline
    {Nevena Veljkovic}  & {ISM AP}                    & Model 1 (ppa,php)                         &                       \\
                                       &                                            & Model 2 (ppa,php,ge)                      &                       \\
                                       &                                            & Model 3 (ppa,php,ge)                      &                       \\ \hline
    Ricardo Vencio                    & SIFTER-T                     & Model 1 (spa,ml,ho)                   s    & \cite{Almeida2015}    \\ \hline
    {J\"org Vogel}     & {APRICOT}     & Model 1 (ho,hmm,ppa,pp)                  &                       \\
                                      &                              & Model 2 (ho,hmm,ppa,pp)                  &                       \\ \hline
    {Slobodan Vucetic} & {MS-kNN}      & Model 1 (ml,sa,ge)                       & {\cite{Lan2013}} \\
                                      &                              & Model 2 (ml,sa,ge)                       &                       \\
                                      &                              & Model 3 (ml,sa,ge)                       &                       \\ \hline
    {Zheng Wang}       & {PANDA}       & Model 1 (spa,ppa,ph,or,pa,ho)            &                       \\
                                      &                              & Model 2 (spa,ppa,ph,or,pa,ho)            &                       \\
                                      &                              & Model 3 (spa,ppa,ph,or,pa,ho)            &                       \\ \hline
    Mark Wass                         & CombFunc                     & Model 1 (spa,sa,ml,ge,pi)                & \cite{Wass2012}       \\ \hline
    N/A~\ddag                         & Blast2GO                     & Model 1 (sa)                             & \cite{Conesa2005}     \\ \hline
\end{longtable}
\end{center}

\noindent * SIFTER is expected to work well on microbial proteins \\
\dag This is a joint group of Stefano Toppo and Aalt-Jan van Dijk\\
\ddag Blast2GO predictions were downloaded from the website https://www.blast2go.com one week before the prediction deadline and
converted into appropriate submission format by the CAFA organizers
} 

\begin{table}[!htbp]
    \caption{Keywords}
  \begin{tabular}{|c|l|c|l|}
    \hline
    \textbf{Code} & \textbf{Keyword}            & \textbf{Code} & \textbf{Keyword} \\ \hline
    sa            & sequence alignment         & sta           & structure alignment \\ \hline
    spa           & sequence-profile alignment & cm            & comparative model \\ \hline
    ppa           & profile-profile alignment  & pps           & predicted protein structure \\ \hline
    ph            & phylogeny                  & dp            & \textit{de novo} prediction \\ \hline
    sp            & sequence properties        & ml            & machine learning \\ \hline
    php           & physicochemical properties & gne           & genome environment \\ \hline
    pp            & predicted properties       & op            & operon \\ \hline
    pi            & protein interactions       & or            & ortholog \\ \hline
    ge            & gene expression            & pa            & paralog \\ \hline
    ms            & mass spectrometry          & ho            & homolog \\ \hline
    gi            & genetic interactions       & hmm           & hidden Markov model \\ \hline
    ps            & protein structure          & cd            & clinical data \\ \hline
    lt            & literature                 & gd            & genetic data \\ \hline
    gc            & genomic context            & nlp           & natural language processing \\ \hline
    sy            & synteny                    & ofi           & other functional information \\ \hline
  \end{tabular}
\end{table}
\newpage
